\RequirePackage{fix-cm}
\documentclass[twocolumn,epjc3]{svjour3}  
\smartqed  
\RequirePackage{graphicx}
\RequirePackage[colorlinks,citecolor=blue,urlcolor=blue,linkcolor=blue]{hyperref} 
\usepackage{color}
\usepackage{ifthen}
\usepackage{multirow}
\newboolean{makefigures}                
\setboolean{makefigures}{true}          
\usepackage{upgreek}
\newcommand{\ctsper}      {cts/(keV$\cdot$kg$\cdot$yr)}

\newcommand{\tctsper}     {{$10^{-2}$~cts/(keV$\cdot$kg$\cdot$yr)}}

\newcommand{\dctsper}     {{$10^{-3}$~cts/(keV$\cdot$kg$\cdot$yr)}}
\newcommand{\pIIbi}       {{$10^{-3}$~cts/(keV$\cdot$kg$\cdot$yr)}}
\newcommand{\vctsper}     {{$10^{-4}$~cts/(keV$\cdot$kg$\cdot$yr)}}

\newcommand{\kgyr}        {{kg$\cdot$yr}}

\newcommand{\cum}         {{m$^3$}}
\newcommand{\mubq}        {{$\upmu$Bq}}
\newcommand{\mum}         {{$\upmu$m}}
\newcommand{\mus}         {{$\upmu$s}}

\newcommand{\gam}         {$\gamma$}

\newcommand{\qbb}         {{$Q_{\beta\beta}$}}
\newcommand{\thalfzero}   {${T^{0\nu}_{1/2}}$}
\newcommand{\thalftwo}    {${T^{2\nu}_{1/2}}$}

\newcommand{\onbb}        {{$0\nu\beta\beta$}}
\newcommand{\nnbb}        {{$2\nu\beta\beta$}}
\newcommand{\twonu}       {{$2\nu\beta\beta$}}

\newcommand{\fgesix}      {\mbox{$f_{76}$}}
\newcommand{\fgesixi}     {\mbox{$f_{76,i}$}}
\newcommand{\actmass}     {\mbox{$M_{act}$}}

\newcommand{\factmass}    {\mbox{$f_{av}$}}

\newcommand{\factvoli}    {\mbox{$f_{av,i}$}}

\newcommand{\up}          {\rule{0mm}{5mm}}

\newcommand{\etal}        {\textit{et al.}}

\newcommand{\gerda}       {\textsc{Gerda}}

\newcommand{\GERDA}       {\mbox{\textsc{Gerda}}}  
\newcommand{\igex}        {\textsc{Igex}}

\newcommand{\hdm}         {\textsc{HdM}}

\newcommand{\geant}       {\textsc{Geant4}}

\newcommand{\mage}        {\textsc{MaGe}}

\newcommand{\gesix}       {{$^{76}$Ge}}
\newcommand{\gess}        {{$^{76}$Ge}}

\newcommand{\thzza}       {{$^{228}$Th}}

\newcommand{\kvn}         {{$^{40}$K}}
\newcommand{\kvz}         {{$^{42}$K}}

\newcommand{\Rn}          {$^{222}$Rn}
\newcommand{\Ra}          {$^{226}$Ra}
\newcommand{\Po}          {$^{210}$Po}

\newcommand{\Bi}          {$^{214}$Bi}

\newcommand{\Th}          {$^{228}$Th}
\newcommand{\Tl}          {$^{208}$Tl}

\newcommand{\Co}          {$^{60}$Co}
\newcommand{\exposure}    {\mbox{$\cal E$}}

\newcommand{\effpsd}      {\mbox{$\varepsilon_{psd}$}}
\newcommand{\effres}      {\mbox{$\varepsilon_{res}$}}
\newcommand{\efffep}      {\mbox{$\varepsilon_{fep}$}}

\newcommand{\efffepi}     {\mbox{$\varepsilon_{fep,i}$}}
\journalname{Eur. Phys. J. C}
\begin{document}

\title{The background in the $0\nu\beta\beta$ experiment \mbox{\sc{Gerda}}
}

\titlerunning{The \textsc{Gerda} background }   

\author{
M.~Agostini\thanksref{TUM} \and
M.~Allardt\thanksref{DD} \and
E.~Andreotti\thanksref{GEEL,TU} \and
A.M.~Bakalyarov\thanksref{KU} \and
M.~Balata\thanksref{ALNGS} \and
I.~Barabanov\thanksref{INR} \and
M.~Barnab\'e Heider\thanksref{HD,TUM,nowCAN} \and
N.~Barros\thanksref{DD} \and
L.~Baudis\thanksref{UZH} \and
C.~Bauer\thanksref{HD}
% ...
}

\authorrunning{the \textsc{Gerda} collaboration}

\thankstext{nowCAN}{\emph{Present Address:} CEGEP St-Hyacinthe,
 Qu{\'e}bec, Canada}
\thankstext{nowLNGS}{\emph{Present Address:} INFN  LNGS, Assergi, Italy}
\thankstext{alsoFI}{\emph{also at:} Universit{\`a} di Firenze, Italy}
\thankstext{alsoMIPT}{\emph{also at:} Moscow Inst. of Physics and Technology,
  Russia} 
\thankstext{alsoIUN}{\emph{also at:} Int. Univ. for Nature, Society and
    Man ``Dubna''} 
\thankstext{nowSJU}{\emph{Present Address:} Shanghai Jiaotong University,
  Shanghai, China} 
\thankstext{nowUNC}{\emph{Present Address:} University North Carolina, Chapel
  Hill, USA} 
\thankstext{AQU}{\emph{Present Address:} University of L'Aquila, Dipartimento
        di Fisica, L'Aquila, Italy}
\thankstext{corrauthor}{\emph{Correspondence},
                                email: bela.majorovits@mpp.mpg.de}
\institute{
INFN Laboratori Nazionali del Gran Sasso, LNGS, Assergi, Italy\label{ALNGS} \and
Institute of Physics, Jagiellonian University, Cracow, Poland\label{CR} \and
Institut f{\"u}r Kern- und Teilchenphysik, Technische Universit{\"a}t Dresden,
      Dresden, Germany\label{DD} \and
Joint Institute for Nuclear Research, Dubna, Russia\label{JINR} \and
Institute for Reference Materials and Measurements, Geel,
     Belgium\label{GEEL} \and
Max Planck Institut f{\"u}r Kernphysik, Heidelberg, Germany\label{HD} \and
Dipartimento di Fisica, Universit{\`a} Milano Bicocca,
     Milano, Italy\label{MIBF} \and
INFN Milano Bicocca, Milano, Italy\label{MIBINFN} \and
Dipartimento di Fisica, Universit{\`a} degli Studi di Milano e INFN Milano,
    Milano, Italy\label{MILUINFN} \and
Institute for Nuclear Research of the Russian Academy of Sciences,
    Moscow, Russia\label{INR} \and
Institute for Theoretical and Experimental Physics,
    Moscow, Russia\label{ITEP} \and
National Research Centre ``Kurchatov Institute'', Moscow, Russia\label{KU} \and
Max-Planck-Institut f{\"ur} Physik, M{\"u}nchen, Germany\label{MPIP} \and
Physik Department and Excellence Cluster Universe,
    Technische  Universit{\"a}t M{\"u}nchen, Germany\label{TUM} \and
Dipartimento di Fisica e Astronomia dell{`}Universit{\`a} di Padova,
    Padova, Italy\label{PDUNI} \and
INFN  Padova, Padova, Italy\label{PDINFN} \and
Physikalisches Institut, Eberhard Karls Universit{\"a}t T{\"u}bingen,
    T{\"u}bingen, Germany\label{TU} \and
Physik Institut der Universit{\"a}t Z{\"u}rich, Z{\"u}rich,
    Switzerland\label{UZH}
}

\date{Received: date / Accepted: date}

\maketitle

\begin{abstract}
 The GERmanium Detector Array (\GERDA) experiment at the Gran Sasso
 underground laboratory (LNGS) of INFN is searching for neutrinoless double
 beta (\onbb) decay of \gesix. The signature of the signal is a monoenergetic
 peak at 2039~keV, the \qbb\ value of the decay.  To avoid bias in the signal
 search, the present analysis does not consider all those events, that fall in
 a 40~keV wide region centered around \qbb.  The main parameters needed for
 the \onbb\ analysis are described.

 A background model was developed to describe the observed energy spectrum.
 The model contains several contributions, that are expected on the basis of
 material screening or that are established by the observation of
 characteristic structures in the energy spectrum.  The model predicts a flat
 energy spectrum for the blinding window around \qbb\ with a background index
 ranging from 17.6 to 23.8$\cdot$\pIIbi. A part of the data not considered before has
 been used to test if the predictions of the background model are
 consistent. The observed number of events in this energy region is consistent
 with the background model. The background at \qbb\ is dominated by close
 sources, mainly due to \kvz, \Bi, \thzza, $^{60}$Co and $\alpha$ emitting
 isotopes from the \Ra\ decay chain. The individual fractions depend on the
 assumed locations of the contaminants. It is shown, that after
 removal of the known \gam\ peaks, the energy spectrum can be fitted in an
 energy range of 200~keV around \qbb\ with a constant background. This gives a
 background index consistent with the full model and uncertainties of the same  
 size.

\end{abstract}
\section{Introduction}
 \label{sec:intro}

 Some even-even nuclei are energetically forbidden to decay via single $\beta$
 emission, while the decay via emission of two electrons and two neutrinos is
 energetically allowed. The experimentally observed neutrino accompanied double 
 beta (\twonu) decay is
 a second order weak process with half lives of the order of
 10$^{18-24}$~yr~\cite{2vbb_half_lives}. The decay process without neutrino
 emission, neutrinoless double beta (\onbb) decay, is of fundamental relevance
 as its observation would imply lepton number violation indicating physics
 beyond the standard model of particle physics.  The
 \gerda\ experiment~\cite{gerda_tec} is designed to search for  \onbb\ decay in the
 isotope \gesix. This process is identified by a monoenergetic line in the
 energy sum spectrum of the two electrons at 2039~keV \cite{qbb}, the \qbb-value of the
 decay.  The two precursor experiments, the Heidelberg Moscow (\hdm) and the
 International Germanium EXperiment (\igex), have set limits on the half live
 \thalfzero\ of \onbb\ decay \thalfzero$>$\,1.9$\cdot10^{25}$~yr~\cite{hdm} and
 \thalfzero$>$\,1.6$\cdot10^{25}$~yr~\cite{igex} (90\,\% C.L.), respectively.  A
 subgroup of the \hdm\ experiment claims to have observed \onbb\ decay with a
 central value of the half life of
 \thalfzero\,=\,1.19$\cdot10^{25}$~yr~\cite{claim}.  This result was later refined
 using pulse shape discrimination (PSD)~\cite{claim2006} yielding a half life of
 \thalfzero\,=\,2.23$\cdot10^{25}$~yr. Several inconsistencies in the latter
 analysis have been pointed out in Ref.~\cite{bernhard}.

 The design of the \gerda\ apparatus for the search of  \onbb\ decay
 follows the suggestion to operate high purity germanium (HPGe) detectors
 directly in a cryogenic liquid that serves as cooling medium and
 simultaneously as ultra-pure shielding against external
 radiation~\cite{heusser}.  \gerda\ aims in its Phase~I to test the
 \hdm\ claim of a signal and, in case of no confirmation, improve
 this limit by an order of magnitude in Phase~II of the experiment.

 Prerequisites for rare-event studies are $(i)$ extremely low
 backgrounds, usually expressed in terms of a background index (BI) measured
 in \ctsper, and $(ii)$ large masses and long measuring times, expressed as
 exposure~\exposure. Reducing the background and establishing a radio-pure
 environment is an experimental challenge. Proper analysis methods must be
 applied to guarantee an unbiased analysis.  The \gerda\ collaboration has
 blinded a region of \qbb$\,\pm$\,20~keV during the data taking
 period~\cite{gerda_tec}.  During this time, analysis methods and background
 models have been developed and tested. The latter is described in this paper
 together with other parameters demonstrating the data quality.

\begin{figure}[h!]
\begin{center}
\ifmakefigures%
  \includegraphics[width=1.0\columnwidth]{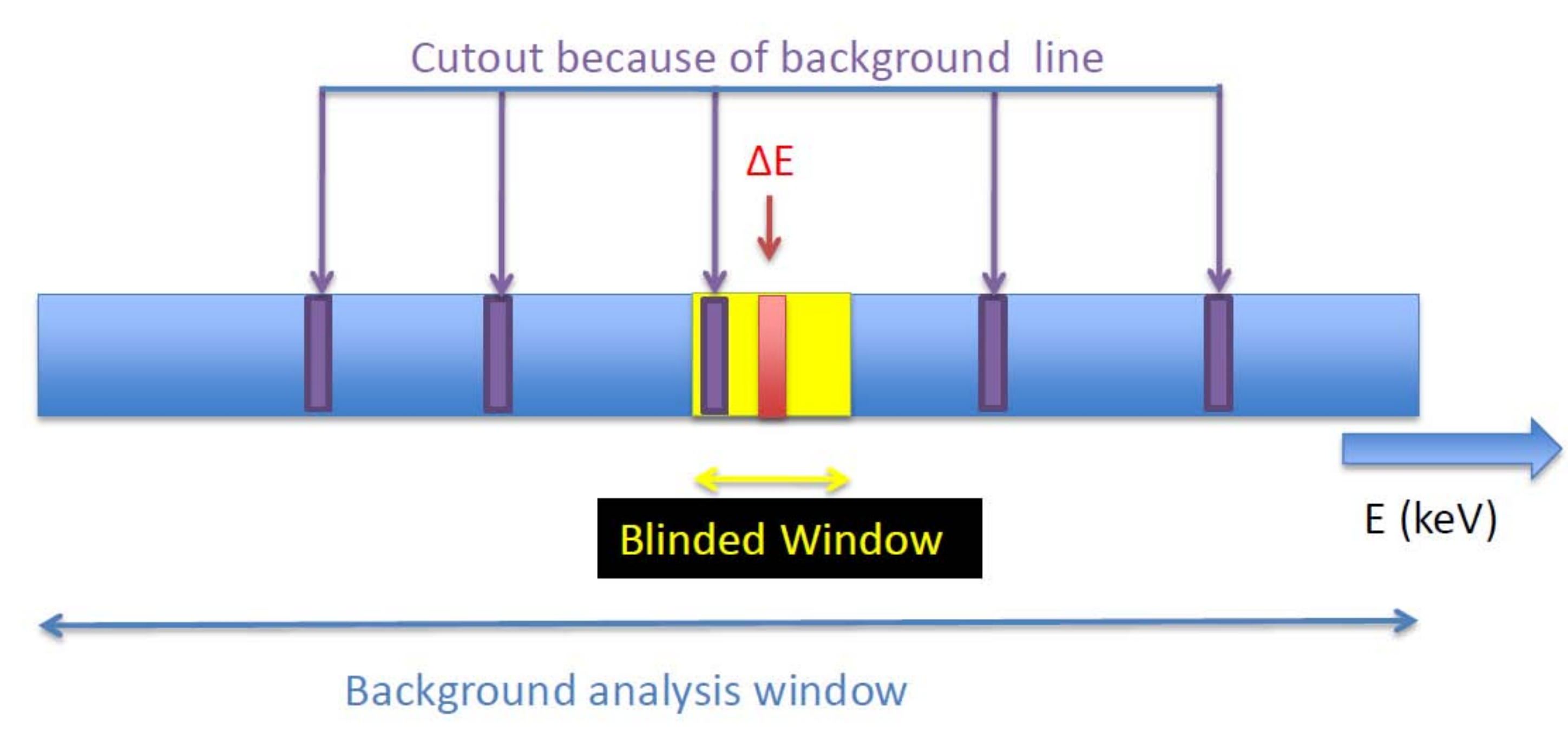}
\fi%
\caption{ \label{fig:windows}
         Representation of energy spectra for definition of the energy windows used in the blind analysis.
}
\end{center}
\end{figure}

 The raw data are converted into energy spectra. If the energies of individual
 events fall within a range \qbb\,$\pm$\,20~keV, these events are stored during
 the blinding mode in the backup files only. They are not converted to the data
 file that is available for analysis. 
 This blinding window is schematically
 represented in Fig.~\ref{fig:windows} by the yellow area, including the red
 range. After fixing the calibration parameters and the background 
 model, the blinding window was partially opened except the peak range 
 at \qbb, indicated in red in Fig.~\ref{fig:windows}.
The blue range 
 covers the energies from 100~keV to 7.5~MeV. The data from this energy range
 were available 
for analysis all the
 time. The observable $\gamma$ lines can be used to identify background
 sources.
A range between 1930 keV and 2190~keV was then used to determine the BI. The energy regions around significant $\gamma$ lines are excluded in the latter, 
as shown schematically in Fig ~\ref{fig:windows}.

Data were taken until 
until May 2013.
 These data provide the exposure~\exposure\ for Phase~I. The data used in this
 analysis of the background are a subset containing data taken until March
 2013.  

 The extraction of the background model is described in detail in this
 paper. In the process, the necessary parameters are defined for the upcoming
 \onbb\ analysis. An important feature is the stable performance of the
 germanium detectors enriched in \gesix; this is demonstrated for the complete
 data taking period (sec.~\ref{sec:exp}).

 The data with 
exposure \exposure\ is used to interpolate the
 background within $\Delta E$.  The expectation for the BI is given in this
 paper before unblinding the data in the energy range $\Delta E$, the region
 of highest physics interest.  

The paper is organized as follows: after
 presenting the experimental details, particularly on the detectors used in
 Phase~I of the \gerda\ experiment, coaxial and BEGe type
 (sec.~\ref{sec:exp}), the spectra and the identified background sources will
 be discussed (sec.~\ref{sec:spectradata} and~\ref{sec:back}). These are the
 basic ingredients for the background decomposition for the coaxial detectors
 (sec.~\ref{sec:modeling}) and for the BEGe detectors
 (sec.~\ref{sec:begemodel}). The models work well for both types of
 detectors. After cross checks of the background model
 (sec.~\ref{sec:cross-checks}) the paper concludes with the prediction for the
 background at \qbb\ and the prospective sensitivity of \gerda\ Phase~I
 (sec.~\ref{sec:extrapol}).

\section{The experiment}
 \label{sec:exp}

 This section briefly recalls the main features of the \gerda\ experiment. The 
 main expected background components are briefly summarized. Due
 to the screening of the components before installation, the known inventory
 of radioactive contaminations can be estimated. Finally, the stable
 performance of the experiment is demonstrated and the data selection cuts are
 discussed.
\subsection{The hardware}

\begin{table*}[t]
\begin{center}
\caption{\label{tab:recval}
   Main parameters for the HPGe detectors employed in the \GERDA\ experiment:
   isotopic abundance of the isotope $^{76}$Ge, \fgesix,  total mass~$M$,
   active mass \actmass,  active volume fraction \factmass\ and the
   thickness of the  effective n$^+$ dead layer, $d_{dl}$.
 }
\vspace*{2mm}
\begin{tabular}{l|l||c||c|r||r}
detector     &\multicolumn{1}{c||}{\fgesix}& $M$& $\actmass(\Delta\actmass)$   
  & \factmass$(\Delta\factmass_t)$
   &\multicolumn{1}{c}{$d_{dl}$}\\
     & &\multicolumn{1}{r||}{g} & g & &    \multicolumn{1}{c}{mm} \\
\hline
\multicolumn{6}{c}{enriched coaxial detectors}\\
\hline
ANG~1 $^\dagger$)  &0.859(29) &  ~958 &  795(50) &0.830(52)&1.8(5)\\
ANG~2 &0.866(25) & 2833 & 2468(145) & 0.871(51)&2.3(7)\\
ANG~3 &0.883(26) & 2391 & 2070(136) & 0.866(57)&1.9(7)\\
ANG~4 &0.863(13) & 2372 & 2136(135) & 0.901(57)&1.4(7)\\
ANG~5 &0.856(13) & 2746 & 2281(132) & 0.831(48)&2.6(6)\\
RG~1  &0.855(15) & 2110 & 1908(125) & 0.904(59)&1.5(7)\\
RG~2  &0.855(15) & 2166 & 1800(115) & 0.831(53)&2.3(7)\\
RG~3 $^\dagger$)   &0.855(15) & 2087 & 1868(113) & 0.895(54)&1.4(7)\\
\hline                   
\multicolumn{6}{c}{enriched BEGe detectors}\\
\hline                   
\up%
GD32B &0.877(13) &  717 & 638(19) &0.890(27)&1.0(2)\\
GD32C &0.877(13) &  743 & 677(22) &0.911(30)&0.8(3)\\
GD32D &0.877(13) &  723 & 667(19) &0.923(26)&0.7(2)\\
GD35B &0.877(13) &  812 & 742(24) &0.914(29)&0.8(3)\\
GD35C $^\dagger$) &0.877(13) &  635 & 575(20) &0.906(32)&0.8(3)\\
\hline
\multicolumn{6}{c}{natural coaxial detectors}\\
\hline
GTF~32 $^\dagger$) &0.078(1)  & 2321 & 2251(116) &0.97(5)~~~&0.4(8)\\
GTF~45 $^\dagger$) &0.078(1)  & 2312 &  &&\\
GTF~112&0.078(~1) & 2965 &  &&\\
\hline
\end{tabular}
\end{center}
$^\dagger$) not used in this analysis
\end{table*}

\begin{figure}[b]
\begin{center}
  \includegraphics[height=85mm]{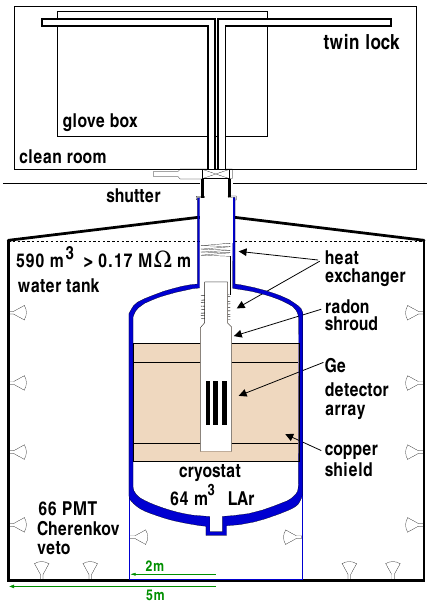}
\caption{ \label{fig:exp}
 Schematic drawing of the main components of the \gerda\ experiment. For
 details see Ref.~\cite{gerda_tec}.
}
\end{center}
\end{figure}

 The setup of the \gerda\ experiment is described in detail in
 Ref.~\cite{gerda_tec}. \gerda\ operates high purity germanium (HPGe) detectors
 made from material
 enriched to about 86\,\% in \gesix\ in liquid argon (LAr) which serves both as
 coolant and as shielding. A schematic view is given in Fig.~\ref{fig:exp}. A
 stainless steel cryostat filled with 64~\cum\ of LAr is located inside a water 
 tank
 of 10~m in diameter. Only very
 small amounts of LAr are lost as it is cooled via a heat exchanger by liquid
 nitrogen. The 590~\cum\ of high purity ($>$\,0.17~M$\Omega$m) water
 moderate ambient neutrons and $\gamma$ radiation. It is instrumented with 66
 photo multiplier tubes (PMT) and operates as a  {C}herenkov muon veto to further
 reduce cosmic induced backgrounds to insignificant levels for the
 \gerda\ experiment.  Muons traversing through the opening of the cryostat
without reaching water are
 detected by plastic scintillator panels on top of the clean room. 

 Three coaxial or five BEGe detectors are mounted into each of the four strings
 which are lowered through a lock separating the clean room from the
 cryostat. The detector strings with coaxial detectors are housed in
 60~\mum\ thin-walled copper containers permeable to LAr - called mini shroud
 in the following - with a distance of a few mm from the detector outer
 surfaces.  A 30~\mum\ thin copper cylinder - called radon shroud in the
 following - with a diameter of 75~cm encloses the detector array.  
 A picture of a detector string can be found in \cite{gerda_tec}.
 The custom
 made preamplifiers are operated in LAr at a distance of about 30~cm from the
 top of the detector array. The analog signals are digitized by 100 MHz FADCs.

\begin{figure}[bH!]
\begin{center}
 \includegraphics[width=0.6\columnwidth]{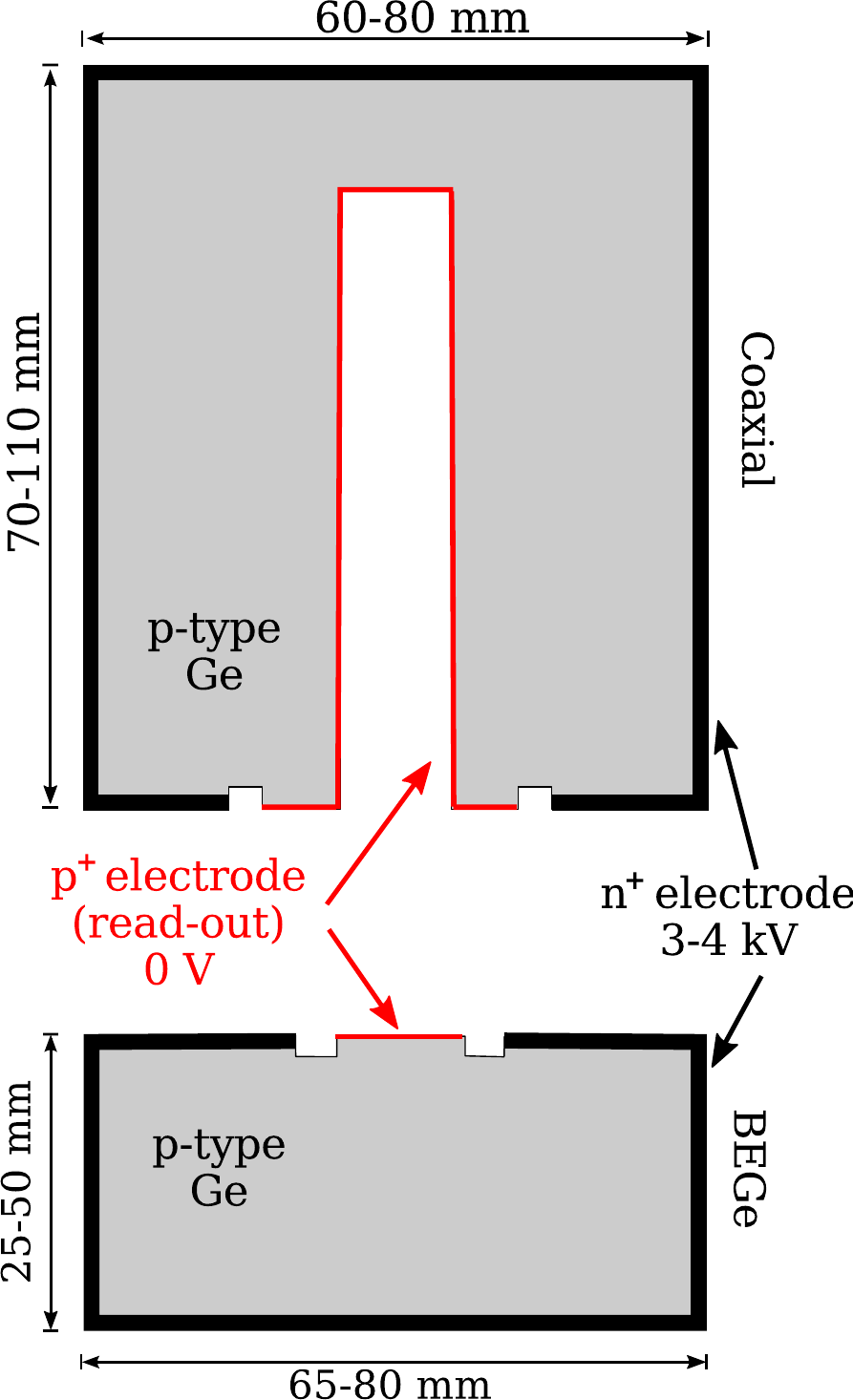}
  \caption{ \label{fig:detector_schematic}
    Schematic sketch of a coaxial HPGe detector (top) and a BEGe detector
    (bottom) with their different surfaces and dead layers (drawings not to
    scale).
}
\end{center}
\end{figure}

 All eight of the reprocessed coaxial germanium detectors from the \hdm\ and the
 \igex\ experiments~\cite{hdm,igex} were deployed on November 9 2011,
 together with three detectors with natural isotopic abundance. A schematic
 drawing of the coaxial detector type is shown in
 Fig.~\ref{fig:detector_schematic}, top. Two enriched detectors (ANG~1 and
 RG~3) developed high leakage currents soon after the start of data taking and were not
 considered in the analysis. RG~2 was taking data for about one year before it
 also had to be switched off due to an increase of its leakage current.  In
 July 2012, two of the low background coaxial HPGe detectors with natural isotopic abundance, 
 GTF~32 and GTF~45, were replaced by five enriched Broad Energy Germanium
 (BEGe) detectors, which follow the Phase~II design of \gerda\ (see
 Fig.~\ref{fig:detector_schematic}, bottom). The geometries and thus the pulse
 shape properties of the two types of detectors differ
 as discussed in Ref.~\cite{gerda_psd}. 
One of these BEGe detectors (GD35C) showed
 instabilities during data taking and was not used for further analysis. The
 most relevant properties of all the germanium detectors are compiled in
 Table~\ref{tab:recval}. Note, that the numbers for dead layers $d_{dl}$ are
 to be interpreted as effective values, because their determination by  
 comparison of count rates and Monte Carlo (MC) predictions depends on the precision of the 
 model and the geometries \cite{marik_thesis}.

\subsection{Expected background sources}

An important source of background is induced by cosmic radiation. Muon induced background events are efficiently vetoed by  
identification of {C}herenkov light emitted by muons when they pass the water tank. The number of long lived cosmogenically produced isotopes, especially $^{68}$Ge and $^{60}$Co are minimized by minimization of the time above ground during processing of the detectors and the structural materials.

Further contributions stem from radioactivity included in the detector and structural materials or the surrounding environment, i.e. the rocks of the laboratory.
The selection of materials has
been described in \cite{gerda_tec}. The most important activities are discussed
in the next section.

Background from $^{42}$Ar present in LAr was found during GERDA commissioning to be more significant than anticipated. The $\beta$ decay of its progeny  $^{42}$K 
can contribute to the background at \qbb\ if the decay happens near detector surfaces. For \gerda\ Phase~I coaxial detectors this background was significantly reduced by implementation of the mini shrouds. However, for the BEGe detectors this remains an important background due to their thinner surface n$^+$ dead layer.

Another source of background stems from
the calibration sources that have a typical initial activity of about 10-20~kBq. When in parking position they are well shielded. Due to an accident during commissioning the experiment, one 20 kBq $^{228}$Th calibration source 
fell to the bottom of the cryostat.
The BI expected from this source is around \pIIbi, thus, significantly less then the Phase~I BI goal. Hence, the calibration source was 
left inside the LAr cryostat. It will be removed during the upgrade of the experiment to its second phase.

A significant fraction of the background is induced by contaminations of
bulk materials and surfaces with nuclei from the $^{238}$U and $^{232}$Th decay chains.
The $^{238}$U decay chain can be subdivided into three sub decay chains:
$^{238}$U to $^{226}$Ra, $^{226}$Ra to $^{210}$Pb and $^{210}$Pb to $^{206}$Pb, due to isotopes with half lives significantly longer than the live time of the experiment. Only the two latter sub decay chains are relevant in the following.
The noble gas $^{222}$Rn (T$_{1/2}$\,=\,3.8 days) plays a special role, 
as it can further break the $^{226}$Ra to $^{210}$Pb chain due to its volatility.
Whenever activities of $^{214}$Bi are quoted it is assumed that the chain is in secular 
equilibrium between $^{226}$Ra and $^{210}$Pb inside metallic materials, while for non metallic materials the equilibrium can be broken at $^{222}$Rn.

\subsection{Known inventory from screening}
\begin{table*}[t]
\begin{center}
\caption{\label{tab:thorium}
     Gamma ray screening and $^{222}$Rn emanation measurement results for
     hardware components. The activity of the mini shroud was derived from
     ICP-MS measurement assuming secular equilibrium of the $^{238}$U decay
     chain. Estimates of the BI at \qbb\ are based on efficiencies obtained by
     MC simulations~\cite{lenz_thesis,gerda_background} of the \gerda\ setup.
} 
\vspace*{2mm}
\begin{tabular}{llcccccc}
component          & units   &$^{40}$K & $^{214}$Bi\&$^{226}$Ra & $^{228}$Th  & $^{60}$Co  & $^{222}$Rn & BI\\
 & & & & & & &  \pIIbi\\
\hline
\multicolumn{7}{l}{close sources: up to 2~cm from detectors} & \\
Copper det. support&\mubq/det.& $<$\,7       & $<$\,1.3      & $<$\,1.5      &
&            & $<$\,0.2~~~~\\
PTFE det. support  &\mubq/det.& 6.0 (11)  &0.25 (9)&0.31 (14)&         &            & 0.1\\
PTFE in array      & \mubq/det& 6.5 (16)& 0.9 (2) &             &         &            & 0.1\\
mini shroud        &\mubq/det.&            &  22 (7)   &             &         &            & 2.8\\
Li salt            &mBq/kg    &            &   17(5)   &             &         &    & $\approx$ 0.003$\dagger$ \\
\hline
\multicolumn{6}{l}{ medium distance sources: 2 - 30~cm from detectors}\\
CC2 preamps        &\mubq/det.& 600 (100)& 95 (9)    & 50 (8)    &         &            & 0.8\\
cables and\\
suspension         & mBq/m    &1.40 (25)&0.4 (2)  &0.9 (2)  &76 (16)&            & 0.2\\
\hline
\multicolumn{6}{l}{distant sources: further than 30~cm from detectors}\\
cryostat           & mBq      &            &             &             &
&54.7 (35)& $<$\,0.7\\
copper of cryostat & mBq      &  $<$\,784    &  264 (80)   &  216 (80)   &
   288 (72)  &   &\multirow{2}{*}{$\left.\rule{0mm}{3mm}\right]~<$\,0.05~~~~~}\\
steel of cryostat  & kBq      &  $<$\,72     &  $<$\,30      & $<$\,30       &   475   &            & \\
lock system        & mBq      &            &             &             &
& 2.4 (3)& $<$\,0.03\\
$^{228}$Th calib. source& kBq &            &            & 20          &             &        & $<$\,1.0\\
\end{tabular}
\end{center}
$^\dagger$) value derived for 1~mg of Li salt absorbed into the surface of each detector\\
\end{table*}

 The hardware components close to the detectors and the components of the
 suspension system have been tested for their radio-purity prior to
 installation~\cite{gerda_tec}.  The hardware parts at close (up to 2~cm) 
 and 
 medium  (up to 30~cm) distance from the detectors have been screened using 
 HPGe screening facilities or ICP-MS
 measurements, while the parts in the lock system have been tested for
 $^{222}$Rn emanation~\cite{emanation}. Some materials proved to have low, but
 measurable, radioactive contaminations.  Table~\ref{tab:thorium} quotes the
 total measured activities and limits of the most significant screened 
 components and their
 expected contribution to the BI close to \qbb.  As the $^{222}$Rn emanation
 rate in the cryostat with its copper lining and the lock system is on the
 order of 60~mBq, some $^{214}$Bi may be expected in the LAr surrounding of the
 detectors. Assuming a homogeneous distribution of \Rn\ in the LAr, this would
 result in a contribution to the BI at \qbb\ of
 7$\cdot$\vctsper.  To reduce this latter contribution to the \GERDA\
 background, the radon shroud was installed around the array with the
 intention to keep \Rn\ transported by LAr convection at sufficient distance from the detectors. 

Additionally, Li salt that is used to dope n$^+$ surfaces of the detectors was screened.
It is not precisely known how much Li diffuses into the crystal. A rough estimation 
assuming an n$^+$ Li doping of 10$^{16}$ Li nuclei per cm$^{3}$ germanium 
results in an overall Li weight per detector of $\approx$\,5\,$\mu$g which leads to negligible
background contributions even if it is assumed that the $^{226}$Ra contamination diffuses  into the germanium with the same efficiency as Li.

 The measured activities in the hardware components within 2~cm from the
 detectors lead to a total contribution to the BI of
 $\approx$\,3$\cdot$\dctsper\ using efficiencies obtained by MC 
 simulations~\cite{lenz_thesis,gerda_background}.  From the medium 
 distance contributions
 $\approx$\dctsper\ are expected, while the far sources contribute with
 $<$ \dctsper.
 As detailed in Ref.~\cite{gerda_tec} the extrapolated background rates for
 all contaminations were predicted to be tolerable for Phase~I and to yield a
 BI of $<$\,\tctsper.

\subsection{Run parameters and efficiencies}

\begin{figure}[bt!]
\begin{center}
  \includegraphics[width=.8\columnwidth]{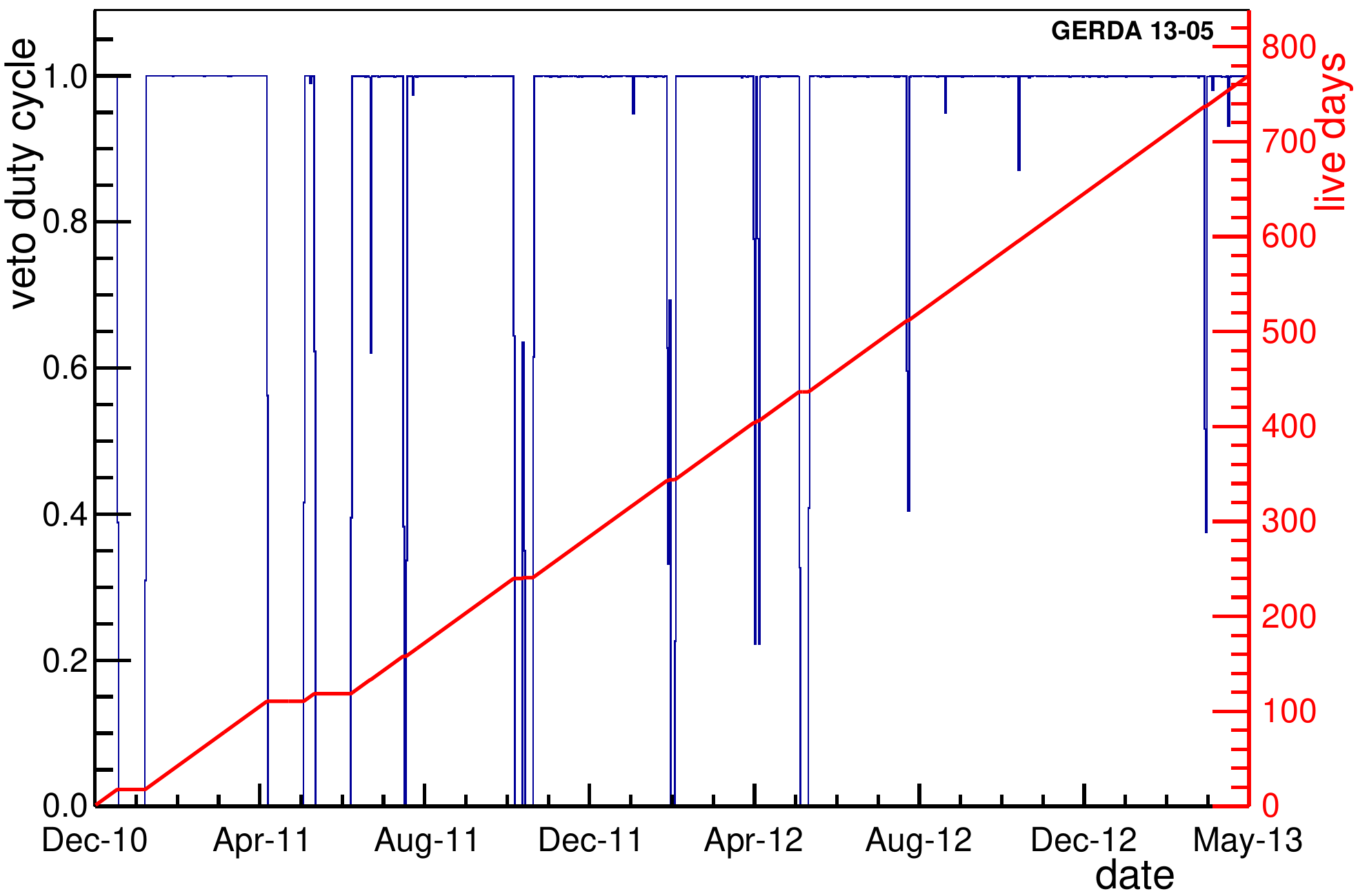}\\[5mm]
  \includegraphics[width=.8\columnwidth]{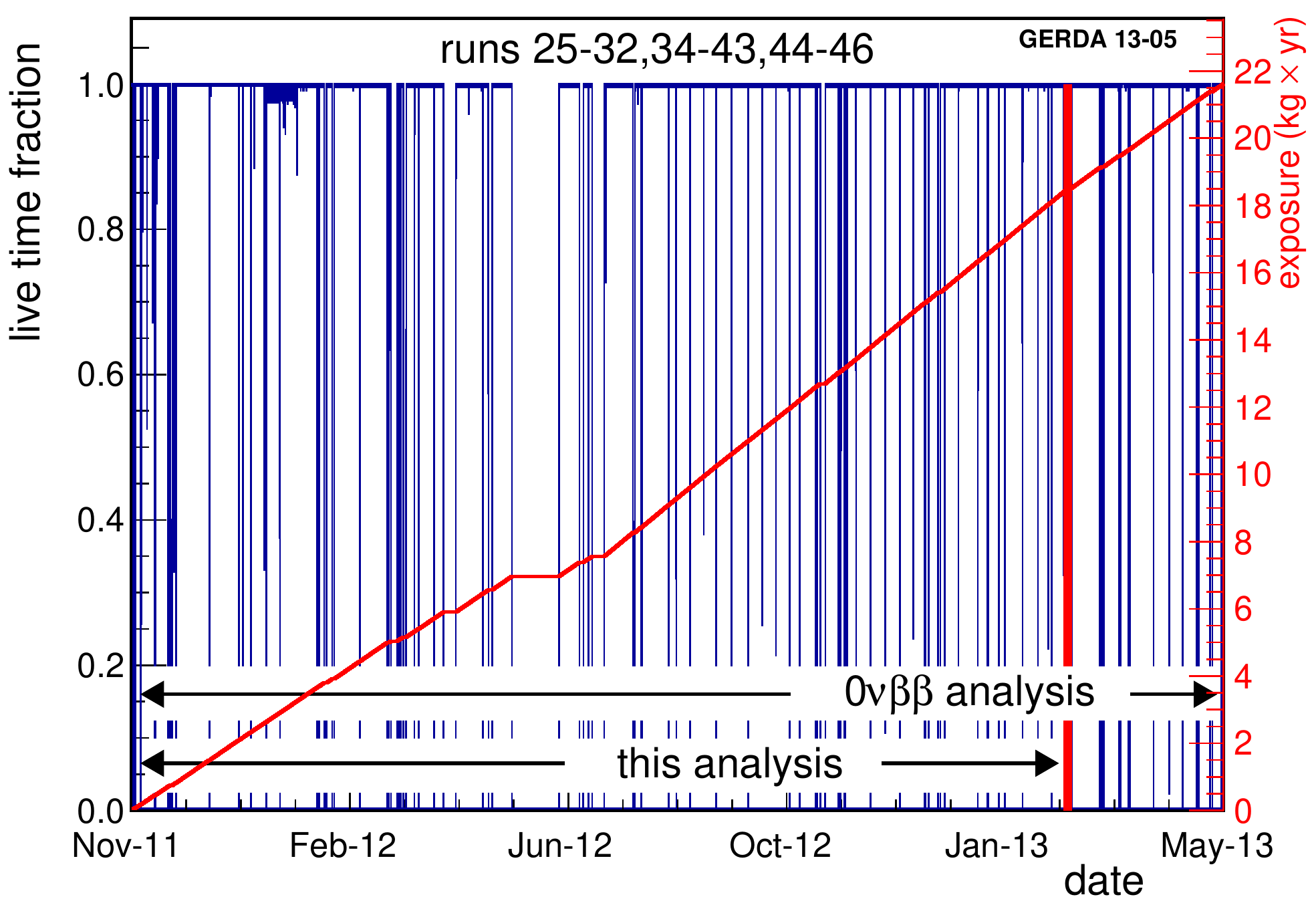}
\caption{ \label{fig:live}
   Live time fraction of the data acquisition for the muon veto (top) and for
   the HPGe detectors (bottom) The spikes in the live time fraction arise from
   the regular calibration measurements. The development of the
   exposure \exposure\ of the enriched detectors (bottom) and the total live
   time of the muon veto system (top) is also shown. The red vertical line
   indicates the end of the data range for the evaluation of the background
   model.  }
\end{center}
\end{figure}

 The muon veto system started operation in December 2010 and ran up to May
 21 2013, when the data taking for the \onbb\ analysis was stopped.  Its
 stable performance is shown in the top graph of Fig.~\ref{fig:live}.  The
 interruptions were due to the test and installation of the plastic panel in
 April/May 2011 and due to short calibrations.  The probability that a muon
 induced event in the detectors is accompanied by a signal in the veto 
 (overall muon rejection efficiency) is $\varepsilon_{\mu
   r}\,=\,0.991^{+0.003}_{-0.004}$, reducing the contribution of the muons to the BI to
 $<$\,10$^{-3}$~\ctsper\ \cite{luciano_muons}. No evidence for delayed coincidences
 between $\mu$ veto events and germanium events was found.

The
 bottom graph in Fig.~\ref{fig:live} demonstrates the live time fraction of data
 taking. The interruption in May 2012 was due to temperature instabilities in the \gerda\ clean room, while the interruption in
July 2012 was due to the insertion of five
 Phase~II type BEGe detectors. 
 The analysis presented here considers data
 taken until March 3, 2013, corresponding to a live time of 417.19~days and an
 exposure of \exposure\,=\,16.70~\kgyr\ for the coaxial detectors; the
 four BEGe detectors acquired between 205 and 230~days of live time each,
 yielding a total exposure of \exposure\,=\,1.80~\kgyr.  The end of run~43 in March is marked
 by the red vertical line in Fig~\ref{fig:live}, bottom.

 The data have been processed using algorithms and data selection
 procedures~\cite{cuts,paolo} implemented in the \gerda\ software
 framework~\cite{gelatio}.  A set of quality cuts, described in detail in
 Ref.~\cite{cuts}, is applied to identify and reject unphysical events,
 e.g. generated by discharges or by electromagnetic noise. The cuts take into
 account several parameters of the charge pulse, such as rise time, baseline
 fluctuations and reconstructed position of the leading edge. The cuts also
 identify events having a non-flat baseline, e.g. due to a previous pulse
 happening within a few hundreds of \mus. Moreover, events in which two
 distinct pulses are observed during the digitization time window (80~\mus)
 are marked as pile-up and are discarded from the analysis.  
 From the total number of triggers roughly  
 91\,\% are kept as physical events. 
 Due to the very low
 counting rate, the \gerda\ data set has a negligible contamination of
 accidental pile-up events and the selection efficiency for genuine \onbb\
 events is hence practically unaffected by the anti pile-up cuts. Similarly,
 the loss of physical events above 1~MeV due to mis-classification by the
 quality cuts is less than 0.1\,\%.

 The linearity and the long term stability of the energy scale as well as the
 energy resolution given as full width at half maximum (FWHM) were checked
 regularly with \thzza\ sources.
 Between calibrations the stability of the gain of the preamplifiers was
 monitored by test pulses induced on the test inputs of the preamplifiers.
 Whenever unusual fluctuations on the preamplifier response were observed,
 calibrations were performed.  
  The linearity of the preamplifier has been
 checked using test pulses up to an energy range of 6~MeV. It was found that
 between 3 and 6~MeV the calibration has a precision of better than 10~keV;
  above  6~MeV some channels exhibit larger non-linearity.

 Physical events passing the quality cuts are excluded from the analysis if
 they come in coincidence within 8~\mus\ with a valid muon veto signal (muon
 veto cut) or if they have energy deposited in more than one HPGe detector
 (anti-coincidence cut). The anti-coincidence cut does not further affect the
 selection efficiency for \onbb\ decays, since only events with full
 energy deposit of 2039~keV are considered. The dead time induced by the muon
 veto cut is practically zero as  the rate of
 (9.3\,$\pm$\,0.4)$\cdot$10$^{-5}$~/s of events coincident between germanium
 detectors and the \gerda\ muon veto system is very low.

 The stability of the energy scale was checked by the time dependence of the
 peak position for the full energy peak at 2614.5~keV from the
 \thzza\ calibration source. The maximal shifts are about 2~keV with the
 exception of 5~keV for the GD32B detector.  The distributions of the shifts are
 fitted by a Gaussian with FWHM amounting to 1.49~keV for the coaxial and to
 1.01~keV for the BEGe detectors. The respective uncertainties are smaller
 than 10\,\%. The shifts are tolerable compared to the energy resolution.

 To obtain the energy resolution at \qbb\ the results from the calibration
 measurements are interpolated to the energy \qbb\ using the standard
 expression FWHM\,= $\sqrt{a^2+b^2\cdot E}$~\cite{helmer}.  The energy
 resolution during normal data taking is slightly inferior to the resolution
 during calibration measurements. The resulting offset was determined by
 taking the difference between the resolution of the $^{42}$K line and the
 interpolated resolution determined from calibrations. The scaled offset is added to the
 resolution at \qbb\  expected from calibration measurements.  The FWHM of all enriched detectors at
 2614.5~keV is determined to be between 4.2 and 5.8~keV for the coaxial detectors
 and between 2.6 and 4.0~keV for the BEGes. The resolutions are stable in time to
 within 0.3~keV for the BEGes and to within 0.2~keV for the coaxial detectors.
 The resolutions of all relevant enriched detectors are shown in Table~\ref{tab:fwhm}.

\begin{table}[b]
\begin{center}
\caption{ \label{tab:fwhm}
 Energy resolutions (FWHM) in keV of the enriched detectors at \qbb. For
 definition of the data sets see sec.~\ref{sec:datasets}. 
}
\vspace*{2mm}
\begin{tabular}{ll|ll}
detector         &FWHM [keV] & detector         &FWHM [keV] \\
\hline
  \multicolumn{2}{c|}{\it SUM-coax} &  \multicolumn{2}{c}{\it SUM-bege} \\
\hline
ANG~2 &5.8 (3) & GD32B & 2.6 (1)\\
ANG~3 &4.5 (1) & GD32C & 2.6 (1)\\
ANG~4 &4.9 (3) & GD32D & 3.7 (5)\\
ANG~5 &4.2 (1) & GD35B & 4.0 (1)\\
RG~1  &4.5 (3) & &\\
RG~2  &4.9 (3) & &\\
\hline
mean  coax &4.8 (2) & mean  BEGe & 3.2 (2)\\
\end{tabular}
\end{center}
\end{table}
\begin{figure*}[t]
\begin{center}
\vspace*{-1mm}
\includegraphics[width=1.8\columnwidth]{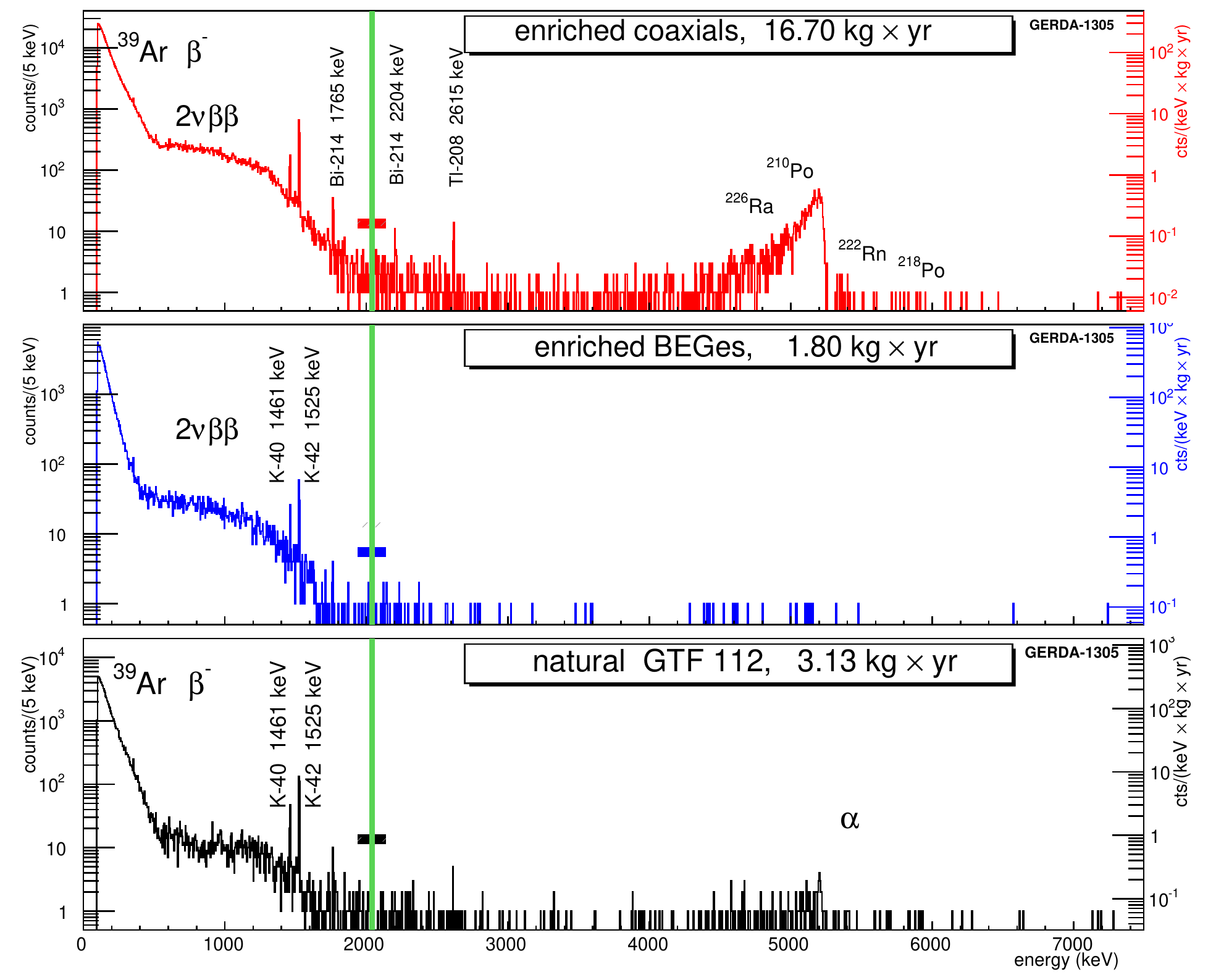}
\caption{\label{fig:spectra-ge}
     Spectra taken with all the enriched coaxial (top) and BEGe (middle) and
     a non-enriched (bottom) detector.  The blinding window of \qbb\,$\pm$\,20~keV
     is indicated as green line. The bars in the color of the histogram
     represent the 200~keV region from which the BI of the dataset is
     determined.
} 
\end{center}
\end{figure*}

 The total exposure~\exposure\ used for the upcoming \onbb\ analysis is given
 by the sum of products of live time $t_i$ and total mass $M_i$, where the
 index~$i$ runs over the active detectors.  
 For the evaluation of \thalfzero, the acceptance of PSD cuts, \effpsd, the
 efficiencies \effres\ to find the \onbb\ within the analysis window $\Delta
 E$ and the detection efficiency of the \onbb\ decay \efffep\ are needed.  The
 energy of \onbb\ events is assumed to be Gaussian distributed with a mean
 equal to the $Q_{\beta\beta}$ value. An exposure averaged efficiency is
 defined as
\begin{equation}
\label{equ:efficiency}
\langle\varepsilon\rangle =
          \frac{\sum_i \factvoli\fgesixi M_i t_i\ \efffepi}{\exposure} \,\,,
\end{equation}
where \factvoli\ is the active volume fraction  and \fgesixi\ the enrichment
fraction of the individual detector $i$. 

 With $N_A$, the Avogadro number, $m_{enr}$ the molar mass of the germanium and 
 $N$ the number of observed counts the half life reads
\begin{equation}\label{eq:thalf}
\mbox{\thalfzero} = \frac{\ln{2}\cdot N_A}{m_{enr}}\,\, \frac{\cal E}{N}\, \,
\langle\varepsilon\rangle\,\,\effpsd \, \, \effres ~~.
\end{equation}

\section{Background spectra and data sets}
  \label{sec:spectradata}

 The main objective of \onbb\ experiments is the possible presence of a peak at
 \qbb.  All other parts of the energy spectrum can be considered as
 background.  As detectors have their own history and experienced different
 surroundings their energy spectra might vary. Furthermore, the experimental
 conditions might change due to changes of the experimental setup. Thus, a
 proper selection and grouping of the data can optimize the result. This
 selection is performed on the ``background data'' and will be applied in the
 same way to
 the ``\onbb\ data''.

\begin{table*}[t]
\begin{center}
\caption{\label{tab:datasets}
       Data sets, the detectors considered therein and their
       exposures~\exposure\ are listed for the data used for this analysis and
       the upcoming \onbb\ analysis. \exposure\ is calculated from the total
       detector mass.
}
\vspace*{2mm}
\begin{tabular}{l|l|rr}
data set & detectors & \multicolumn{2}{c}{exposure \exposure} \\
         &           &  this analysis & \onbb\  analysis      \\
         &           &  \multicolumn{2}{c}{\kgyr}      \\
\hline
{\it SUM-coax}    & all enriched coaxial           & 16.70   & 19.20 \\
{\it GOLD-coax}   & all enriched coaxial           & 15.40   & 17.90 \\
{\it SILVER-coax} & all enriched coaxial           &  1.30   &  1.30 \\
{\it GOLD-nat}    & GTF~112                        &  3.13   &  3.98 \\
{\it GOLD-hdm}    & ANG~2, ANG~3, ANG~4, ANG~5     & 10.90   & 12.98 \\
{\it GOLD-igex}   & RG~1, RG~2                     &  4.50   &  4.93 \\ 
{\it SUM-bege}    & GD32B, GD32C, GD32D, GD35B     &  1.80   &  2.40 \\
\end{tabular}
\end{center}
\end{table*}
\subsection{Background spectra}
\label{sec:spectra}

 Fig.~\ref{fig:spectra-ge} compares the energy spectra in the range 
 from 100~keV to 7.5~MeV obtained from the three
 detector types: ($i$) the enriched coaxial detectors (top), ($ii$) the
 enriched BEGe detectors (middle) and ($iii$) the coaxial low background detector GTF~112
 (bottom) with natural isotopic abundance. 

 Some prominent features can be identified. The low energy part up to 565~keV is
 dominated by  $\beta$ decay of cosmogenic $^{39}$Ar in all spectra. Slight
 differences in the spectral shape between the coaxial and BEGe type detectors result
 from  differences in detector geometry and of the n$^+$ dead layer thickness. Between 600 and 1500~keV the spectra of the
 enriched detectors exhibit an enhanced continuous spectrum due to
 \nnbb\ decay~\cite{2vbb}. In all spectra, $\gamma$ lines from the decays of
 $^{40}$K and $^{42}$K can be identified, the spectra of the enriched coaxial
 detectors contain also lines from $^{60}$Co, $^{208}$Tl, $^{214}$Bi,
 $^{214}$Pb and $^{228}$Ac.  A peak-like structure appears around 5.3~MeV in
 the spectrum of the enriched coaxial detectors.  This can be attributed to
 the decay of $^{210}$Po on the detector p$^+$ surfaces.  Further peak like
 structures at energies of 4.7~MeV, 5.4~MeV and 5.9~MeV can be attributed to
 the $\alpha$ decays  on the detector
 p$^+$ surface of $^{226}$Ra, $^{222}$Rn and $^{218}$Po, respectively.  These events are discussed in more detail
 below.
There are no hints  for contamination of the detector p$^{+}$ surfaces with isotopes from the $^{232}$Th decay chain in the data analyzed here.
 
\begin{figure}[b]
\begin{center}
   \includegraphics[width=1.0\columnwidth]{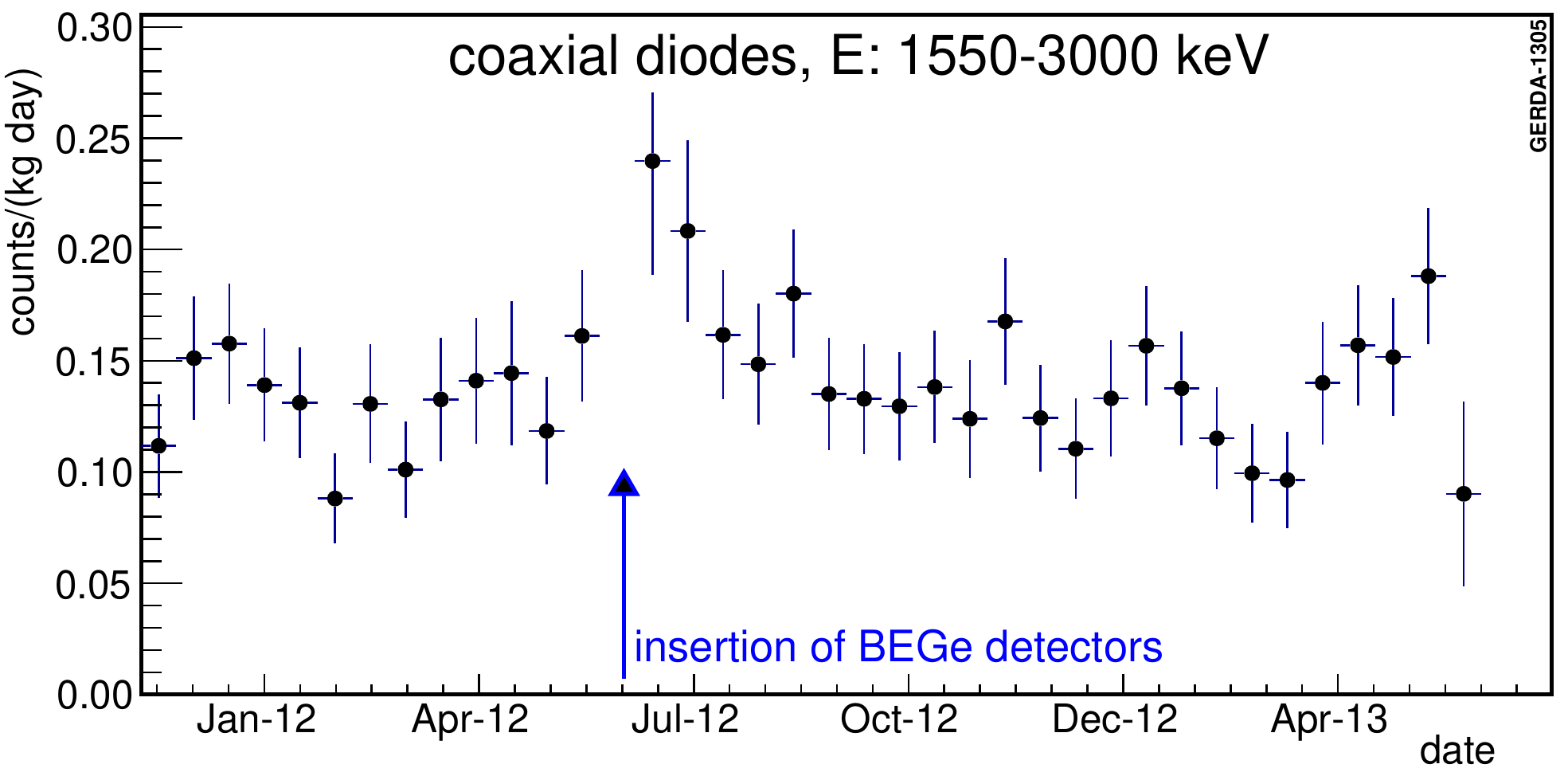}
\caption{\label{fig:bkg_stability}
      Time distribution of background rate of the enriched coaxial detectors
      in the energy range between 1550 and 3000~keV in 15 day intervals. An
      increase of the BI after BEGe deployment in July 2012 is clearly
      visible.
}
\end{center}
\end{figure}

 The observed background rate of the coaxial enriched detectors in the energy
 region between 1550 and 3000~keV in 15 calendar day intervals is displayed in
 Fig.~\ref{fig:bkg_stability}.  The data are corrected for live time. Apart
 from the time period directly after the deployment of the BEGe detectors to
 the \gerda\ cryostat in July 2012, the rate in this energy region was stable
 within uncertainties over the whole time period.

\subsection{Data sets}
\label{sec:datasets}

 For further analysis of the background contributions the data are divided
 into different subsets based on the observed BI near \qbb.  In the energy
 region between 600 and 1500~keV, the spectrum of the enriched detectors is
 dominated by \twonu\ decays. Thus, characteristic $\gamma$ lines expected from
 known background contributions  might be visible only with the natural 
 GTF~112 detector.

 Data taken with enriched coaxial detectors in runs that were not affected by
 the experimental performance such as drift in gain stability, deterioration
 of energy resolution etc. are contained in the {\it SUM-coax} data set. The
 energy spectrum of this data set is shown in Fig.~\ref{fig:spectra-ge}, top. It
 has an overall exposure of 16.70~\kgyr\ (see also Table~\ref{tab:datasets}).
 The higher BI observed after the deployment of the BEGe detectors dropped to
 the previous level after approximately 30~days as shown in
 Fig.~\ref{fig:bkg_stability}.  Hence, the coaxial data are split: the {\it
   SILVER-coax} data set contains data taken during the 30~days after the BEGe
 detector deployment. The {\it GOLD-coax} data set contains the rest of the data.  The detectors from the \hdm\ and \igex\ experiments
 have different production, processing and cosmic ray exposure history. A
 different background composition could be expected, despite their common
 surface reprocessing before insertion into the \gerda\ experiment. Indeed, $^{210}$Po
 $\alpha$-contaminations are most prominent on detectors from the
 \hdm\ experiment (see Table~\ref{tab:alpha_rates}).  The {\it GOLD-coax} data
 set is therefore divided into two subsets {\it GOLD-hdm} and {\it GOLD-igex}
 to verify the background model on the two subsets individually.  The {\it
   SUM-bege} data set contains the data taken with four out of the five
 Phase~II BEGe detectors.  The {\it GOLD-nat} data set contains data taken
 with the low-background detector GTF~112 of natural isotopic composition.

 The data sets used in this analysis, the detectors selected and the
 exposures~\exposure\ of the data used in this analysis 
 and  separately for the
 upcoming \onbb\ analysis are listed in Table~\ref{tab:datasets}.

\section{Background sources and their simulation}
  \label{sec:back}

 The largest fraction of the \gerda\ Phase~I exposure was taken with the
 coaxial detectors from the \hdm\ and \igex\ experiments.  Thus, the
 background model was developed for these detectors first. Some preliminary
 results were presented in Ref.~\cite{NesliLRT}. 

 Background components that were identified in the energy spectra (see
 sec.~\ref{sec:spectra}) or that were known to be present in the vicinity of
 the detectors (see Table~\ref{tab:thorium}) were simulated using the
 \mage\ code~\cite{mage} based on \geant~\cite{geant}.  The expected BIs due to
 the neutron and muon fluxes at the LNGS underground laboratory have been
 estimated to be of the order 10$^{-5}$~\ctsper \cite{georg} and
 10$^{-4}$~\ctsper~\cite{luciano_muons} in earlier works. These contributions
 were not considered in this analysis. Also other potential background sources
 for which no direct evidence could be found were not taken into
 consideration.

 It should be mentioned that some isotopes can cause peaks at or close to the
 \qbb\ value of $^{76}$Ge.  All known decays that lead to $\gamma$ emission with
 $\sim$2040~keV either have very short half lives or have significant
 other structures (peaks) that are not observed in the
 \gerda\ spectra. Three candidates are \gesix~\cite{georg}, which can
 undergo neutron capture, $^{206}$Pb~\cite{mei}, which has a transition that
 can be excited by inelastic neutron scattering and $^{56}$Co that decays with
 a half life of 77~days.  None of the strong prompt $\gamma$ lines at 470, 861, 4008
 and 4192~keV from neutron capture on \gesix\ could be identified.  In case of
 inelastic neutron scattering off $^{206}$Pb, peaks would be expected to
 appear at 898, 1705 and 3062~keV. These are not observed. In case of a
 $^{56}$Co contamination peaks would be expected at 1771, 2598 and 3253~keV,
 none of which is observed.  Hence, these sources are not considered in the
 following for the simulation of the background components.

 The \gerda\ Phase~I detectors and the arrangement of the germanium
 detector array with four detector strings (`array' in Table~\ref{tab:mcs})
 were implemented into the \mage\ code.  Simulations of contaminations of 
 the following hardware components were performed (see Fig.~\ref{fig:exp} and
 Ref.~\cite{gerda_tec}): inside the germanium, on the p$^+$ and n$^+$ surfaces
 of the detectors (see Fig. \ref{fig:detector_schematic}) , in the liquid argon close to the p$^+$ surface,
 homogeneously distributed in the LAr, in the detector assembly representing contaminations in or on the detector holders and their components, 
the  mini shroud,  the radon shroud and the heat exchanger.  
Note, that the thicknesses of the detector assembly components, the shrouds and the heat exchanger are significantly smaller than the mean free path of the relevant simulated $\gamma$ particles in the given material, thus, no significant difference can be expected between the resulting spectra of bulk and surface contaminations. 
 Various (DL) thicknesses were considered.  The n$^+$ dead layer
 thicknesses~$d_{dl}({\rm n}^+)$ of the detectors were implemented according
 to the values reported in Table~\ref{tab:recval}. Spectra resulting from
 contaminations on effective p$^+$ dead layer thicknesses $d_{dl}({\rm
   p}^+)$ of 300, 400, 500 and 600~nm were simulated.

 Most of the identified sources for contaminations were
 simulated. However, $\gamma$ induced energy spectra from sources with similar
 distances to the detectors have similar shapes that can not be disentangled
 with the available exposure.  Representatively for
 $\gamma$ contaminations in the close vicinity of the detectors (up to 2~cm
 from a detector) events in the detector assembly were simulated. Spectra due
 to contaminations at medium distances (between 2 and 30~cm), such as the
 front end electronics or the cable suspension system are represented by
 simulations of events in the radon shroud, while spectra resulting from 
 distant sources (further than 30~cm) are represented by simulation of
 contaminations in or on the heat exchanger (see Fig.~\ref{fig:exp}).  The
 contributions of the cryostat and water tank components to the BI have not
 been considered in this analysis. It has been shown in earlier work that they
 contribute to the \gerda\ BI with $<$\,\vctsper\ \cite{gerda_background}.

 The simulated energy spectra were smeared with a Gaussian distribution with
 an energy dependent FWHM width corresponding to the detector resolution.  The
 spectra for this analysis resulting from different
 contaminations in different locations of the experiment are summarized in
 Table~\ref{tab:mcs}.

\begin{table*}[t]
\begin{center}
\caption{\label{tab:mcs}
       Summary of simulated background components for the coaxial
       detectors. For the p$^+$ dead layers~$d_{dl_{{\rm p}^+}}$ the
       thicknesses of 100, 200, 300, ..., 1000~nm were simulated. The
       $^{226}$Ra chain comprises the isotopes $^{226}$Ra, $^{222}$Rn,
       $^{218}$Po, and $^{214}$Po; the $^{222}$Rn chain comprises the isotopes
       $^{222}$Rn, $^{218}$Po, and $^{214}$Po.
}
\vspace*{2mm}
\begin{tabular}{llcc}
source & location & simulation & events simulated\\
\hline
\up
$^{210}$Po       & p$^+$ surface     & single det.,  $d_{dl_{{\rm p}^+}}$ & 10$^{9}$\\
$^{226}$Ra chain & p$^+$ surface     & single det., $d_{dl_{{\rm p}^+}}$ &  10$^{9}$\\
$^{222}$Rn chain & LAr in bore hole  & single det., $d_{dl_{{\rm p}^+}}$ &  10$^{9}$\\
\hline
\up
$^{214}$Bi and   & n$^+$ surface     & single det.         & 10$^{8}$\\
$^{214}$Pb       & mini shroud       &  array            & 10$^{9}$\\
                & detector assembly &  array            & 10$^{8}$\\
                & p$^+$ surface     & single det.          & 10$^{6}$\\
                & radon shroud      & array            & 10$^{9}$\\
                & LAr close to p$^+$ surface& single det.          & 10$^{6}$\\
\hline
\up
$^{208}$Tl and   & detector assembly & array            & 10$^{8}$\\
$^{212}$Bi       & radon shroud      & array            & 10$^{9}$\\
                & heat exchanger    & array            & 10$^{10}$\\
\hline
\up
$^{228}$Ac       & detector assembly &  array             & 10$^{8}$\\
                & radon shroud      &  array             & 10$^{9}$\\
\hline
\up
$^{42}$K         & homogeneous in LAr &   array           &  10$^{9}$ \\
                & n$^+$ surface      & single det.        &  10$^{8}$ \\
                & p$^+$ surface      & single det.      &  10$^{6}$ \\
\hline
\up
$^{60}$Co        & detectors         &   array           & 2.2$\cdot$10$^{7}$\\ 
                & detector assembly &   array            & 10$^{7}$\\ 
\hline
\up
\twonu          & detectors         &   array            & 2.2$\cdot$10$^{7}$\\
\hline
\up
$^{40}$K         & detector assembly&   array           & 10$^{8}$\\
\hline
\end{tabular}
\end{center}
\end{table*}

\subsection{$\alpha$ events from $^{226}$Ra, $^{222}$Rn and $^{210}$Po
 contaminations} 

 Strong contributions
 from $^{210}$Po can be observed in the energy spectra shown in 
 Fig.~\ref{fig:spectra-ge}. No other $\alpha$ peaks with similar
 intensity can be identified. This is indication for a surface $^{210}$Po
 contamination of the detectors. 
 However, there are also hints for other peak like
 structures at 4.7~MeV, 5.4~MeV and 5.9~MeV. These can be attributed to the
 decays of $^{226}$Ra, $^{222}$Rn and $^{218}$Po on p$^+$ detectors surfaces,
 respectively. However, the decay chain is clearly broken at $^{210}$Pb.
 Screening measurements indicate the presence of $^{226}$Ra in
 the vicinity of the detectors, in or on the mini shroud and of $^{222}$Rn in
 LAr. Thus, decays from $^{222}$Rn and its daughters are also expected in
 LAr (see Table~\ref{tab:thorium}).

 Due to the short range of $\alpha$ particles in germanium and LAr of the
 order of tens of \mum,
 only decays occurring on or in the close vicinity (few \mum) of the p$^+$
 surface (assumed dead layer thickness roughly 300~nm) can contribute to the
 measured energy spectrum as the n$^+$ dead layer thickness is roughly 1~mm. Additionally, $\alpha$ decays occurring on the
 groove of the detector (see Fig.~\ref{fig:detector_schematic}) may deposit
 energy in the active volume. For this part of the surface, however, no
 information on the actual dead layer thickness is available.  The energy
 deposited in the active volume of the detector by surface or close to surface
 $\alpha$ particles is very sensitive to the thickness of the dead layer and
 on the distance of the decaying nucleus from the detector surface.

 All $\alpha$ decays in the $^{226}$Ra to $^{210}$Pb sub-decay chain and the
 $^{210}$Po decay have been simulated on the p$^+$ detector surface
 separately.  Additionally, the decays in the chain following the $^{226}$Ra
 decay were simulated assuming a homogeneous distribution in a volume
 extending up to 1~mm from the p$^+$ surface in LAr.

\begin{figure*}[t]              
\begin{center}
  \includegraphics[width=0.9\columnwidth]{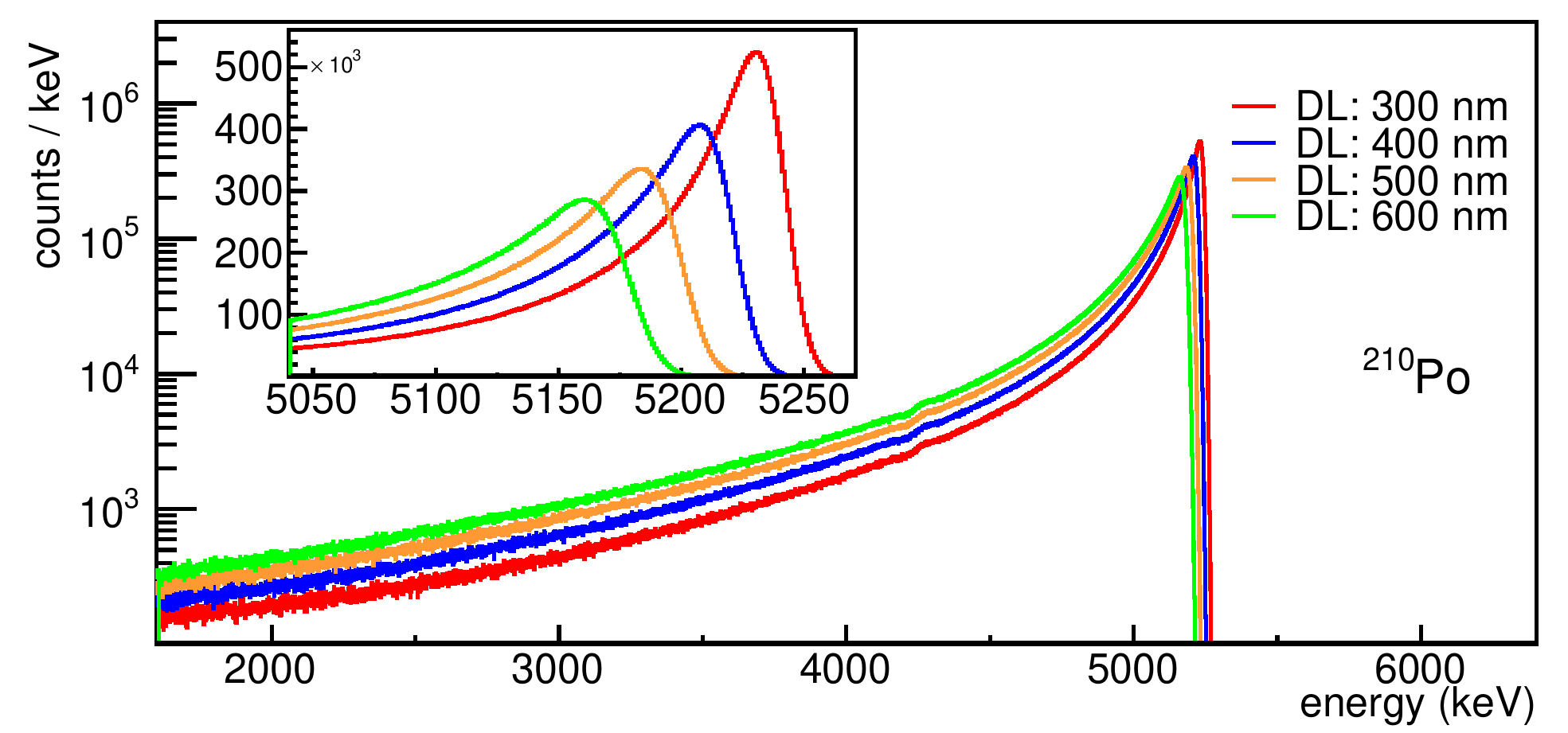}
\hspace*{10mm}
  \includegraphics[width=0.9\columnwidth]{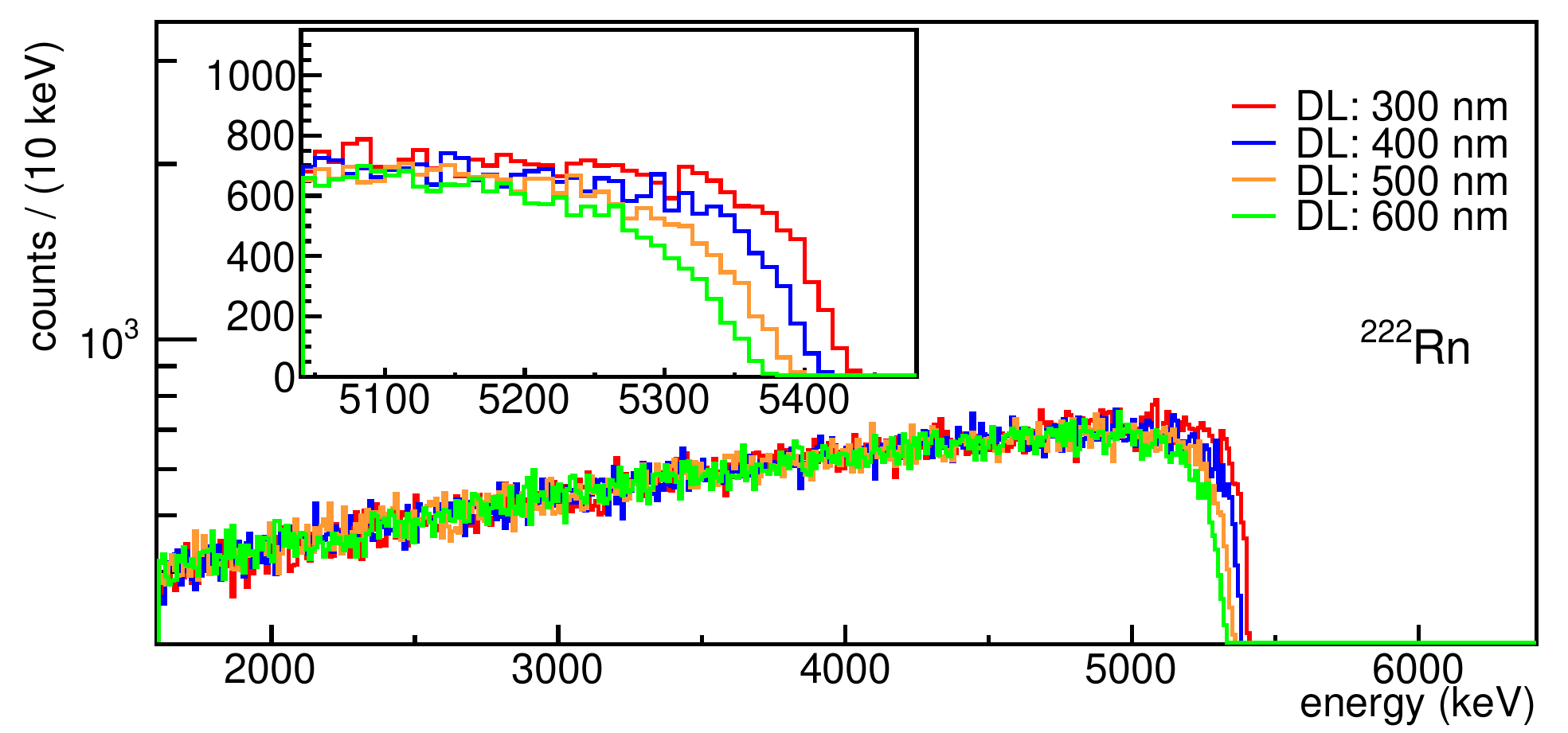}
\caption{ \label{fig:mc_alphas}
   Simulated spectra resulting from $^{210}$Po decays on the p$^+$ surface
  (left) and from $^{222}$Rn in LAr close to the p$^+$ surface
  (right) for different dead layer (DL) thicknesses. The spectra are scaled arbitrarily for visual purposes.
  }
\end{center}
\end{figure*}

 The resulting spectral shapes for $^{210}$Po on the p$^+$ detector surface
 and for \Rn\ in liquid argon are displayed in Fig.~\ref{fig:mc_alphas}.  The
 individual decays on the p$^+$ surface result in a peak like structure with
 its maximum at slightly lower energies than the corresponding $\alpha$ decay
 energy with a quasi exponential tail towards lower energy. The decays
 occurring in LAr close to the p$^+$ surface result in a broad spectrum without
 any peak like structure extending to lower energies. $\alpha$ decays of the
 other isotopes result in similar spectral shapes with different maximum
 energies.

 \subsection{$^{214}$Bi and $^{214}$Pb}

 The screening measurements indicate that the $^{226}$Ra daughters
 $^{214}$Bi and $^{214}$Pb are present in the vicinity of the detector
 array. Additionally, these isotopes are also expected on the detector p$^+$
 surface and in its close surrounding resulting from the $^{226}$Ra contamination of
 the detector surfaces.  The spectra expected from decays of $^{214}$Bi and
 $^{214}$Pb in or on the radon shroud,
 the mini shroud, the detector assembly, on the n$^+$ and p$^+$
 surfaces and in LAr inside the bore hole (BH) of the detector to represent
 decays close to the p$^+$ detector surfaces have been simulated.  
$^{214}$Bi and $^{214}$Pb are the only isotopes  in the $^{226}$Ra to $^{210}$Pb  chain decaying by $\beta$ decay accompanied by emission of high energy $\gamma$ particles. Except for contaminations of the p$^+$ surfaces that have been described in the previous section, the mean free paths of the $\alpha$-particles emitted in the decay chain are much smaller in LAr and germanium than the distance of the contamination sources from the detector active volume. Hence,  only the decays of $^{214}$Bi and $^{214}$Pb  can contribute to the background in the energy region of interest (RoI). These isotopes are assumed to be in equilibrium.

 The spectral shapes obtained by the simulation of $^{214}$Bi and $^{214}$Pb
 decays in the detector assembly, on the p$^+$ surface and inside the bore
 hole of the detector are shown in Fig.~\ref{fig:mc_shapes}.  The
 spectral shapes resulting from decays in or on the detector assembly components, the mini shroud and
 on the n$^+$ surface turn out to be very similar. Hence, these three are
 treated together and are represented by the spectrum obtained for $^{214}$Bi and
 $^{214}$Pb decays inside the detector assembly. The spectral shape from decays in the
 radon shroud exhibits a much lower peak to continuum ratio at lower energies
 (E$<$\,1500~keV), while for higher energies the spectral shape is similar to
 the one obtained from simulations of decays in the detector assembly.  For $^{214}$Bi
 and $^{214}$Pb decays on the p$^+$ surface and, to some extent also inside
 the LAr of the bore hole, the peak to continuum ratio is much reduced at
 higher energies because of the sensitivity to the electrons
 due to the thin p$^+$ dead layer.

\subsection{$^{228}$Ac and $^{228}$Th}
 \label{ssec:acth}

 The presence of $^{228}$Th is expected from screening measurements in the
 front end electronics and the detector suspension system. The characteristic
 $\gamma$ line at 2615~keV can be clearly identified in the spectra of the
 enriched coaxial detectors and the detector with natural isotopic abundance
 shown in Fig.~\ref{fig:spectra-ge}.  Possible locations for $^{228}$Th
 contaminations are the detector assembly and the mini shrouds in the close
 vicinity, the radon shroud and the heat exchanger of the LAr cooling system
 at the top of the cryostat.

As only negligible hints of $^{232}$Th or $^{228}$Th  surface contaminations were observed,  $\alpha$-decays resulting from the $^{232}$Th decay chain are not considered in the following.

 No significant top - bottom asymmetries in the count rates of the
 \Tl\ and \Bi\ \gam\ lines could be observed. This indicates that the 
 front end electronics and suspension system above the detector array (medium 
 distance sources) and the calibration source 
 at the bottom of the cryostat (far source) 
 are not the main background contributions.

 As $^{228}$Ac and $^{228}$Th do not necessarily have to be in equilibrium,
 the two parts of the decay chain were simulated separately.  From the
 sub-decay chain following the $^{228}$Th decay only the contributions from
 the $^{212}$Bi and $^{208}$Tl decays were simulated, as theses are the only
 ones emitting high energetic $\gamma$ rays and electrons that can reach the
 detectors.

 The resulting spectral shapes are shown in Fig.~\ref{fig:mc_shapes}.  For
 $^{228}$Th contaminations in or on the radon shroud and the heat exchanger the
 continuum above the 2615~keV line is suppressed, while for sources in or on the
 detector assembly and the mini shroud the continuum above the 2615~keV  
 peak can be
 significant due to summation of two $\gamma$ rays or of an emitted electron
 and a $\gamma$ particle.

 The simulated spectral shapes resulting from $^{228}$Ac decays differ mainly
 for energies E$<$\,1~MeV. Generally, the peak to continuum ratio is higher for
 the spectrum obtained with contamination inside the detector assembly. 

\subsection{$^{42}$Ar}
 \label{ssec:ar}

\begin{figure*}[t]             
\begin{center}
\includegraphics[width=0.9\columnwidth]{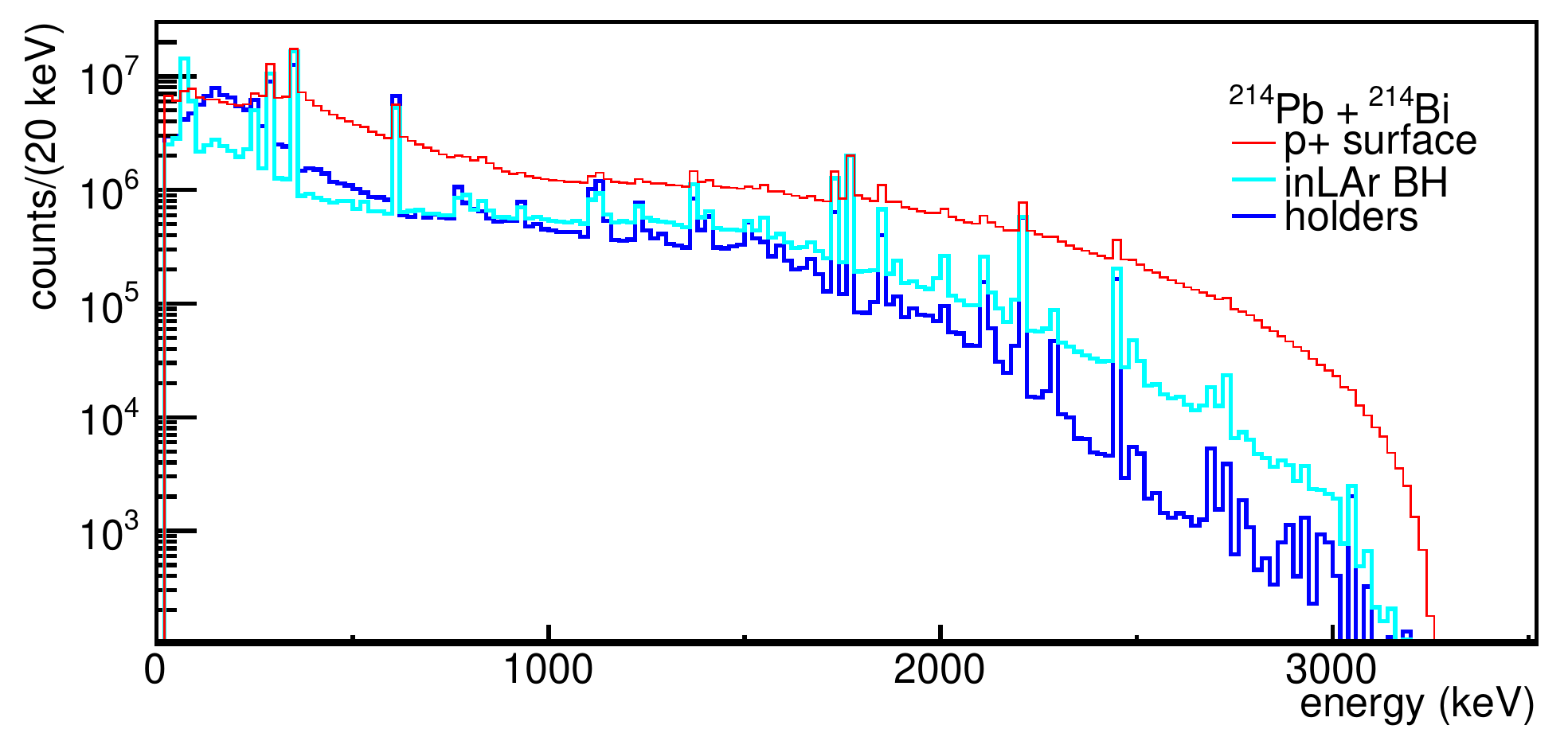}\hspace*{5mm}
\includegraphics[width=0.9\columnwidth]{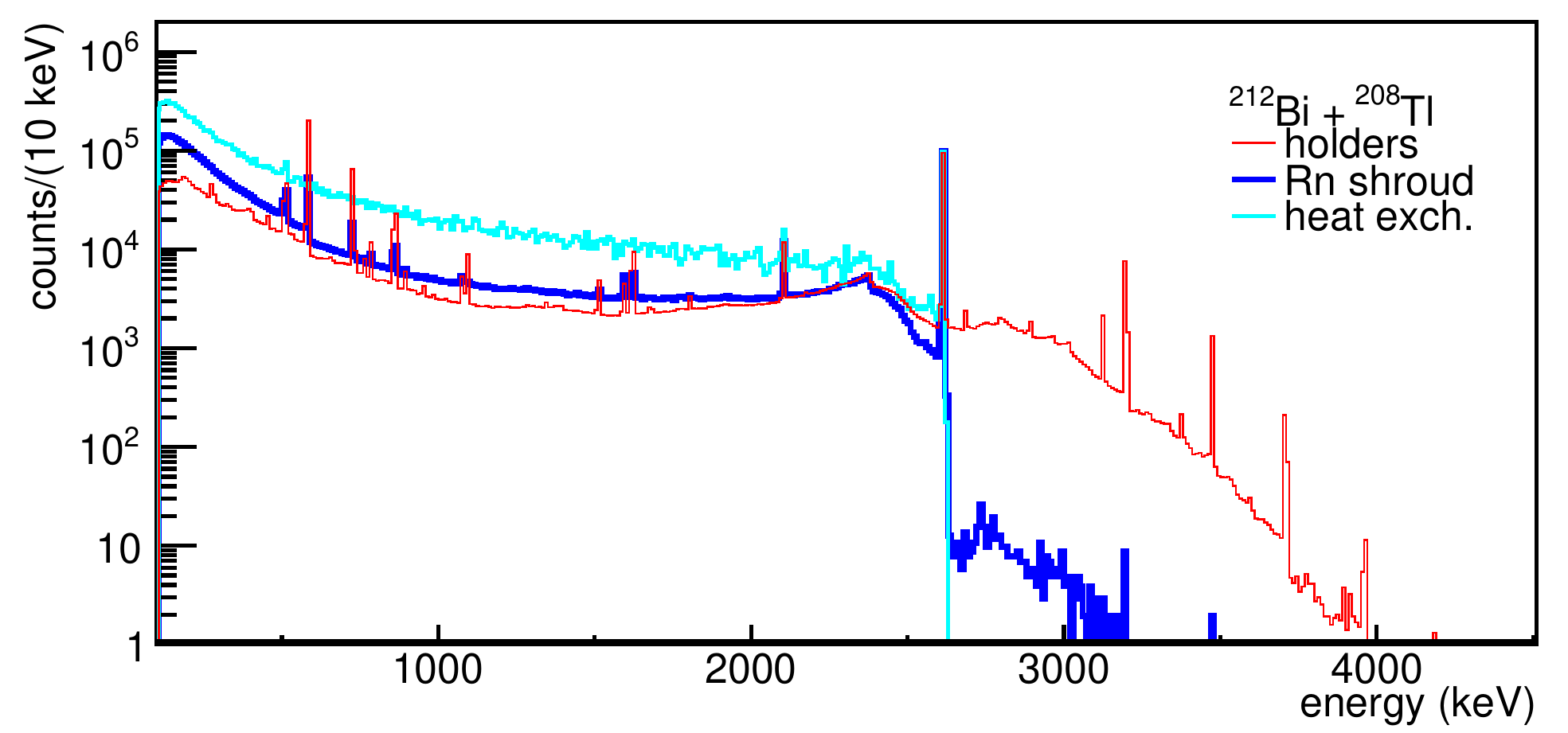}\\
\includegraphics[width=0.9\columnwidth]{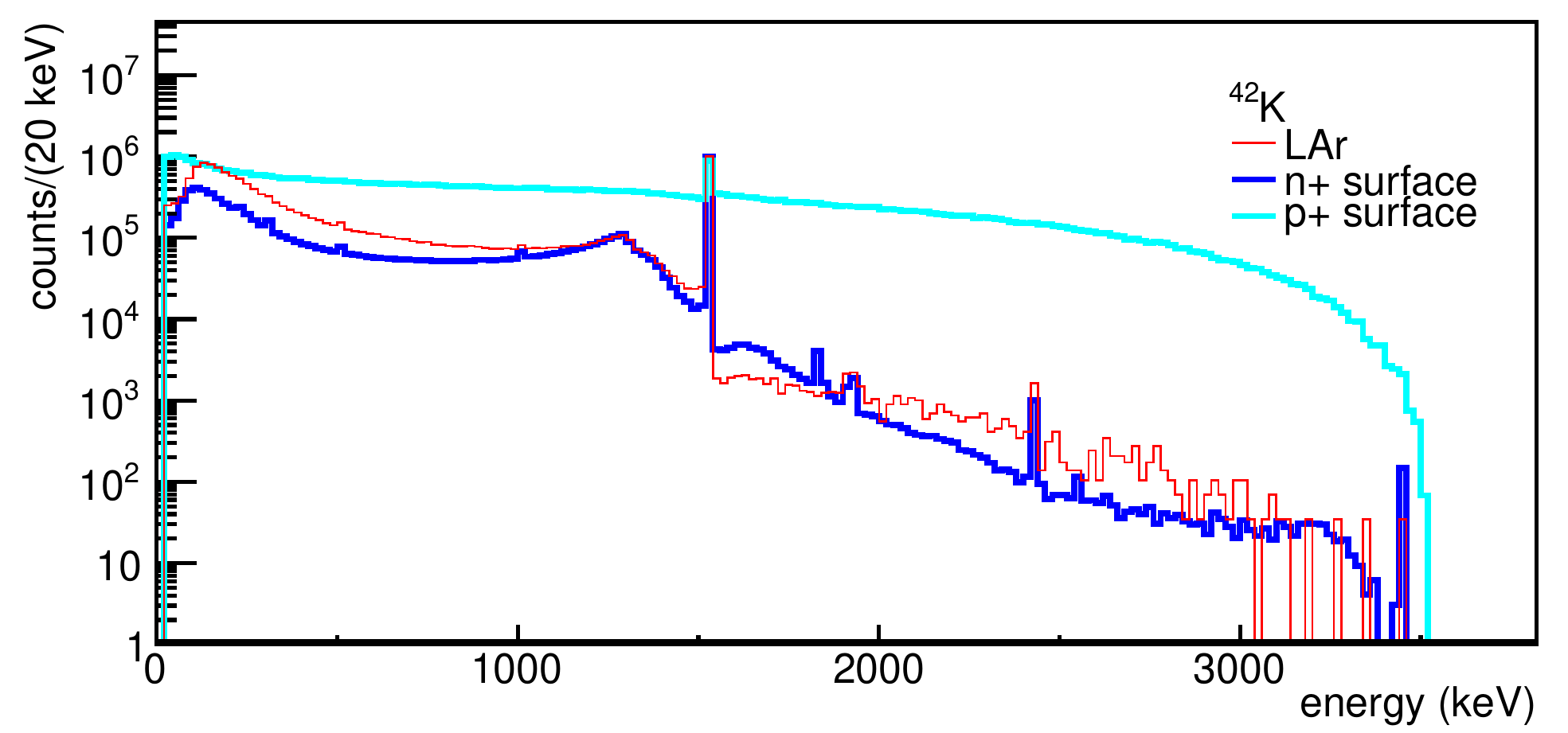}\hspace*{5mm}
\includegraphics[width=0.9\columnwidth]{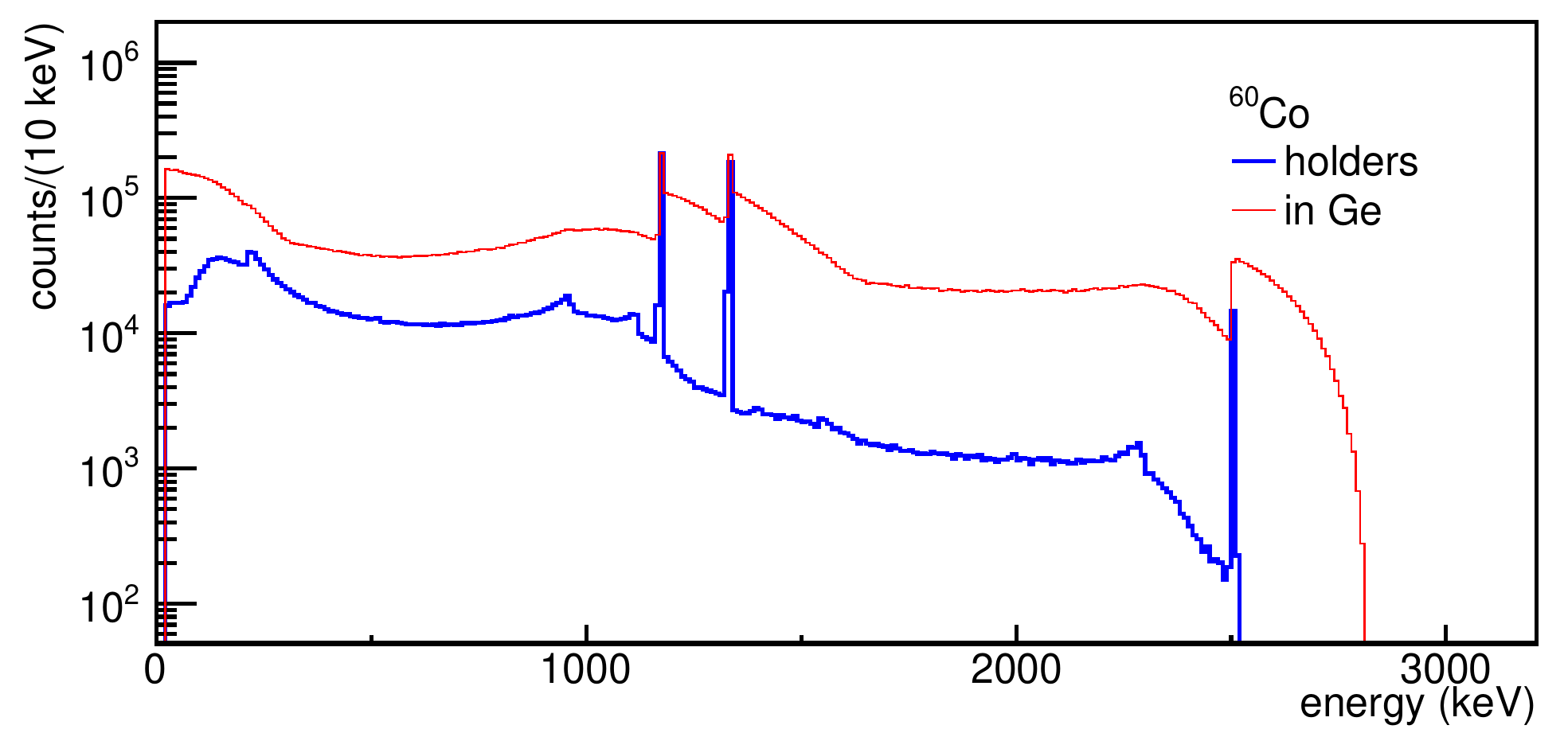}
\caption{ \label{fig:mc_shapes}
      Simulated spectra for different background contributions at different
      source locations. Spectra are scaled arbitrarily for visual purposes.
      \Bi\ and $^{214}$Pb on the p$^+$ detector surface and in LAr of the bore
      hole (BH) close to the p$^+$ surface (upper left), \Th\ in the detector
      detector assembly, the radon shroud and on the heat exchanger (upper right)
      $^{42}$K homogeneous in LAr, on the n$^+$ and on the p$^+$ surfaces
      (lower left) and $^{60}$Co in the detector assembly and in the germanium
      (lower right).
 }
\end{center}
\end{figure*}

 While the distribution of $^{42}$Ar is homogeneous inside LAr, the short
 lived ionized decay product $^{42}$K (T$_{1/2}$\,= 12.3 hours) can have a
 significantly different distribution due to drifts of the $^{42}$K 
 ions inside the
 electric fields that are 
 present near the detectors.  Spectra for
 three $^{42}$K distributions have thus been simulated: ($i$) homogeneous in
 LAr in a volume of 6.6~m$^3$ centered around the full detector array, ($ii$)
 on the n$^+$ and ($iii$) on the p$^+$ detector surface of the detectors. 
The n$^+$  surface has a thickness comparable to the absorption length of the electrons emitted in the $^{42}$K decays  in germanium. In the $^{42}$K n$^+$ surface simulation a 1.9 mm dead layer thickness was used, typical for the coaxial detectors. The resulting spectral shape is similar to the one obtained for $^{42}$K homogeneous distribution in LAr.
Also, as the spectral shape is not expected to vary strongly between the detectors, $^{42}$K on the p$^+$ surface was simulated only for a single detector. In this case a much higher contribution at high energies is present due to the electrons in the $^{42}$K decay. Consequently a much lower 1525~keV peak to continuum ratio is expected. The simulated spectral shapes are shown in Fig.~\ref{fig:mc_shapes}.



\subsection{$^{60}$Co}
 \label{ssec:co}

 Two simulated spectral shapes were used for the background decomposition, one
 for \Co\ in the detector vicinity and one for \Co\ inside the detectors.  The
 resulting spectral shapes are shown in Fig.~\ref{fig:mc_shapes}.  The peak to
 continuum ratios are significantly different due to the electron emitted in
 the decay that can only deposit energy in the detector for a
 contamination in the close vicinity of the detector 
 but is shielded efficiently by the liquid argon for
 contaminations further away.

\subsection{\twonu\ decay}
 \label{ssec:twonu}

 The spectral shape induced by \twonu\ decay of \gess\ was simulated with a
 homogeneous distribution of \gess\ inside each individual detector. Decays
 inside the active volume and the dead layer of the detectors were simulated
 separately and summed later weighted by their mass fractions.  The spectral
 shape of the electrons emitted in \twonu\ decay as reported in
 Ref.~\cite{tretyak_2nbb_shape} and implemented in the DECAY0 event generator
 was used.

\subsection{\kvn}
  Screening measurements revealed that contaminations with \kvn\ are expected
  in the detector assembly, the nearby front end electronics, and in the near
  part of the detector suspension system. The 1460~keV $\gamma$ line intensity
  is, within uncertainty, the same for the individual detectors. There is no
  reasonable explanation for an isotropic distant source.  Hence, in the
  analysis it is assumed that the $^{40}$K contamination is in the detector
  assembly.

\section{Background decomposition}
  \label{sec:modeling}

 A global model that describes the background spectrum was obtained by fitting
 the simulated spectra of different contributions to the measured energy
 spectrum using a Bayesian fit. A detailed description of the statistical
 method is given in sec.~\ref{ssec:statistic}.

 First, the high energy part of the spectrum was analyzed. Above 3.5~MeV,
 the Q-value of \kvz, the main contribution to the energy spectrum is expected
 to come from $\alpha$ decays close to or on the thin dead layers on the
 detector p$^+$ surfaces. The time distribution of the events above 3.5~MeV
 gives information on the origin of those events.  The event rate distribution
 was therefore analyzed a priori to check the assumptions on the sources of
 $\alpha$ induced events. The spectrum in the 3.5 - 7.5~MeV region was then
 analyzed by fitting it with the simulated spectra resulting from $\alpha$
 decays from isotopes in the \Ra\ decay chain. The spectral model developed
 for the $\alpha$ induced events also allows one to gain information on the
 background due to \Bi\ in the \Ra\ decay chain.

 Subsequently, the energy range of the fit was enlarged to include  as 
 much data as possible for a higher accuracy of the model, including \qbb\
 of \gess. Background components discussed in sec.~\ref{sec:back} are
 used in the spectral analysis.  A minimum fit was performed by taking only a
 minimum amount of well motivated close background sources into account.  
In a
 maximum fit further simulated background components, representing also medium distance and distant sources, were added to the
 model.  The analysis was repeated for the different data sets and the obtained
 global models were used to derive the activities of the different
 contaminations.

\subsection{Statistical analysis}
\label{ssec:statistic}
 The analyses of both event rate and energy distributions were carried out by
 fitting binned distributions.  The probability of the model and its
 parameters, the posterior probability is given from Bayes theorem as
\begin{equation}
\label{eq:posterior} 
P(\vec{\lambda}|\vec{n}) = \frac{P(\vec{n}|\vec{\lambda}) P_{0}(\vec{\lambda})}
{\int P(\vec{n}|\vec{\lambda}) P_{0}(\vec{\lambda}) d\vec{\lambda}} ,
\end{equation}
 where P($\vec{n}|\vec{\lambda}$) denotes the likelihood and
 $P_{0}(\vec{\lambda})$ the prior probability of the parameters. The
 likelihood is written as the product of the probability of the data given the
 model and parameters in each bin
\begin{equation} 
P(\vec{n}|\vec{\lambda}) = \prod_{i} P(n_{i}|\lambda_{i})
= \prod_{i} \frac{e^{-\lambda_{i}} \lambda_{i}^{n_{i}} }{ n_{i}! } ,
\end{equation} 
 where $n_{i}$ is the observed number of events and $\lambda_{i}$ is the
 expected number of events in the i-th bin.

 For the analysis of the event rate distributions, the expected number of
 events, $\lambda_{i}$, is corrected for the live time fraction. For example,
 when fitting the event rate distribution with an exponential function, the
 expectation is written as
\begin{equation} \lambda_{i}
 = \epsilon_{i} \int_{(i-1)\Delta{t}}^{i\Delta{t}} \!
 {N_{0} \cdot e^{-\ln2 \, {t}/T_{1/2}}\,dt } \end{equation}  
 where $\epsilon_{i}$ is the value in the $i$-th bin of the live time fraction
 distribution, $\Delta{t}$ is the bin width, $N_{0}$, the initial event rate
 and $T_{1/2}$, the half life.

 The spectral analysis was done by fitting the spectra of different
 contributions obtained from the MC simulations to the observed
 energy spectrum. The expected number of events in the $i$-th bin, $\lambda_{i}$
 is defined as the sum of the expected number of events from each model
 component in the $i$-th bin and is written as
\begin{equation} \label{eq:expectation} 
\lambda_{i} = \sum_{M}{\lambda_{i,M}},
\end{equation} 
 where $M$ corresponds to the simulated background components considered in
 the fit.  The expectation from a model component in the $i$-th bin is defined
 as
\begin{equation} \label{eq:expectM}
\lambda_{i,M} = N_{M} \int_{\Delta{E_{i}}} { f_{M}(E) dE } 
\end{equation}
 where $f_{M}(E)$ is the normalized simulated energy spectrum of the component
 $M$ and $N_{M}$ is the scaling parameter, i.e., the integral of the spectrum
 in the fit window.

 Global fits of the experimental spectra and fits of the event rate
 distributions were performed according to the procedure described above and
 using the Bayesian Analysis Toolkit BAT~\cite{BAT}. The prior probabilities of
 the parameters $P_{0}(\vec{\lambda}_{M})$ are given as a flat distribution
 if not otherwise indicated.

\subsection{$\alpha$ event rate analysis}

 The $\alpha$ induced events in the energy range 3.5 - 7.5~MeV are expected to
 mainly come from $\alpha$ emitting isotopes in the \Ra\ decay chain, which
 can be broken at $^{210}$Pb and at \Po\ with half lives of 22.3~yr and
 138.4~days, respectively. An analysis of the time distribution is therefore
 used to infer the origin of these events.

\begin{figure}[t]
\begin{center}
\includegraphics[width=0.95\columnwidth]{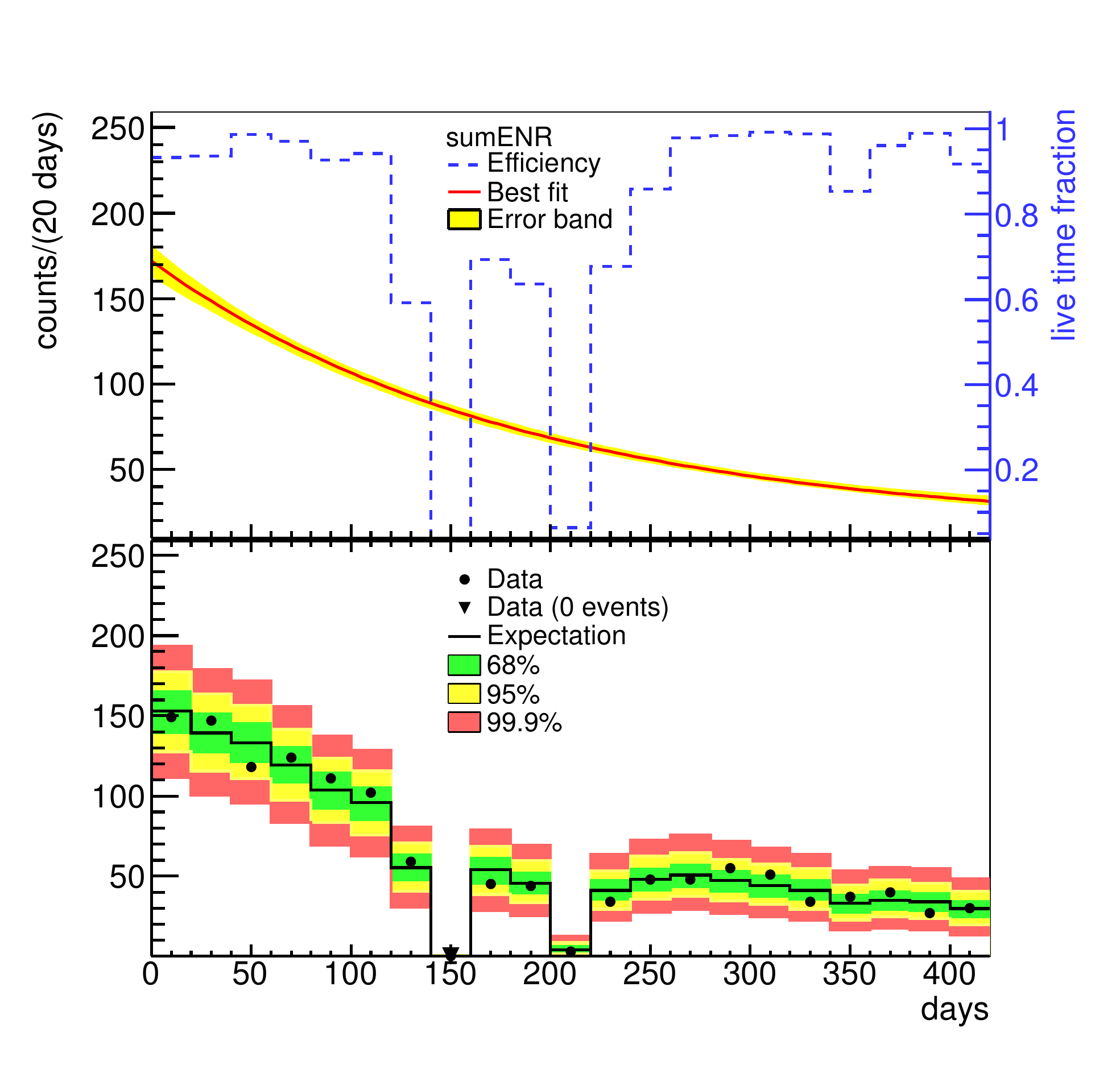}
\includegraphics[width=0.95\columnwidth]{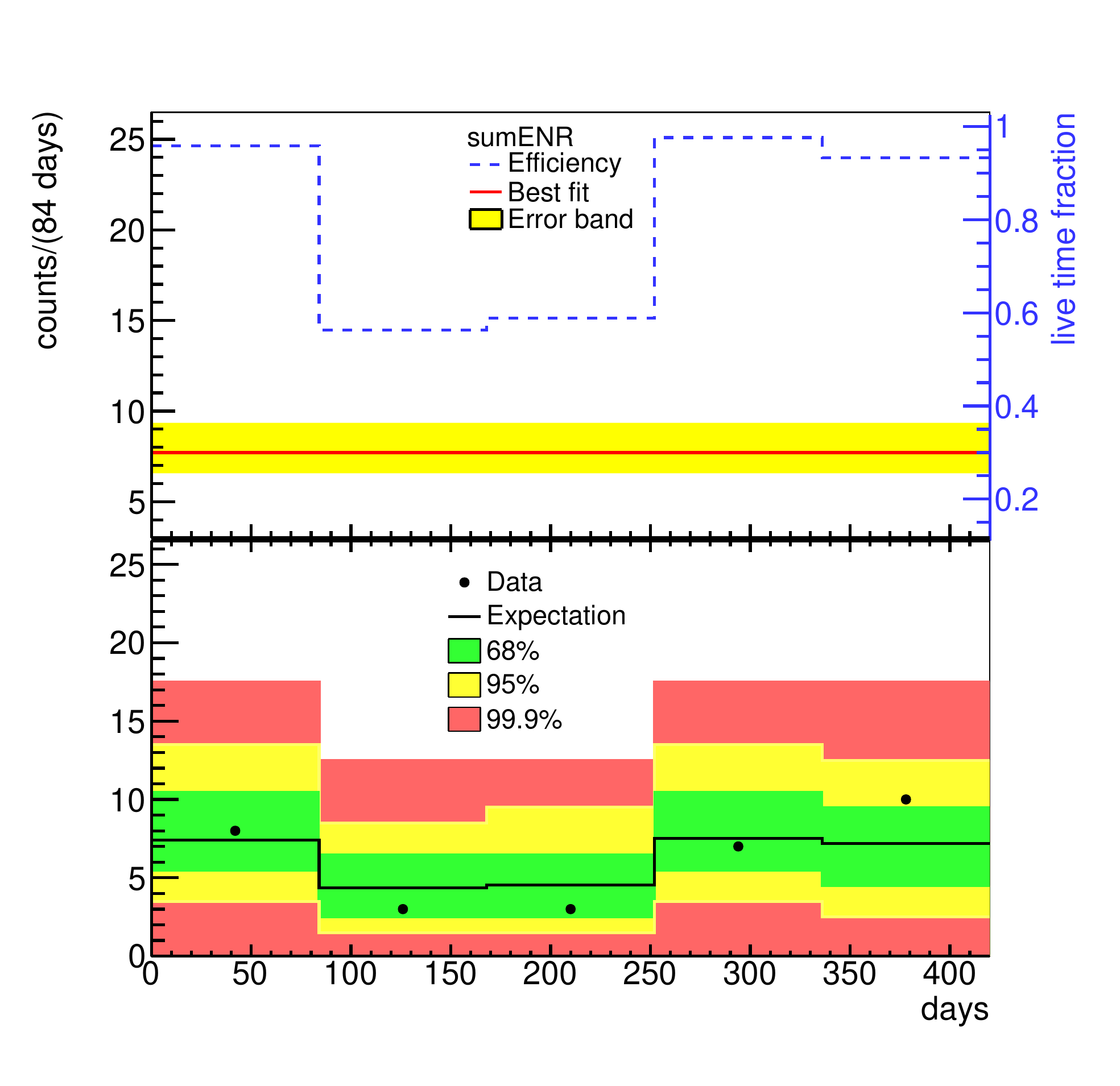}
\caption{ \label{fig:decay_times}
   Results of fitting the event rate distributions for events in the 3.5
   - 5.3\,MeV range with an exponential plus constant rate model (top) and
   for the events in the 5.3 - 7.5~MeV range fitted with a constant rate
   model (bottom). The upper panels show the best fit model (red lines) with
   68\,\% uncertainty (yellow bands) and the live time fraction distribution of
   the experiment (dashed blue line). The lower panels show the observed
   number of events (markers) and the expected number of events (black
   line) due to the best fit. The smallest intervals of 68\,\%, 95\,\% and
   99.9\,\% probability for the expectation are also shown in green, yellow and
   red regions, respectively~\cite{RA}.
}
\end{center}
\end{figure}

 If only \Po\ is present as a contamination, the count rate in the energy
 region from 3.5 to 5.3~MeV (see Fig.~\ref{fig:decay_times}) should decrease
 with a decay time expected from the half life of \Po. Whereas an initial
 $^{210}$Pb surface contamination would cause an event rate in this energy
 interval appearing constant in time, as the half life of $^{210}$Pb is much
 longer than the life time of the experiment. The event rate at energies
 E\,$>$\,5.3 MeV should appear constant in time in case events originate from the
 decay of \Ra\ with a half life of T$_{1/2}$\,=\,1600~yr and its short lived
 daughters.

 The {\it GOLD-coax} data set was analyzed using the statistical method
 described in sec.~\ref{ssec:statistic}. For the energy region 3.5 to 5.3~MeV
 two models were fitted to the event rate distributions: an exponentially
 decreasing rate and an exponentially decreasing plus a constant rate. For the
 rate of events with E\,$>$\,5.3 MeV only a constant rate was fitted. The live
 time fraction as a function of time is taken into account in the
 analysis. A strong prior probability distribution for the half life parameter,
 $P_0$(T$_{1/2}$), is given as a Gaussian distribution with a mean value of
 138.4~days and a standard deviation of 0.2~days for both models, to check the
 assumption of an initial \Po\ contamination. The analysis was also performed by
 giving a non-informative prior, a flat distribution, on the half life
 parameter.  For the energy range between 3.5 and 5.3~MeV both models describe
 the distribution adequately. The exponentially decreasing plus constant rate
 model is, however, clearly preferred as it results in  a nine times higher  
 p-value of 0.9.  The  rates derived from the fit are
 (0.57\,$\pm$\,0.16)~cts/day for the constant term and (7.9\,$\pm$\,0.4)~cts/day for
 the initial rate, exponentially decreasing with a half life of
 (138.4\,$\pm$\,0.2)~days, according to the fit performed with the strong prior on
 the half life. The fit with non informative prior on the half life parameter
 results in a half life of (130.4\,$\pm$\,22.4)~days, which is in very good
 agreement with the half life of \Po. The constant event rate model for the
 events with E\,$>$\,5.3 MeV gives a very good fit as well and results in a count
 rate of (0.09\,$\pm$\,0.02)~cts/day.

 Fig.~\ref{fig:decay_times} shows the observed event rate for both energy
 regions and the expectation due to the best fit event rate model together with
 the smallest intervals containing 68\,\%, 95\,\% and 99.9\,\% probability for
 the expectation. The live time fraction for data taking was varying with time
 and is also shown. The observed exponential decay rate clearly shows that the
 majority of the observed $\alpha$ events come from an initial
 \Po\ contamination on the detector surfaces. The results of the time analysis
 show agreement with the assumed origins of the events.

\subsection{Background components}
\subsubsection{$^{226}$Ra decay chain on and close to the detector surface}

 As demonstrated in Fig.~\ref{fig:mc_alphas} the maximal energy in the peak
 like structure resulting from an $\alpha$ decay on the detector surface is
 very sensitive to the dead layer thickness. The $^{210}$Po peak structure
 around 5.3~MeV with high statistics in the {\it GOLD-coax} data set was used
 to determine the effective dead layer model.  Spectra from $^{210}$Po
 $\alpha$ decay simulations on the p$^+$ surface with different dead layer
 thicknesses were used to fit the spectrum in the energy region dominated by
 the $^{210}$Po peak, i.e. between 4850 and 5250~keV.  The weight of each
 spectrum was left as a free parameter. A combination of the spectra for 300,
 400, 500 and 600~nm dead layer thicknesses describes the observed peak
 structure well and results in a very good fit, whereas a spectrum with a
 single dead layer thickness does not give a sufficiently good fit.  Consequently the derived dead layer model was used for the later fits of $\alpha$ induced spectra. Spectra with
 lower (down to 100~nm) and higher dead layer thicknesses (up to 800~nm) give
 insignificant contributions, if at all, to the overall spectrum.

\begin{figure}[t!]
\begin{center}
\includegraphics[width=0.95\columnwidth]{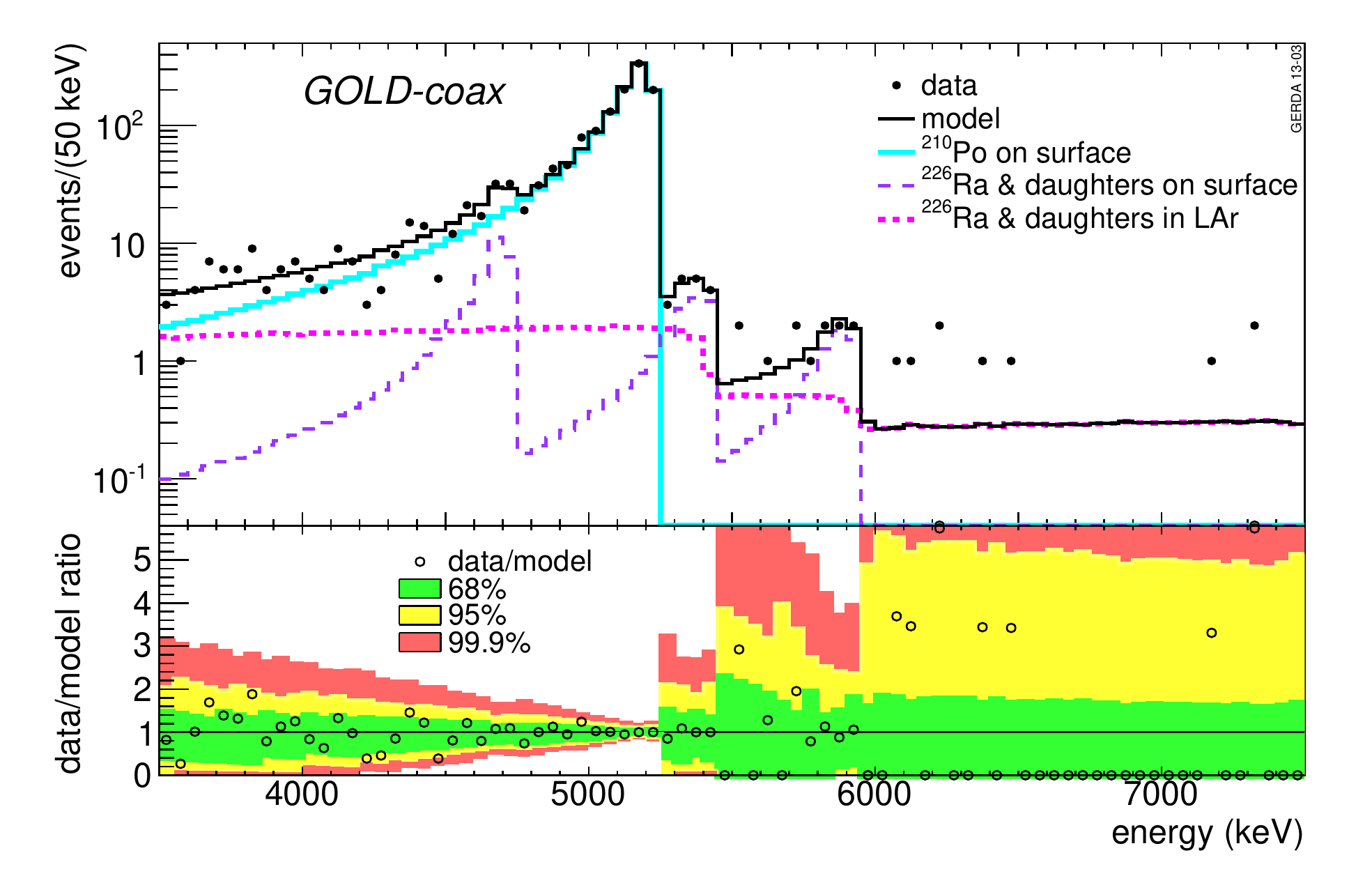}\\
\includegraphics[width=0.95\columnwidth]{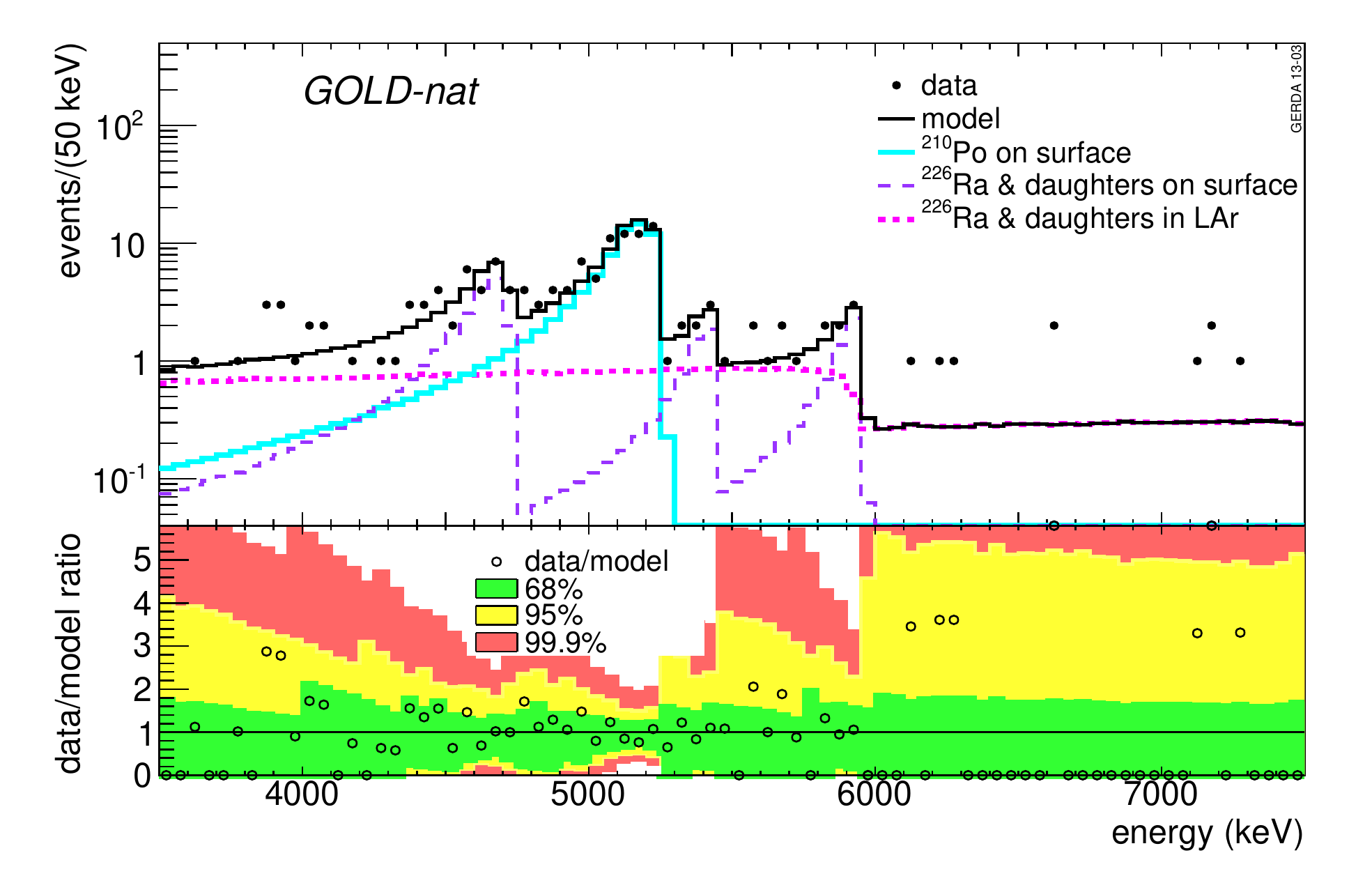}
\end{center}
\caption{\label{fig:alpha_fits}
      The upper panels show the best fit model (black histogram) and observed
      spectrum (black markers) for the {\it GOLD-coax} (upper plots) and the
      {\it GOLD-nat} (lower plots) data sets. Individual components of the
      model are shown as well.  The lower panels show the ratio of data and
      model and the smallest intervals of 68\,\% (green), 95\,\% (yellow) and
      99.9\,\%\,(red) probability for the model expectation.
}
\end{figure}

 In order to describe the whole energy interval dominated by $\alpha$-induced 
 events, the simulated spectra of $\alpha$ decays of $^{210}$Po as well as from
 $^{226}$Ra and its short lived daughter nuclei on the p$^+$ surface and in
 LAr (see Table~\ref{tab:mcs}) were used to fit the energy spectrum between
 3500 and 7500~keV.  The number of events in the considered energy range was left as a free
 parameter for each component. The same analysis was repeated for different data sets.  The best
 fit model together with the individual contributions and observed spectrum
 for the {\it GOLD-coax} and {\it GOLD-nat} data sets are shown in
 Fig.~\ref{fig:alpha_fits}. While the surface decays alone can successfully
 describe the observed peak structures, they could not describe the whole
 spectrum. A contribution from an approximately flat component, like the spectra
 from $\alpha$ decays in LAr, is needed in the model.  All the data sets, and
 even the spectrum of individual detectors can be
 very well described by this model. Therefore, no further possible
 contributions from other $\alpha$ sources are considered in the analysis of
 this energy range.

 The number of expected events in the whole energy range (0.1 - 7.5~MeV) from each component
 of the model are listed in Table~\ref{tab:alpha_rates}.  In each subsequent
 decay in the chain the number of events measured is systematically reduced
 with respect to the mother nuclei for p$^+$ surface decays.  Due to having
 few events above 5.3~MeV, mostly only a limit could be derived for the
 components of decays in LAr.  Nevertheless, a similar systematic decrease can 
 be observed. The model for all data sets and also the model for individual
 detectors show the same effect. This could be explained by the removal of the 
 mother nucleus from the surface by recoil with $\approx$100~keV.
 Since the range of $\alpha$ particles in LAr and in Ge (DL) is only
 few \mum, the detection efficiency of the $\alpha$ particle emitted by the
 isotope that has recoiled from the surface can be reduced.  However, the
 recoil of the nuclei away from the surface will practically not effect the
 detection efficiencies of $\beta$ particles or $\gamma$ rays, since they have
 significantly greater penetration depths in LAr.  Therefore, decays of
 \Bi\ in LAr in the vicinity of the surface (\mum) and decays directly on the
 p$^+$ surface are expected to have very similar detection
 efficiencies and result in the same spectral shapes within uncertainties. Thus, for the rate of
 \Bi\ decaying on the detector p$^+$ surface a rate equal to the one of the
 \Ra\ decays on the p$^+$ surface as obtained by the $\alpha$ model is
 assumed. In the
 background model this expectation is accounted for by putting a Gaussian prior
 probability on the number of \Bi\ events on the p$^+$ surface as 
 described in the following section.

\begin{figure}[h!]
\begin{center}
\includegraphics[width=0.9\columnwidth]{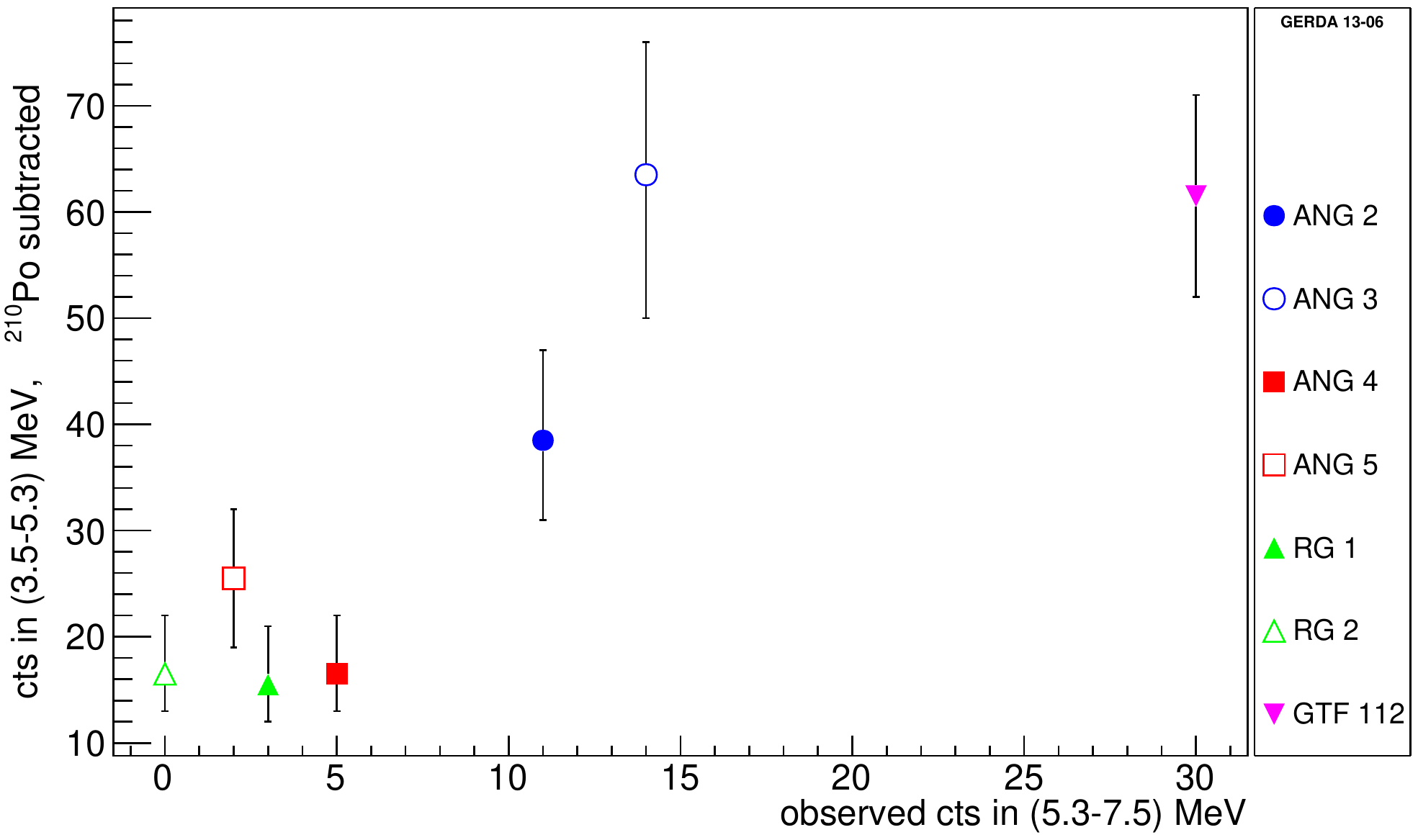}
\caption{\label{fig:alpha_correlation}
Number of observed events with energies $>$\,5.3 MeV versus the number
of events in the interval  3.5 - 5.3 MeV with the $^{210}$Po
contribution subtracted according to the model for individual 
detectors.
 }
\end{center}
\end{figure}

 Fig.~\ref{fig:alpha_correlation} shows the number of observed events with
 energies $>$\,5.3~MeV versus the number of expected events from the $\alpha$
 model excluding the contribution from \Po\ in the energy range between 3.5
 and 5.3~MeV for individual detectors. A correlation between the two
 numbers is visible, supporting the assumption that the events between 3.5 and
 5.3 MeV that are not due to \Po\ and events above 5.3 MeV are originating
 from the same source.

\begin{table*}[t]
\begin{center}
\caption{\label{tab:alpha_rates}
       Number of events in the whole energy range (0.1 - 7.5~MeV) from
       each component of the $\alpha$ model obtained for different data sets.
       Shown are the mode and the smallest 68\,\% probability intervals or
       90\,\% quantiles of the marginalized distributions of the parameters. Note that while the marginalized distributions result in upper limits, the global best fits result in positive contributions.}
\begin{tabular}{l|rcrcrcrc}
\hline
     & \multicolumn{2}{c}{\it GOLD-coax} &  \multicolumn{2}{c}{\it GOLD-nat} & \multicolumn{2}{c}{\it GOLD-hdm} &  \multicolumn{2}{c}{\it GOLD-igex}\\
     & \multicolumn{8}{c}{number of counts in the spectrum}\\
\hline
$^{210}$Po p$^+$   & 1355& [1310,1400] & 76.5  &[66,88] &1285.5 & [1240,1320] & 74.5 &[65,86]\\
\hline
$^{226}$Ra p$^+$   &  50.5& [36.0,65.0]& 27.5   &[20,36] & 46.5 & [35,62] & 8.5 &[5,13]\\
$^{222}$Rn p$^+$   &  24.5& [18,33]    & 13.5   &[9,20]  & 23.5 & [17,32] & 6.5 &[3,10]\\
$^{218}$Po p$^+$   &  13.5& [9.0,19.0] & 15.5   &[10,20] & 13.5 & [9,19]  &     &$<$\,6\\
$^{214}$Po p$^+$   &      & $<$\,10      &        & $<$\,11  &      & $<$\,9    &     &$<$\,7\\
\hline
$^{226}$Ra LAr  &	   & $<$\,159.0 &         &$<$\,45   &      & $<$\,148  &      & $<$\,26 \\
$^{222}$Rn LAr  &          & $<$\,64    &         &$<$\,25   &      & $<$\,52   &      & $<$\,10\\
$^{218}$Po LAr  &          & $<$\,30    &         &$<$\,26   &      & $<$\,30   &      & $<$\,6\\
$^{214}$Po LAr  & 19.5     & [10, 29] & 16.5    & [8,27] & 14.5 & [8,25]  &      & $<$\,5\\
\end{tabular}
\end{center}
\end{table*}

\subsubsection{Additional relevant background components}
  The energy spectrum is described from 570~keV (from energies above the
  Q-value of the $^{39}$Ar decay) up to 7500~keV by considering all the
  background sources in the model that are expected to be present in the setup
  -- namely, \twonu\ decay of \gesix, $^{40}$K, $^{60}$Co, $^{228}$Th, $^{228}$Ac,
  $^{214}$Bi, $^{42}$K and $\alpha$-emitting isotopes in the $^{226}$Ra decay
  chain.

 A minimum model was defined to fit the spectrum with a minimum number of
 expected contributions from sources  close to the detector array.  It contains the following components: the \twonu\ 
 spectrum, $^{40}$K, $^{60}$Co, $^{228}$Th, $^{228}$Ac and $^{214}$Bi in or on the
 detector assembly, $^{42}$K distributed homogeneously in LAr and the best fit
 $\alpha$ model. For $^{60}$Co in germanium a flat prior probability
 distribution is given to the number of expected events, 
 allowing a maximum contribution of
 2.3~\mubq\ initial activity. This upper boundary is derived from the known
 activation history of the detectors with the assumption of 4~nuclei/(kg~day)
 cosmogenic production rate~\cite{marik_thesis}. \Bi\ on the p$^+$ surface is
 given a Gaussian prior probability due to the expected $^{226}$Ra activity on
 p$^+$ surface derived from the $\alpha$ model, with a mean equal to the
 marginalized mode and a $\sigma$ corresponding to the smallest 68\,\% interval
 of the parameter.

 The maximum model considers additional contributions.
These are: $^{42}$K on the p$^+$ and
 n$^+$ surfaces of the detectors, $^{228}$Th in or on the radon shroud and the
 heat exchanger, $^{228}$Ac and $^{214}$Bi in or on the radon shroud and $^{214}$Bi
 in the LAr close to p$^+$ surfaces of the detector. The number of expected 
 events  from all the
 additional components are left as free parameters, i.e. they are not given any 
 informative priors. 

\subsection{Fit results}
 The minimum and the maximum fits were performed in the energy range from 570
 to 7500~keV with a 30~keV binning using the statistical method given in
 sec.~\ref{ssec:statistic}.

\begin{figure*}[t]
\begin{center}
\includegraphics[width=0.95\columnwidth]{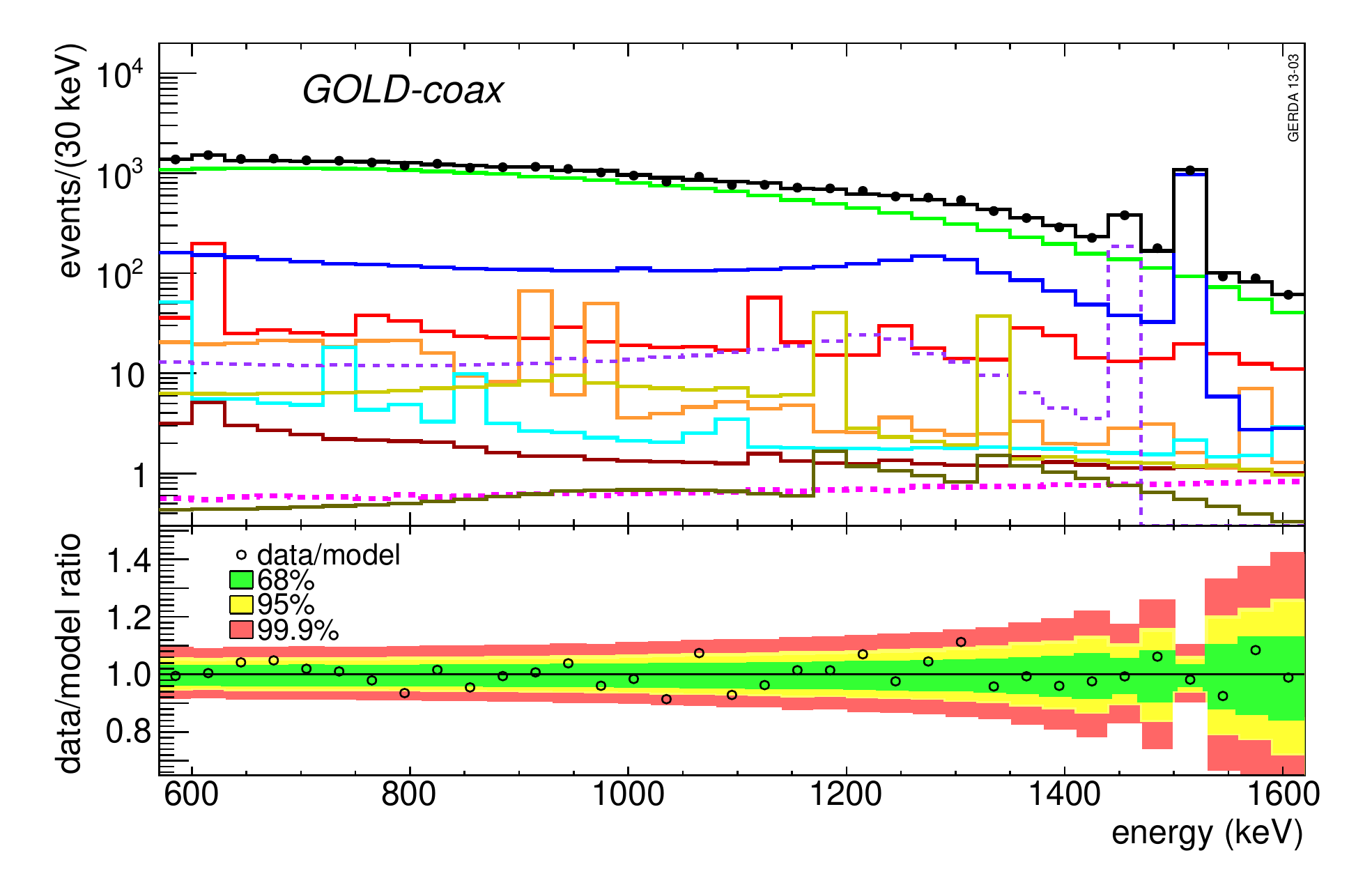}
\includegraphics[width=0.95\columnwidth]{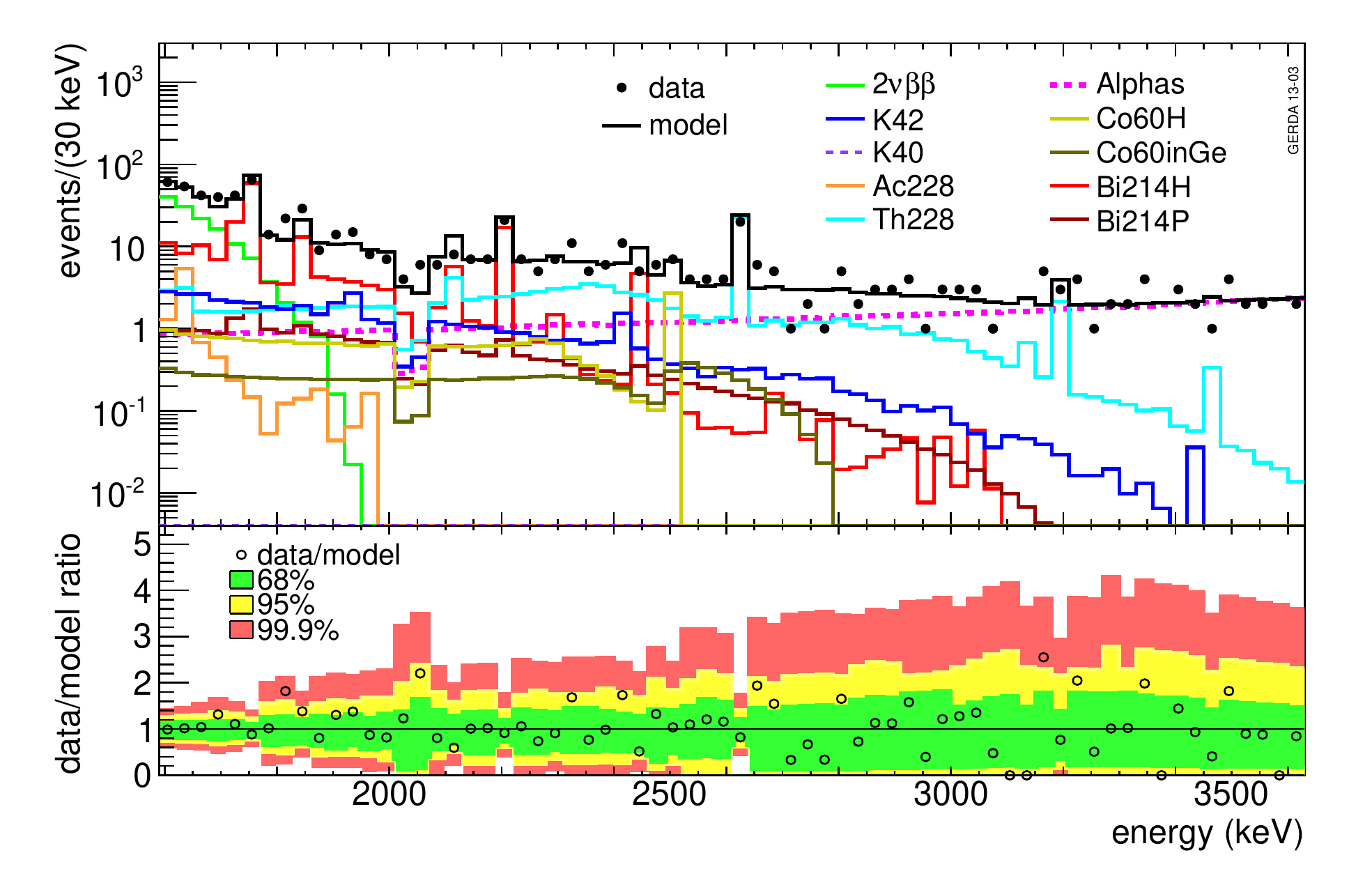}
\caption{\label{fig:min_fit_coax}
           Background decomposition according to the best fit minimum model of
           the {\it GOLD-coax} data set. The lower panel in the plots shows
           the ratio between the data and the prediction of the best fit model
           together with the smallest intervals of 68\,\% (green band),
           95\,\% (yellow band) and 99.9\,\% (red band) probability for the
           ratio assuming the best fit parameters.
}
\end{center}
\end{figure*}

\begin{figure*}[t]
\begin{center}
\includegraphics[width=0.95\columnwidth]{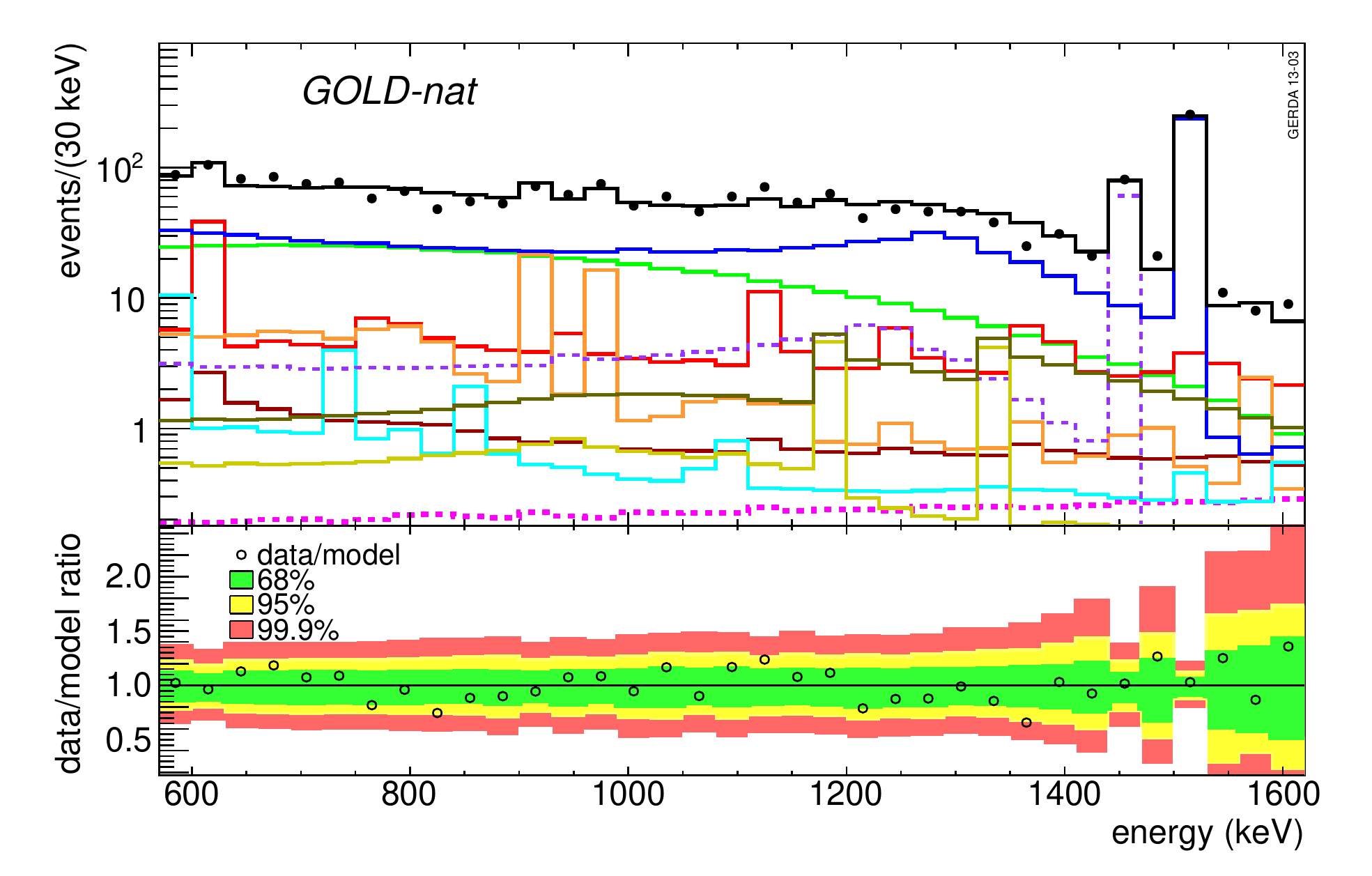}
\includegraphics[width=0.95\columnwidth]{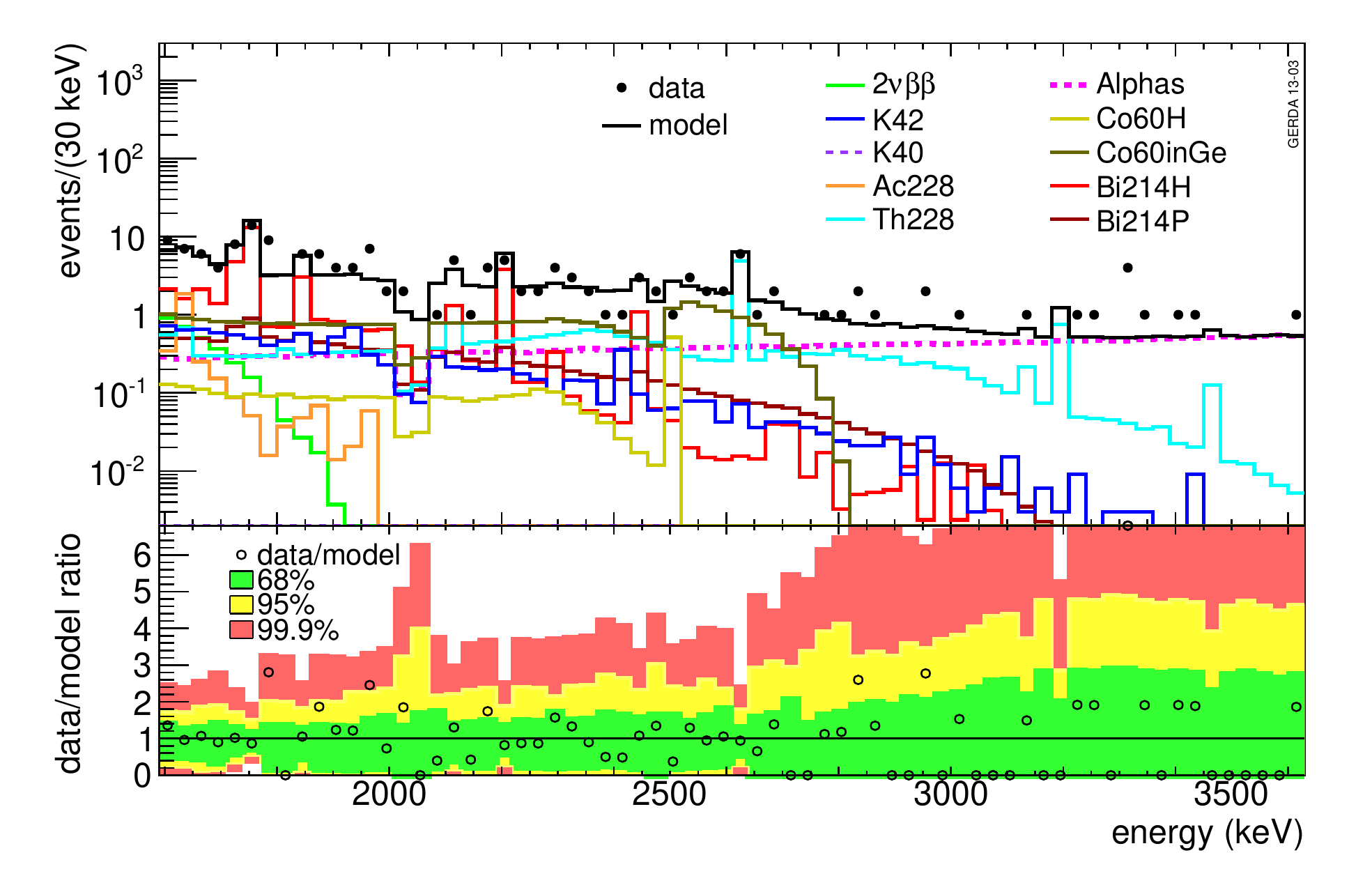}
\caption{  \label{fig:min_fit_nat}
           Same as in Fig.~\ref{fig:min_fit_coax} for the  {\it GOLD-nat} 
           data set. 
}
\end{center}
\end{figure*}

 Both models describe the data very well resulting in reasonable p-values with no clear preference for one model.  Figs.~\ref{fig:min_fit_coax} and
 \ref{fig:min_fit_nat} show the minimum model fit to the {\it GOLD-coax} and
 the {\it GOLD-nat} data sets in the energy regions between 570 and 1620~keV
 and between 1580 and 3630~keV. The lower panels in the plots show the data to
 model ratio (markers) and the smallest intervals containing 68\,\%, 95\,\% and
 99.9\,\% probability for the ratio assuming the best fit parameters in green,
 yellow and red bands, respectively~\cite{RA}. The data is within reasonable
 statistical fluctuations of the expectations.


 From the minimum and the maximum fit models, activities of contaminations of
 some of the components with different radioactive isotopes have been derived
 and are summarized in Table~\ref{tab:fits_results_golden}.

\begin{table*}[t]
\begin{center}
\caption{\label{tab:fits_results_golden}
         Activities of the individual contaminations of different hardware
         components derived from the global models of different data sets. 
         The location of the sources is also indicated. The
         numbers are according to the best fit model.  The uncertainty
         interval obtained as the smallest 68\,\% interval
         of the marginalized distributions of the
         parameters are given as well.  Limits are given with 90\,\% C.L. Also the activities as derived from
         the coincident spectra (see sec.~\ref{sec:cross-checks}) are shown.
         }
\begin{tabular}{r|l|l|cc|c|c}
source  &location &         & \multicolumn{2}{c|}{\it GOLD-coax} &{\it GOLD-nat} & {\it GOLD-coax}\\
                  &  &  units    &  minimum   &  maximum  & minimum & coincident\\ 
\hline
$^{40}$K  ~$^c)$& det. assembly            & \mubq/det. & 152[136,174]   & 151[136,174]       & 218[188,259]    & 252[164,340]\\

$^{42}$K  ~$^c)$ & LAr             & \mubq/kg     & 106[103,111]   & 91[72,99]          & 98.3[92,108]       & 168[150,186]\\

$^{42}$K   ~$^c)$ & p$^+$ surface   & \mubq        &                & 11.6[3.1,18,3]     &                  &     \\

$^{42}$K  ~$^c)$ & n$^+$ surface   & \mubq        &                &  4.1[1,2,8.5]      &                  & \\

$^{60}$Co  ~$^c)$ & det. assembly           & \mubq/det. & 4.9[3.1,7.3]   &  3.2[1.6,5.6]      & 2.6[0,6.0]      & 5.0[2.5,7.5] $^\star$)\\

$^{60}$Co  ~$^c)$ & germanium      & \mubq        & $>$\,0.4 $^\dagger$)& $>$\,0.2 $^\dagger$)&  6[3.0,8.4]      &\\

$^{214}$Bi  ~$^c)$ & det. assembly          & \mubq/det. & 35[31,39]      & 15[3.7,21.1]       & 34.1[27.3,42.1]  & 40[28,52]  \\

$^{214}$Bi  ~$^c)$& LAr close to p$^+$ & \mubq/kg     &                & $<$\,299.5   &\\

$^{214}$Bi ~$^m)$& radon shroud        &      mBq     &                & $<$\,49.9     &\\

$^{214}$Bi  ~$^c)$ & p$^+$ surface & \mubq        & 2.9[2.3,3.9]  $^\dagger$) & 3.0[2.1,4.0] $^\dagger$) &1.6[1.2,2.1] $^\dagger$)& \\

$^{228}$Th  ~$^c)$ & det. assembly       & \mubq/det. & 15.1[12.7,18.3]&  5.5[1.8,8.8]      & 15.7[10.0,25.0]  & 9.4[7.9,10.9]\\

$^{228}$Ac  ~$^c)$ & det. assembly     & \mubq/det. & 17.8[10.0,26.8]&  $<$\,15.7     & 25.9[16.7,36.7]  & 33[18,48]\\

$^{228}$Th ~$^m)$& radon shroud     &   mBq        &                &  $<$\,10.1     &                  &\\

$^{228}$Ac  ~$^m)$& radon shroud     &    mBq       &                & 91.5[27,97]        &                  &\\

$^{228}$Th ~$^f)$& heat exchanger&    Bq        &                & $<$\,4.1\\

\end{tabular}
\end{center}
source distance: ~$^c)$ close ($<$\,2~cm); ~$^m)$ medium (2-30~cm); ~$^f)$ far ($>$\,30~cm)\\
$^\dagger$) prior: discussed in the text\\
$^\star$) single: Obtained from coincident spectrum with histogram entries for
      each detector event separately.
\end{table*}

 A comparison of the resulting activities in
 Table~\ref{tab:fits_results_golden} with the known inventory of radio
 contaminations shown in Table~\ref{tab:thorium} shows that all contaminations
 expected from screening are seen in the background spectra.  However, the
 activities identified by screening measurements are not sufficient to explain
 the total background seen. The minimum model describes the background
 spectrum well without any medium distance and distant contaminations. Also if
 medium distance and distant sources are added, the largest fraction to the
 background comes from close sources, especially on the p$^+$ and n$^+$ surfaces. Note, that the activity obtained for
 $^{42}$K and hence for the $^{42}$Ar contamination of LAr is higher than the
 previously most stringent limit reported in Ref.~\cite{42Ar}.

 In the maximum model, strong correlations are found between several
 background sources. Contaminations of $^{42}$K on the n$^+$ surface and in
 LAr can not be distinguished. Similarly, the model has no distinction power
 between contaminations of the radon shroud, the heat exchanger and the
 detector assembly with \Bi, $^{228}$Th and $^{228}$Ac.  This explains the differences 
 of the derived activities in the two models.  The main difference between
 minimum and maximum models is the number of events on the p$^+$ surface of
 the detectors.

 A fit of the background model has also been made to the {\it SILVER-coax}
 data set. Its overall spectral shape can only be described sufficiently well,
 if either an additional $^{42}$K contamination of the p$^+$ surface and/or an
 additional \Bi\ contamination of the LAr is assumed. Both additional
 contaminations seem plausible after a modification of the experimental
 surrounding like the insertion of BEGe detectors to the cryostat.

\section{Background model for BEGe detectors}
   \label{sec:begemodel}

 An equivalent procedure as for the coaxial detectors was used to model the
 energy spectrum observed for the {\it SUM bege} data set.  Since the exposure
 collected with the BEGe detectors is much smaller than for the coaxial data
 set, only a qualitative analysis is possible for this data set. 
 The lower mass of the BEGe detectors with respect to the coaxial
 detectors reduces the detection efficiency for full energy peaks.
 Hence, fewer $\gamma$ lines are positively identified in the BEGe
 spectrum.  This makes it even more difficult to establish and to constrain  
 possible background components.

 The contributions to the BEGe background model were simulated using an
 implementation of the \gerda\ Phase~I detector array containing the three
 coaxial detector strings and an additional string with the five BEGe
 detectors.  The n$^{+}$ dead layer thicknesses used in the MC are
 listed in Table~\ref{tab:recval}.  The effective p$^{+}$ dead layer thickness
 was set to 600~nm.

 The minimum model contributions were considered.
 Additionally, two contributions were added to the BEGe model: $^{68}$Ge decays 
 in germanium
 and $^{42}$K decays on the n$^+$ surface.  A contribution from $^{68}$Ge is
 expected due to cosmogenic activation above ground, analogously to
 $^{60}$Co in germanium.  Due to the rather short half life of 271~d the
 $^{68}$Ge contribution can be neglected for the coaxial detectors, which have
 been stored underground for several years.  For the newly produced BEGe
 detectors, however, these decays and the subsequent decay of $^{68}$Ga have
 to be taken into account.  The contribution from $^{42}$K decays on the n$^+$
 surface, on the other hand, is enhanced with respect to the coaxial detectors
 due to the thinner dead layer and has to be taken into account for the model.
 The n$^+$ surface dead layer is partially
 active~\cite{aguayo_psa}, which in particular affects the detection
 efficiency for surface $\beta$ interactions. Thus, the MC simulation
 used for $^{42}$K on n$^+$ surface included an approximation of this effect.
 40\,\% of the dead layer thickness (as stated in Tab. 1) has been modeled with
 zero charge collection efficiency, the other part with a linearly increasing  
 charge collection efficiency from 0\,\% to 100\,\%.

 The contributions of $^{60}$Co and $^{68}$Ge to the model are limited to
 0.05~cts/day and 0.32~cts/day, respectively. The upper values for these 
 cosmogenically produced isotopes are derived from an assumed activation rate
 for these isotopes according to Ref.~\cite{cosmogenic_activation} and the 
 known histories of exposure to cosmic rays of the individual detectors.

\begin{figure*}[t]
\begin{center}
    \includegraphics[width=0.95\columnwidth]{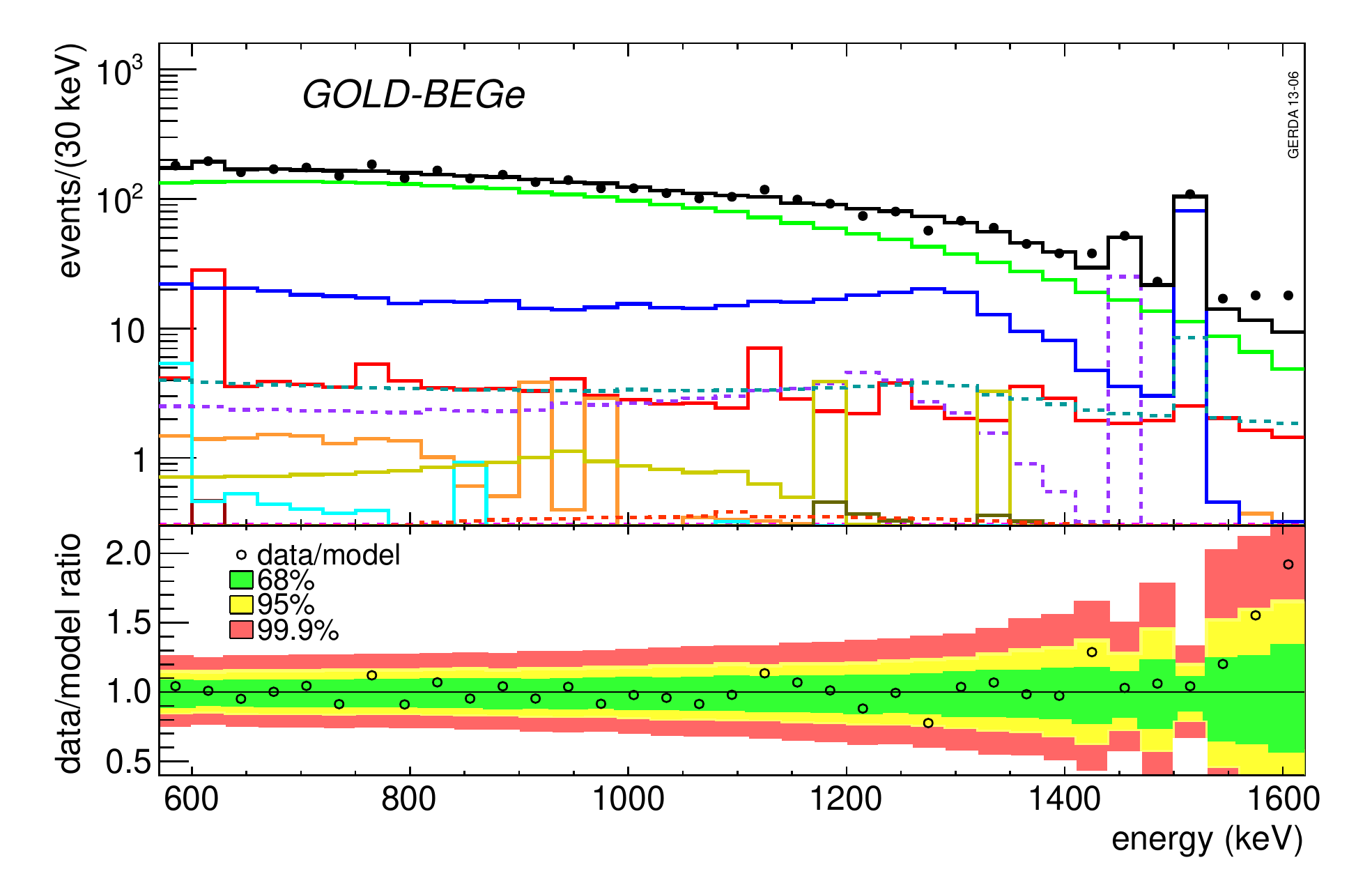}
    \includegraphics[width=0.95\columnwidth]{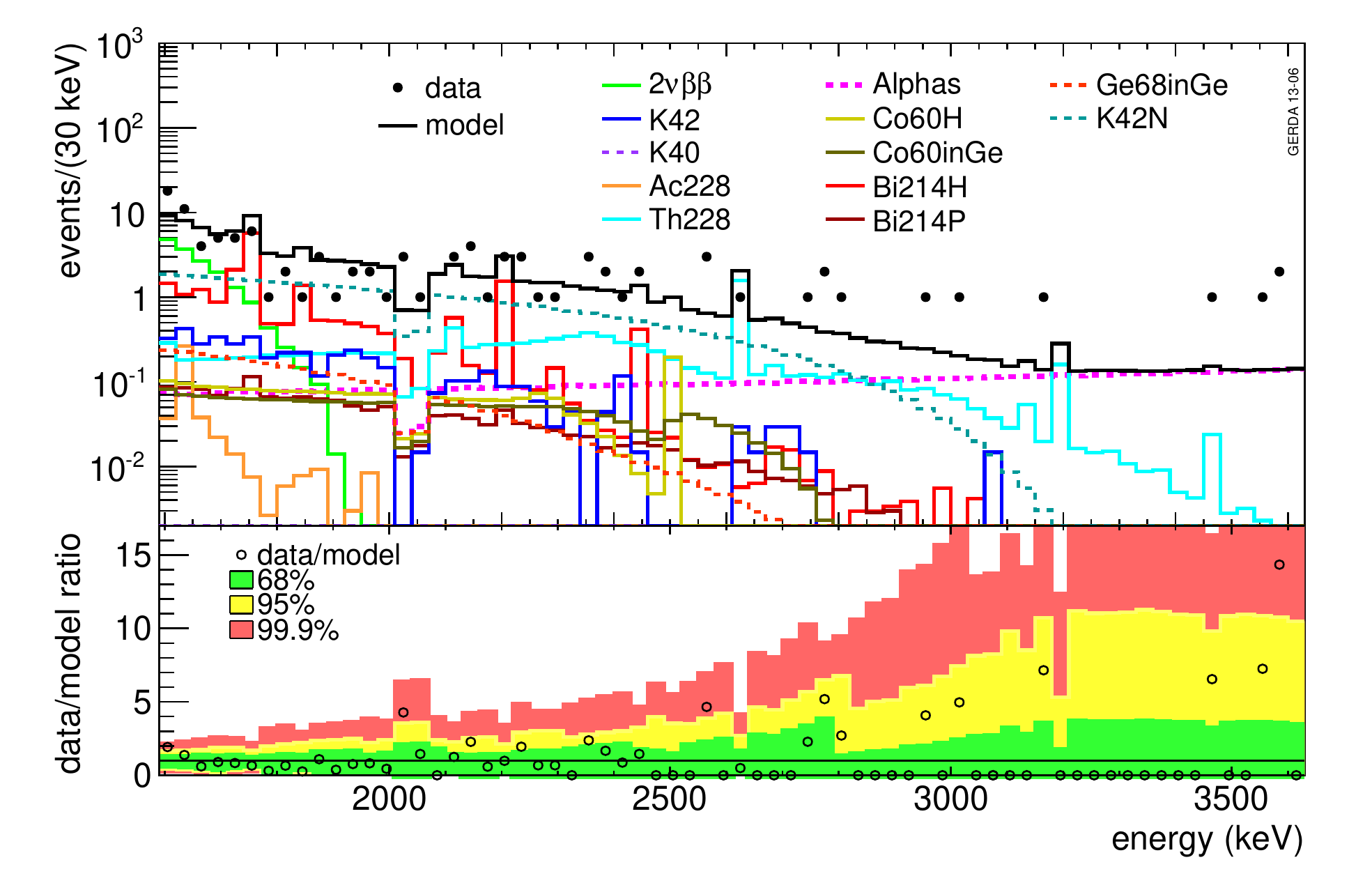}
\caption{\label{fig:BEGe_best_fit}
   Same as in Fig.~\ref{fig:min_fit_coax} for the {\it SUM-bege} data set.
}
\end{center}
\end{figure*}

 The procedure to obtain the best fit was equivalent to the model definition
 of the coaxial detectors.  The best fit model for BEGe detectors is shown in
 Fig.~\ref{fig:BEGe_best_fit}.  Around \qbb\ the largest contribution arise from
 $^{42}$K on the n$^+$ surfaces (see last column of 
 Table~\ref{tab:background_components}).

The presented BEGe background model is consistent with a background
decomposition obtained by pulse shape discrimination of the
data \cite{gerda_psd}.

\section{Cross checks of the background model}
 \label{sec:cross-checks}

 The background model developed has some predictive power that can be checked
 with the available data. This section describes cross checks performed on the
 background model.

\subsection{Half life derived for \twonu\ decay}

 From the best fit models the resulting half life~\thalftwo\ for
 $2\nu\beta\beta$ decay can be extracted.  
\thalftwo\ is calculated 
using the relation

\begin{equation}
{T^{2\nu}_{1/2}} = \frac{\ln2 \cdot N_A}{m_{enr}} \frac{\exposure}{N_{model}} \langle\varepsilon^{2\nu}\rangle~~~~,
\end{equation}
 where $N_{model}$ is the best fit number of \twonu\ decays derived from the individual
 model.  The efficiency $\langle\varepsilon^{2\nu}\rangle$ is given by the weighted detection
 efficiency of \twonu\ decays in the fit range, $\varepsilon_{i}$,

\begin{equation}
\langle\varepsilon^{2\nu}\rangle =\frac{\sum_if_{av,i}f_{76,i}M_it_i\varepsilon_{i}}{\exposure}~~~~.   
\end{equation}
 Table~\ref{tab:2vbb_half_lives} gives the half lives extracted from the
 different background models. All results are consistent with the earlier
 \GERDA\ \twonu\ analysis~\cite{2vbb} within the uncertainties. Note, that a
 three times larger exposure was available for this analysis as compared to the
 analysis in Ref.~\cite{2vbb} while systematic uncertainties are not considered.

\begin{table}[t]
\begin{center}
\caption{\label{tab:2vbb_half_lives}
     \thalftwo\ values derived from different background models. The
     uncertainties are those from the fit parameters 
     and do not include systematic uncertainties.
}
\vspace*{2mm}
\begin{tabular}{lrl}
model & \exposure\ [\kgyr]& \thalftwo $\cdot 10^{21}$yr \\
\hline
\up {\it GOLD-coax} minimum     &15.40 &1.92$^{+0.02}_{-0.04}$\\
\up {\it GOLD-coax} maximum     &15.40 &1.92$^{+0.04}_{-0.03}$\\
\up {\it GOLD-nat} minimum      & 3.13 &1.74$^{+0.48}_{-0.24}$\\
\up {\it SUM-BEGe}              & 1.80 &1.96$^{+0.13}_{-0.05}$\\
\up Analysis in Ref.~\cite{2vbb}& 5.04 &1.84$^{+0.09}_{-0.08~fit}~^{+0.11}_{-0.10~syst}$\\
\end{tabular}
\end{center}
\end{table}

\subsection{Intensities of $\gamma$ lines}
\begin{figure}[h!]
\begin{center}
  \includegraphics[width=0.95\columnwidth]{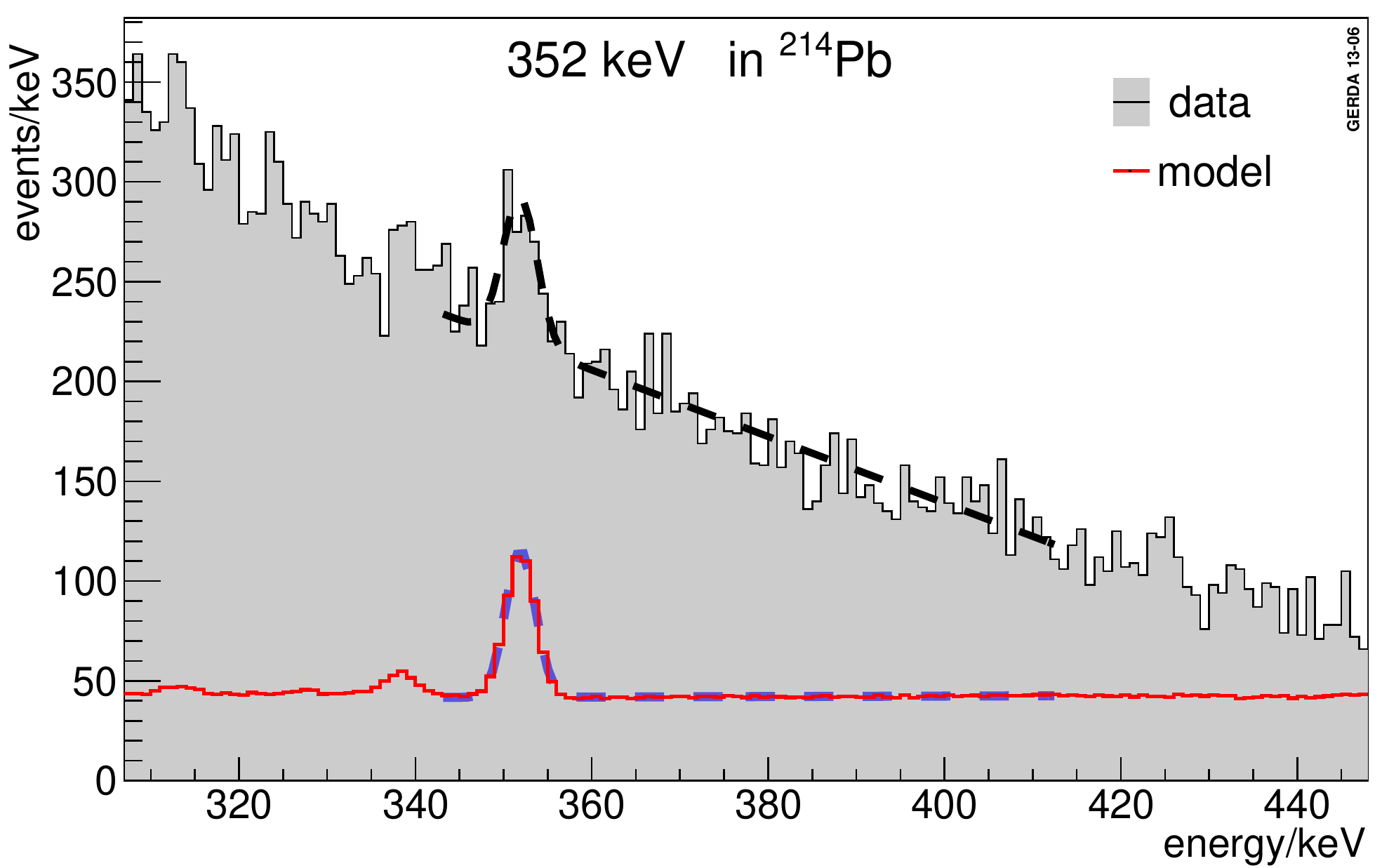}
  \caption{ \label{fig:352keV}
         Energy spectrum of the {\it GOLD-coax} data set (filled histogram)
         and the minimum model prediction (red histogram). The data and the
         model spectrum are fitted with a Gaussian plus linear background
         (dashed lines). 
 }
\end{center}
\end{figure}

 At energies below 600~keV the energy spectrum is dominated by $^{39}$Ar with an
 activity of  $A$\,=\,[1.01\,$\pm$\,0.02(stat) $\pm$\,0.08(syst)]~Bq/kg~\cite{39Ar}
 homogeneously distributed in LAr.  This part of the spectrum has not been
 included into the background 
 fit to avoid uncertainties due to the n$^+$ dead layer
 thickness and the theoretical shape of the beta decay spectrum.  A strong
 $\gamma$ line at 352~keV is, however, expected from decays of $^{214}$Bi in the
 vicinity of the detectors. The intensity of this line depends strongly on
 the distance of the $^{214}$Bi contamination from the detectors. Hence, this
 cross check can give a hint on how realistic the assumed distribution of the
 $^{214}$Bi contamination is.  The minimum (maximum) model predicts 20.1$^{+2.6}_{-2.2}$
 (17.5\,$^{+7.0}_{-13.3}$) counts/(kg$\cdot$yr) in the peak while a fit of a Gaussian plus a linear
 background to the data gives 20.4$^{+4.4}_{-4.2}$ counts/(kg$\cdot$yr) for the {\it GOLD-coax} data
 set.  Fig.~\ref{fig:352keV} shows the energy spectrum of the {\it GOLD-coax} data
 set in the energy region between 310 and 440~keV. The Gaussian plus linear
 background fit to the data as well as the minimum model prediction without
 the $^{39}$Ar contribution dominating the spectrum in this energy region is
shown.
For the {\it GOLD-nat} data set the minimum model prediction of
(22.1$^{+5.2}_{-4.6}$ counts/(kg$\cdot$yr) is also consistent with the observed (25.6 $^{+8.5}_{-7.5}$ 6.2) counts/(kg$\cdot$yr). This cross
 check makes it possible to distinguish between the locations of
 \Bi\ contaminations if it is assumed that the decays of \Bi\ and $^{214}$Pb
 happen at the same location. It excludes the results for $^{214}$Pb
 contamination in or on the radon shroud as the best fit maximum model for the {\it
   GOLD-nat} data set predicts. This model predicts only (4.6$^{+1.4}_{-1.5}$ counts/ (kg$\cdot$yr).
 This cross check confirms the indication from the background model 
 that close sources are responsible for most of the \Bi\ background contribution.

\begin{table*}
\begin{center}
\caption{   \label{tab:backgroundLines}
             Count rates of background \gam\ lines  for the {\it GOLD-coax} and
             {GOLD-nat} data sets obtained by the global analysis (minimum and
             maximum model) and with a fine binned fit of the data.  Upper
             limits correspond to 90\,\%~credibility interval. The central
             value is the mode of the posterior probability distribution
             function and the confidence intervals account for the smallest interval
             containing 68\,\% probability.
}
\begin{tabular}{r|r|ccc|ccc}
iso-& energy & \multicolumn{3}{c|}{\it GOLD-coax} & \multicolumn{3}{c}{\it
  GOLD-nat} \\
\cline{3-8}
tope &\up [keV]~ & \multicolumn{6}{c}{rate~~~ [cts/(kg$\cdot$yr)]}\\[1mm]
\cline{3-8}
&\up        &  Global &  Global  & Fit to&  Global &  Global & Fit to\\
&           & analysis&  analysis & data & analysis &  analysis & data\\
&           &(min. fit)& (max. fit) &  & (min. fit) &  (max. fit) & \\
\hline
$^{40}$K     & 1460.8   &  11.9 [10.6,13.6] & 11.8 [10.6,13.6] & 13.6[12.5,15.0]  & 19.4 [16.7,23.0] & 19.8 [16.7,23.1] & 18.3 [15.7,21.4]\\
$^{42}$K     & 1524.7   &  61.2 [59.3,63.8] & 61.4 [48.5, 66.2] & 60.3[58.1, 62.5]  & 75.0 [70.2, 82.4] & 75.6 [49.1, 78.3] & 73.8 [69.1, 80.1]\\

\hline
$^{60}$Co    & 1173.2 &  2.5 [1.6, 3.7]  &  1.6 [0.6, 2.9]  & 4.2 [2.8, 5.6]  & 2.0 [0.1, 4.6]  & 1.9 [0.1, 3.2]  & $<$\,3.8\\ 
     & 1332.3 &  2.3 [1.4, 3.4]  & 1.5 [0.5, 2.6]  & $<$\,1.6  & 1.9 [0.1, 4.4]  & 1.7 [0.1, 2.8]  & 3.3 [1.6, 5.2]\\
\hline
$^{228}$Ac  & 911.2 &  3.8 [2.2, 5.8]  & 3.4 [1.0, 3.6]  & 3.9 [2.2, 5.6]  & 6.1 [3.9, 8.6]   & 5.4 [0.5, 9.8]  & 4.9 [2.7, 7.3]\\ 
   & 969.0 &    2.7 [1.5, 4.1] & 2.8 [0.8, 2.8]  & 3.5 [1.8, 5.0]   & 4.4 [2.8, 6.2]  & 3.9 [0.4, 7.1]  & 4.8 [2.6, 7.4]\\ 
\hline
$^{208}$Tl & 583.2 & 2.9 [2.5, 3.5]  & 1.3 [0.3, 2.0]  & 3.9 [1.8, 5.7]  & 3.0 [1.8, 4.7]  & $<$\,0.9   & 2.5 [0.4, 4.6]\\
   & 2614.5 & 1.4 [1.2, 1.7]  & 1.0 [0.2, 1.5]  & 1.2 [0.9, 1.5]  & 1.5 [0.9, 2.4]  & 1.5 [0.0, 6.7]  & 1.4 [0.6, 2.3]\\ 
\hline
$^{214}$Pb  & 351.9 & 20.1 [17.9,22.7] & 17.5 [4.2,24.5] & 20.4 [16.2,24.8] & 22.1 [17.7,27.3] & 4.6 [3.1,6.0]  & 25.6 [18.1,34.1]\\
$^{214}$Bi  & 609.3 &   11.2 [10.0, 12.6] & 8.0 [2.0, 11.2]  & 10.0 [8.0, 12.3] & 11.2 [9.0, 13.8]  & 6.5 [4.6, 8.6]  & 7.6 [4.8, 11.0]\\ 
    & 1120.3 &  2.6 [2.3, 2.9]  & 1.8 [0.4, 2.5]   & $<$\,3.1   & 2.6 [2.1, 3.2]  & 3.0 [2.1, 3.9]  & 4.0 [1.8, 6.3]\\
    & 1729.6 &  1.0 [0.9, 1.1]  & 1.0 [0.2, 1.4]  & 0.5 [0.2, 0.9]  & 1.3 [1.0, 1.6]  & 0.8 [0.5, 1.1]  & 0.9 [0.3, 1.]\\ 
    & 1764.5 &  3.6 [3.2, 4.1]  & 2.7 [0.7, 3.8]  & 3.1 [2.6, 3.7]  & 4.1 [3.3, 5.1]  & 3.7 [2.5, 4.9]  & 3.5 [2.4, 5.0]\\ 
    & 1847.4 &  0.6 [0.5, 0.7]  & 0.6 [0.1, 0.8]  & 0.6 [0.3, 1.0]  & 0.7 [0.6, 0.9]  & 0.5 [0.3, 0.6]  & 1.2 [0.5, 2.1]\\ 
    & 2204.2 &  1.0 [0.9, 1.1]  & 0.8 [0.2, 1.1]  & 0.8 [0.5, 1.2]  & 1.2 [1.0, 1.5]  & 1.3 [0.9, 1.7]  & 0.8 [0.2,1.6]\\
    & 2447.9 &  0.3 [0.27, 0.34]  & 0.2 [0.05, 0.3]  & 0.2 [0.1, 0.4]  & 0.3 [0.2, 0.4]  & 0.4 [0.3, 0.5]  & $<$\,1.8\\

\end{tabular}
\end{center}
\end{table*}

 As the fit has been performed with a binning larger than the energy
 resolution of the detectors, the information from the line intensities is not
 maximized in the fitting procedure. Hence, it is instructive to cross check
 the line intensities obtained from fitting the peaks with a Gaussian plus
 linear background in the different data sets with the expectation from the
 models. Table~\ref{tab:backgroundLines} compares the $\gamma$-line
 intensities from the minimum and maximum models to those obtained from a fine
 binned analysis, i.e. a fit to data.  Note, that for some of the $\gamma$
 peaks no fit could be performed due to limited number of events in the peak
 region or the low intensity of the $\gamma$ line compared to the other
 background contributions.  In those cases the number of counts in the
 $\pm$\,3$\sigma$ energy range around the peak positions were used.  The
 background has been estimated according to the continuum seen in the
 $\pm$\,5$\sigma$ side bands at lower and higher energies around the peak. A
 narrower side band is used when there is a second line close to the peak. The
 intensities of the $\gamma$ lines are obtained by marginalizing the posterior
 probability distribution of the signal rate.
The uncertainties on the predicted rate by the global models are 
due to the fit uncertainty on the parameters of the model components 
that give contribution to the $\gamma$-ray line. The statistical uncertainties 
due to the simulated number of events is on the order of 0.1\%, 
i.e. negligible compared to the fit uncertainty.
There is excellent agreement between the numbers from the global
 analysis and those from the fine-binned analysis.

\subsection{Stability of the fit}
 To check for stability, the fits were performed using different binnings. As
 the energy resolution of the detectors is around 4.5~keV at \qbb\ and the
 calibration at higher energies E\,$>$\,5~MeV, relevant for the $\alpha$ model is precise to
 about 10~keV, the lowest binning chosen was 10~keV. Also a 50~keV binning was
 performed. The activities of different components derived from the fits with
 different binnings do not vary outside the uncertainties given for the 30~keV
 binning fits.

 Additionally it was checked whether the overall goodness of fit and the
 predicted BI and individual contributions in the region of interest changes
 if biases are introduced to the fits by single strong assumptions on
 individual background components.

 Minimum model fits for the {\it GOLD-coax} and {\it GOLD-nat} data sets were
 performed with the following individual modifications: $^{228}$Th and
 $^{228}$Ac are only in or on the radon shroud; no $^{214}$Bi is present on the p$^{+}$
 surface; $^{42}$K is only on the p$^{+}$ surface; $^{42}$K is present on the
 p$^{+}$ surface; $^{60}$Co is only inside the crystal, $^{60}$Co is only
 inside the detector assembly.  Except for the unrealistic assumption that all $^{42}$K 
 comes from p$^{+}$ surface contaminations all fits describe the measured spectrum reasonably well.

 The prediction for the BI at the region of interest varies by 10\,\%
 between the different models for the  {\it GOLD-coax} data set and
 by 15\,\% for the {\it GOLD-nat} data set.

 The predictions for the activities of the individual components of the
 different models are consistent within the 68\,\% uncertainty range quoted in
 Table~\ref{tab:fits_results_golden}.

\subsection{BiPo coincidences}
 An important contribution to the background model are surface events from
 \Ra\ daughters. These include the decays of $^{214}$Bi and
 $^{214}$Po. $^{214}$Po has a half life of only 164.4~\mus. Hence, a number of
 events is expected where a low energy event from the $^{214}$Bi decay is
 followed by a high energy $\alpha$ event from the $^{214}$Po decay in the
 same detector, a BiPo tag.  The data reduction includes cuts that remove pile
 up events. As the half life of $^{214}$Po is of the order of the decay time
 of the preamplifier signal, the analysis is not tailored to reconstruct
 typical BiPo events. Hence it is difficult to quantify precisely the
 efficiency for this tag. Nevertheless, for the purpose of a qualitative
 statement an order of magnitude guess is made: An efficiency of 50\,\% is
 assumed for the BiPo recognition efficiency. Using the number for the {\it
   GOLD-coax} data set obtained by the marginalized probability density
 function of the fit the number of detected $^{218}$Po surface events is 13.5 being
 reduced to approximately seven events from $^{214}$Bi decays on the surface
 that can lead to energy deposition in the detector active volume (the
 decrease being due to the decay nucleus recoiling away from the surface). With
 an efficiency of the order of 50\,\% to detect the BiPo tag only roughly 3 to
 4 BiPo events are expected. In the {\it GOLD-coax} data set in total 5 events
 have been found (2 in ANG~2, 2 in ANG~3 and 1 in RG~1) that satisfy the
 criteria for a BiPo tag.

\subsection{Recognition of p$^{+}$ events}
 A mono-parametric pulse shape analysis technique for the identification of
 surface interactions on the p$^+$ electrode of coaxial detectors has been
 recently developed and applied on Phase~I data~\cite{matteo_thesis}.  The
 method is based on a cut on the rise time of the charge pulses computed
 between 5\,\% and 50\,\% of the maximum amplitude. The cut level is calibrated
 on experimental data using the pure sample of high-energy $\alpha$-induced
 events.

 The analysis has been extended to the entire {\it GOLD-coax} data set.
 Fixing the cut to accept 95\,\% of the events occurring in the proximity of the
 $p^+$ electrode, in the energy region of interest 43\,\% of the events survive
 the cut.  Part of the events surviving the cut is expected to be due to
 $\gamma$ interactions in the proximity of the $p^+$ surface.  Applying the
 corrections described in Ref.~\cite{matteo_thesis}, the total amount of
 $\alpha$ and $\beta$ induced events on the p$^+$ electrode is estimated to be
 between 15\,\% and 35\,\% of the number of events in the energy region of
 interest.

 This result is consistent with the number of decays on the p$^+$ surface
 predicted by the minimal model (20.5\,$\pm$\,2.7)\,\%, given by the $\alpha$
 emitting isotopes plus $^{214}$Bi. It is slightly lower compared to the
 maximal background model that requires 50\,\% considering $\alpha$,
 $^{214}$Bi and $^{42}$K on the p$^+$ surface.

\subsection{Coincident spectrum}
 As a large fraction of the contaminations are, according to the background model(s),
 located inside the detector array (i.e. in the detector assembly, or in LAr
 close to the surfaces of the detectors), a significant number of events are
 expected to have coincident hits in two detectors. The efficiency to detect
 coincident events is expected to be increased with respect to single $\gamma$
 emitters for decays of isotopes with emission of multiple $\gamma$ rays such as
 $^{42}$K, $^{60}$Co, $^{214}$Bi and $^{208}$Tl. Coincident spectra are, thus,
 sensitive to differences in source locations.  

 A sum coincidence spectrum was produced for the {\it GOLD-coax} data set by
 summing the energies of all detectors in an event
 and filling the corresponding bin of
 the histogram. Also a single coincidence spectrum was produced by filling the
 corresponding bin of the histogram for each individual detector separately.

 The same procedure as for the minimum fit model (see sec.~\ref{sec:modeling})
 was applied to get best fit coincidence models for the single and sum
 spectra. The results for the activities obtained from the minimum model best
 fit to the sum spectrum (except for $^{60}$Co, where the single spectrum was
 used) are summarized in Table~\ref{tab:fits_results_golden}.  The obtained
 activities from coincident and single detector spectra are consistent with
 each other. Note, that the simulations were not tuned for the coincidence
 analysis. The background source distribution was simplified in the
 simulation, while small changes in source location, especially within the
 detector array, can have significant effects on the coincidence
 efficiencies. The fact that the $^{42}$K activity derived from the
 coincidence fit is slightly higher than for the minimum and maximum models of
 the {\it GOLD-coax} and {\it GOLD-nat} data sets may be a hint that the
 distribution of $^{42}$K in LAr is not homogeneous.  The consistency between
 the derived activities from coincident and single detector spectra support
 the result of the background model that the spectrum around \qbb\ is
 dominated by contaminants close to the detectors.

\section{Background prediction at \qbb\ and expected sensitivity for
  {\sc Gerda} Phase~I}
   \label{sec:extrapol}

\subsection{Background prediction at  \qbb}
\label{ssec:backqbb}

\begin{table*}[t]
\begin{center}
\caption{\label{tab:background_components}
         The total BI and individual contributions in 10~keV
         (8~keV for BEGes) 
         energy window around \qbb\ for different models and data sets. Given
         are the values due to the global mode together with the uncertainty
         intervals obtained as the smallest 68\,\% interval
        of the marginalized
         distributions. Limits are given with 90\,\% C.L. For details see the text.
}
\begin{tabular}{ll|rcrc|rc|rc|c}
        &  & \multicolumn{4}{c|}{\it GOLD-coax}&
 \multicolumn{2}{c|}{\it GOLD-nat} & \multicolumn{2}{c}{\it SUM-bege} & Exp. from\\
component & location & \multicolumn{2}{c}{minimum model} &\multicolumn{2}{c}{maximum model} & \multicolumn{2}{c}{minimum model}  & \multicolumn{2}{c}{minimum + n$^+$} & screening\\
          && \multicolumn{8}{c}{  BI ~~~~\dctsper}\\

\hline
\up
Total          &              &18.5 &[17.6,19.3]&21.9& [20.7,23.8]& 29.6 &[27.1,32.7] & 38.1& [32.2,43.3] &\\
\hline
\up
$^{42}$K & LAr homogeneous & 3.0& [2.9,3.1] &  2.6 &[2.0,2.8] & 2.9 &[2.7,3.2] & 2.0& [1.8,2.3] & --\\
$^{42}$K & p$^+$ surface     &    &           &  4.6 &[1.2,7.4] &     &   && & -- \\
$^{42}$K & n$^+$ surface     &    &           &  0.2 &[0.1,0.4] &     &          & 20.8 &[6.8,23.7] & --\\
$^{60}$Co & det. assembly              & 1.4& [0.9,2.1] &  0.9 &[0.3,1.4] & 1.1 &[0.0,2.5] &  &  $<$\,4.7 & --\\
$^{60}$Co & germanium        & 0.6 &  $>$\,0.1 $^\dagger$)  &  0.6  & $>$\,0.1 $^\dagger$)  & 9.2 &[4.5,12.9]& 1.0& [0.3,1.0] & --\\
$^{68}$Ge & germanium        &    &           &      &          &     &          &    & 1.5 ($<$\,6.7) & --\\
$^{214}$Bi & det. assembly        & 5.2& [4.7,5.9] &  2.2 &[0.5,3.1] & 4.9 &[3.9,6.1] & 5.1 &[3.1,6.9] & $\approx$2.8\\
$^{214}$Bi & LAr close to p$^+$     &    &           &  3.1 & $<$\,4.7   &     &     && & $<$\,0.7 \\
$^{214}$Bi & p$^+$ surface   & 1.4& [1.0,1.8] $^\dagger$)&  1.3 &[0.9,1.8] $^\dagger$)& 3.7 &[2.7,4.8] $^\dagger$)& 0.7 &[0.1,1.3] $^\dagger$) & --\\
$^{214}$Bi & radon shroud       &    &           &  0.7  & $<$\,3.5   &     &     && & --\\
$^{228}$Th & det. assembly         & 4.5& [3.9,5.4] &  1.6 &[0.4,2.5] & 4.0 &[2.5,6.3] & 4.2 &[1.8,8.4] & $<$\,0.3\\
$^{228}$Th & radon shroud *)  &    &           &  1.7  & $<$\,2.9   &     &     && & $\approx$1.0\\
$\alpha$ model  & p$^+$ surface and & 2.4&[2.4,2.5]  &  2.4 &[2.3,2.5] & 3.8 &[3.5,4.2] & 1.5 &[1.2,1.8] & --\\
                & LAr close to p$^+$ &&&&&&&&&\\
\end{tabular}
\end{center}
$^\dagger$) prior: discussed in sec.~\ref{sec:modeling}\\
$*$) Representing all distant sources including the heat exchanger, the wall of the steel cryostat and the calibration source at the bottom of the tank. 

\end{table*}

\begin{figure*}[ht!]
\begin{center}
    \includegraphics[width=1.9\columnwidth]{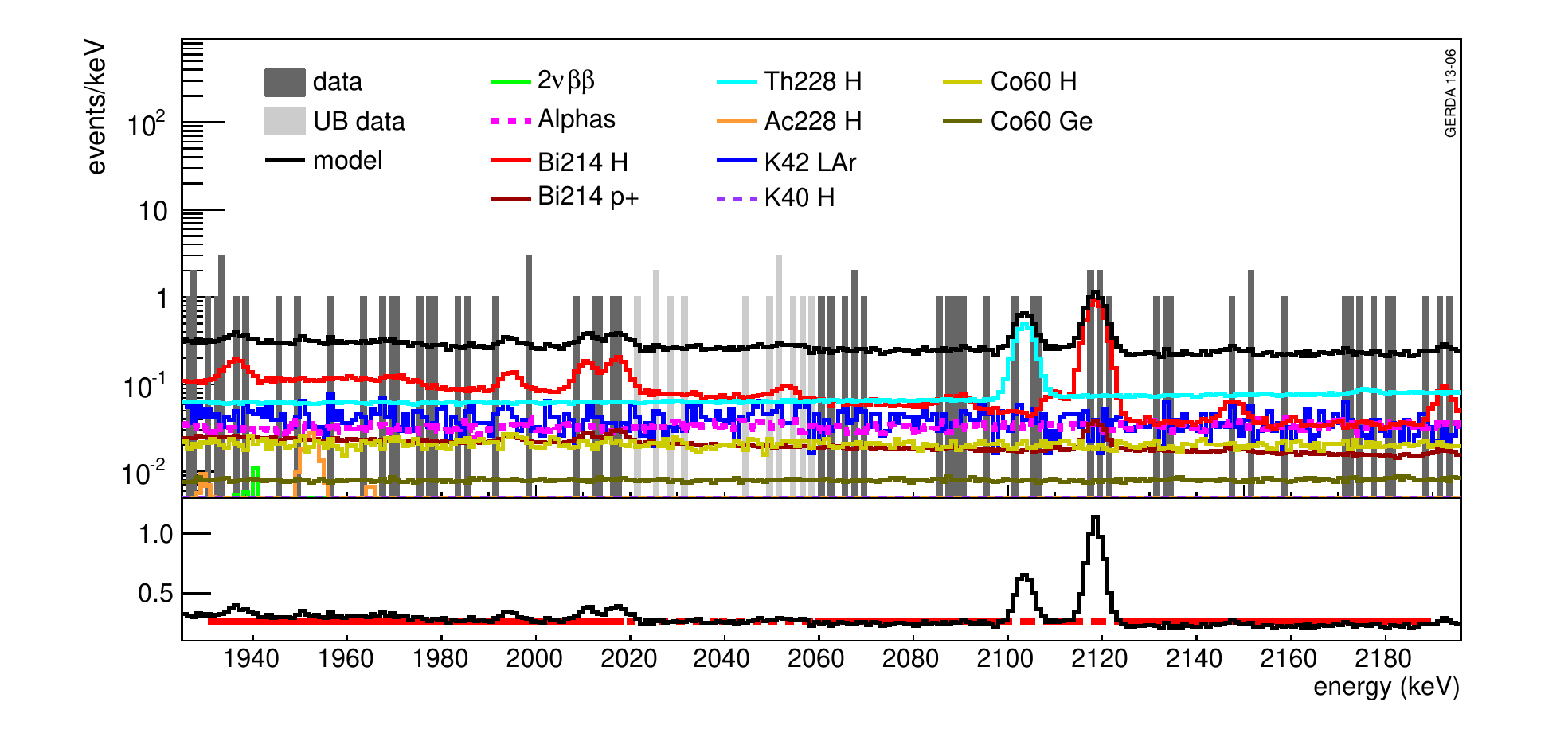}
    \includegraphics[width=1.9\columnwidth]{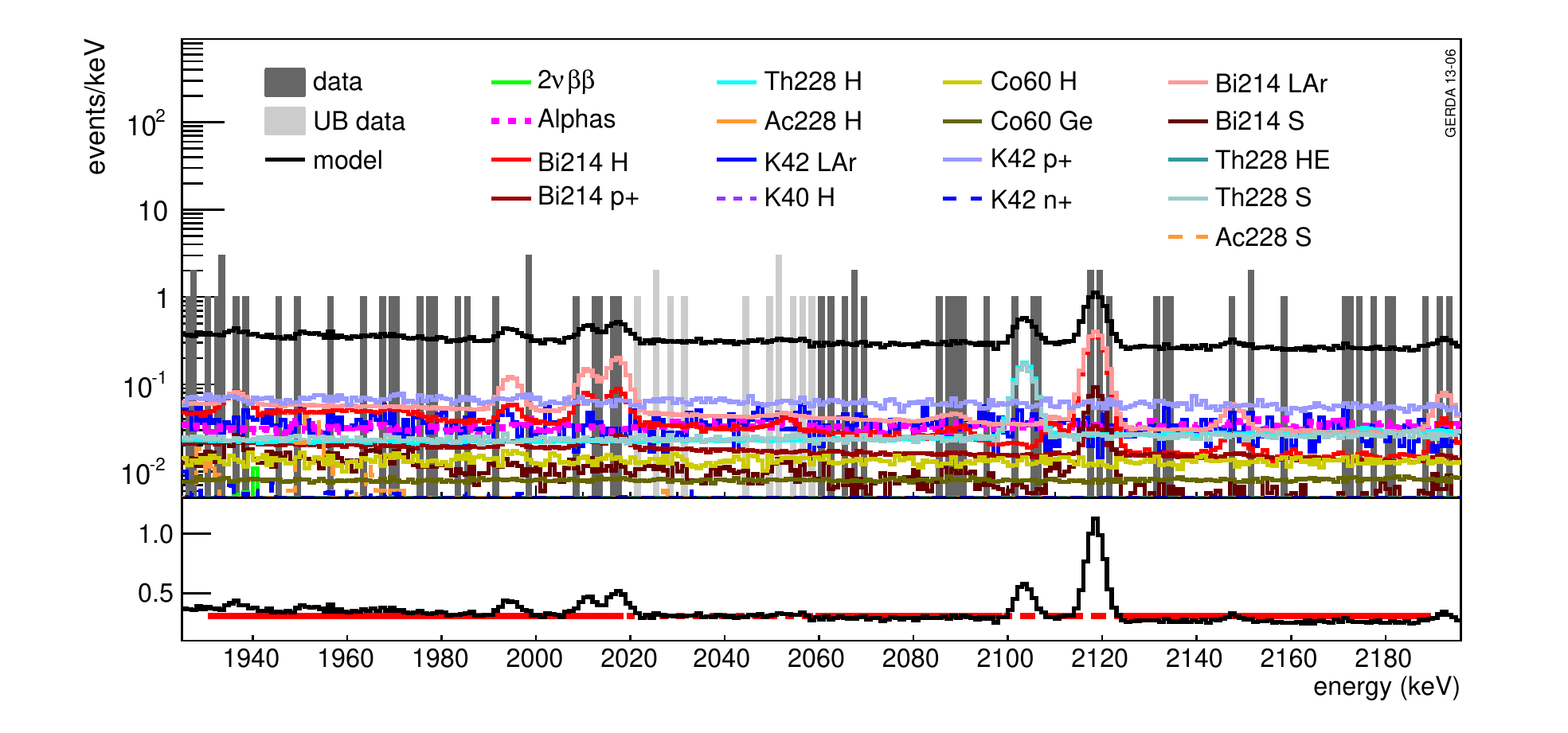}
\caption{\label{fig:0nbb_region}
          Experimental spectrum with minimum (upper plot) and maximum (lower
          plot) models around \qbb\ for the {\it GOLD-coax} data set. The
          upper panels show the individual contributions of the considered
          background sources to the total background spectrum in logarithmic
          scale. The lower panels show the best fit models fitted with a
          constant. In the fit the peak areas predicted by the model and the
          40~keV blinding window are not considered. The light grey shaded
          (unblinded data, UB data) events in the experimental spectrum have
          not been used in the analysis.
}
\end{center}
\end{figure*}

 The background models obtained by global fits in the 570 - 7500~keV region allow
 to predict the individual background contributions and the total background
 at \qbb. Table~\ref{tab:background_components} lists the 
 predictions for the BI from different contributions in a 10~keV window for coaxial
 detectors and in a 8~keV window for BEGe detectors around \qbb\ for different data
 sets. The results obtained from the best fit parameters are quoted together with
the smallest 68\,\% interval of the marginalized distributions of the parameters. 
 If the maximum of the marginalized distribution is at zero a 90\,\% upper limit is given. For the case of an internal $^{60}$Co contamination a 90\,\% lower limit is given, because a higher contamination gives a better fit, however, is constrained by prior knowledge of above ground exposure to cosmic rays.
 According to the models the main contributions to the
 background at \qbb\ are due to the $\alpha$-emitting isotopes in the
 \Ra\ decay chain, \kvz, \Co, \Bi\ and \Th. The fraction with which each
 component contributes depends on the assumed source location.

Tab. \ref{tab:background_components} also lists the BIs expected from the
screening measurements as reported in Tab. \ref{tab:thorium}.
The BIs due to the individual identified components  
do not match well with the BIs derived from the background model, indicating that the
 unidentified close-by  $^{214}$Bi and $^{228}$Th contributions have to be present.

 Fig.~\ref{fig:0nbb_region} shows the best fit minimum and maximum models and
 the individual contributions together with the observed spectrum around
 \qbb\ for the {\it GOLD-coax} data set. The spectral shapes of the best fit
 models are constant around \qbb. No peaks are predicted in the
 blinded regions. This indicates that the BI at \qbb\ can be estimated by
 interpolating the results of a fit to the observed number of events outside
 the signal search window. The window to be used for this estimation is chosen
 as sum of the 1930--2019~keV, 2059--2099~keV, 2109--2114~keV and
 2124--2190~keV intervals for a total width of 200~keV. The BI evaluation
 window excludes the central 40~keV window around \qbb\ and the regions within
 $\pm$\,5~keV from the $\gamma$ lines expected from the background
 model -- namely, single escape peak from \Tl\ at 2104~keV and the
 \Bi\ $\gamma$ line at 2119~keV. The resulting background indices from the
 interpolation are listed in Table~\ref{tab:extrapolated_bi} for different
 data sets together with the prediction of the background models for
 comparison. The lower panels of the plots in Fig.~\ref{fig:0nbb_region}
 demonstrate that the background model can be described by a constant in the
 BI evaluation windows.  The difference in the resulting BI is less than 1\,\%
 if a linear instead of a constant distribution is assumed.  The statistical 
 uncertainty for the
 approximation of the BI by an interpolation is of the same size as the
 systematic uncertainty expected by the model.
 
\begin{table*}
\begin{center}
\caption{\label{tab:extrapolated_bi}
       BI in the RoI 
       as predicted by the minimum and maximum models as well as by
       interpolation from a 200~keV wide
       window around \qbb.
       Comparison of counts in the previously blinded window (width differs
       for different data sets) and model predictions is also given. Values in
       the parentheses show the uncertainty interval.
}
\begin{tabular}{lccc}
\hline 
                      &{\it GOLD-coax} & {\it GOLD-nat}  & {\it SUM-bege}\\
\hline
\up
  & \multicolumn{3}{c}{BI in central region around \qbb\ (10~keV for coaxial, 8~keV for BEGe)}\\
  &  \multicolumn{3}{c}{10$^{-3}$ cts/(kg keV yr)}\\
\hline
interpolation	&	17.5 [15.1,20.1]&30.4 [23.7,38.4] & 36.1 [26.4,49.3]\\
minimum		&	18.5 [17.6,19.3]&29.6 [27.1,32.7] &  38.1 [32.2,43.3] \\
maximum		&	21.9 [20.7,23.8]&37.1 [32.2,39.2] & \\
\hline
\up         & \multicolumn{3}{c}{background counts in the previously blinded energy region}\\
         & 30~keV  & 40~keV   & 32~keV\\
\hline
data		&	13		&	5	   & 2 \\
minimum		&	8.6 [8.2,9.1]	&	3.5 [3.2,3.8]	& 2.2 [1.9,2.5]	\\
maximum		&10.3 [9.7,11.1]	&	4.2 [3.8,4.6]	 & \\
\hline
\end{tabular}
\end{center}
\end{table*}

 The global fits were performed by excluding the central 40~keV region around
 \qbb\, which was completely blinded until May 2013. Thereafter, a 30~keV
 (32~keV) window was opened for analysis by keeping the central 10~keV (8~keV)
 window still blinded for the enriched coaxial (BEGe) detectors. The
 natural detector GTF~112 was completely unblinded. The first step of
 unblinding gives the possibility to compare the model predictions to the
 observed number of events in those regions as a consistency check for the
 model.  Table~\ref{tab:extrapolated_bi} also lists the predicted and observed
 number of events in these energy regions for different data sets. In total 13
 events were observed in the unblinded 30~keV window of the {\it GOLD-coax}
 data set. The predictions in this window were 8.6 events from minimum and
 10.3 events from maximum model. The probability to observe 13 events or more
 given the predictions are 10\,\% and 24\,\%, respectively. In the {\it GOLD-nat} data set 5
 events were found in the 40~keV unblinded window, resulting in a 27\,\%
 probability for the minimum model prediction of 3.5 events and a 41\,\%
 probability for the maximum model prediction of 4.2 events. For the {\it SUM-bege} data set there is a perfect agreement between the observed two events and the expectation of 2.2 events from the model.


 If the additional events seen in the 30~keV unblinded window are included to
 the interpolation the expected BI at \qbb\ increases  to
 19$\cdot$\pIIbi\ for the {\it GOLD-coax} data set.


\subsection{Sensitivity for {\sc Gerda} Phase~I}
\label{ssec:sensitivity}

 Given, for the {\it GOLD-coax} data set, the background prediction of the
 minimum model of 18.5$\cdot$10$^{-3}$ \ctsper\ and the known
 \GERDA\ 17.90~\kgyr\ exposure at the end of Phase~I, the sensitivity for
 the $0\nu\beta\beta$ decay half life~\thalfzero\ was calculated.  
The value of the exposure-averaged total
 efficiency (see eq. \ref{equ:efficiency}) for the {\it GOLD-coax} data set is
 $\langle\varepsilon\rangle\,=\,$0.688.

From the energy spectrum an upper limit on the $0\nu\beta\beta$ signal
  strength $N_{up}$  at specified probability or confidence level can be derived and converted to a half life limit
  $T_{1/2}^{0\nu}$ using
\begin{equation}  
\mbox{\thalfzero} ~~>~~
   \frac{\ln2\cdot N_A}{m_{enr}}\,\, \frac{\cal E}{N_{up}}\, \,
   \langle\varepsilon\rangle\, \,.
\end{equation}

 In order to estimate the limit setting sensitivity without pulse shape analysis, 10$^4$ MC
 realizations of \gerda\ were generated assuming no \onbb\ signal.  For each
 realization, the number of events was allowed to fluctuate according to a
 Poisson distribution with expectation given by the number of predicted
 background events.

 The expected lower limit for \thalfzero\ was estimated by using both, Bayesian
 and Frequentist analyses.  In both analyses the signal and background
 strengths were free parameters. For the Frequentist analysis, the
 \onbb\ decay rates were estimated from a profile likelihood fit to the
 unbinned energy spectrum of each realization. The  90\,\% C.L. lower
 limit  \thalfzero\ $>$\,1.9$\cdot$10$^{25}$~yr (90\,\% C.L.) corresponds to the median of the 90\,\% quantile of the
 profile likelihood.
 In the Bayesian analysis, the 90\,\% probability lower limit for
 \thalfzero\ was calculated as the median of the 90\,\% quantiles of the
 posterior marginalized probabilities p$(T^{0\nu}_{1/2}\big
 |\mbox{spectrum},\bar{H})$, where $\bar{H}$ is the hypothesis that both
 background and $0\nu\beta\beta$ events contribute to the spectrum.  The
 result is \thalfzero\ $>$\,1.7$\cdot$10$^{25}$~yr (90\,\% C.L.).  The difference
 in the numerical values from the Bayesian and Frequentist analysis (which
 have conceptually a different meaning) is mainly due to the behavior of the
 two approaches in the cases  when the
 number of observed counts is smaller than the background expectation. The
 \GERDA\ sensitivity is expected to be about 10\,\% better than calculated from
 the {\it GOLD-coax} data set only, because of the extra exposure available in
 the {\it SILVER-coax} and {\it SUM-bege} data sets. Also the sensitivity
 might further increase by applying pulse shape discrimination techniques to
 the {\it GOLD-coax}, {\it SILVER-coax} and {\it SUM-bege} data sets  \cite{gerda_psd}.

\section{Conclusions}
   \label{sec:conclusion}

 The background measured by the \GERDA\ experiment has been presented in an
 energy range between 100 and 7500~keV. It has been demonstrated that stable
 low background data taking with the innovative technique of operating bare
 HPGe detectors in a cryogenic liquid is possible over a time period of about
 1.5~yr.  More than 20~\kgyr\ of data have been acquired by the
 \GERDA\ experiment with six enriched coaxial detectors of a total mass of
 14.6~kg and with four enriched BEGe detectors of a total mass of 3.0~kg.
  A background model has been developed  with the  $\Delta E$\,=\,40~keV  blinded that allows
 to predict the BI in this energy range. The predictions of the models have been  
 tested for consistency on a 30~keV (32~keV) range for the coaxial (BEGe) 
 detectors, 
 while the central $\Delta E$\,=\,10~keV region of interest  and $\Delta E$\,=\,8~keV 
 region for the BEGe detectors around \qbb\ was
 still blinded.   The model describes the background in an
 energy range from 570 to 7500~keV well. The only significant background
 contributions in \GERDA\  originate from decays of $^{42}$K in the LAr bath, from  \Bi\ in
 the detector assembly, from residual \Rn\ dissolved in LAr, from \thzza\ and $^{60}$Co in the
 detector assembly, and from surface $\alpha$ particles. The largest
 contributions come from contaminants located close to the detectors. Several
 cross checks confirm the validity of the background model.  
The 68\,\% credibility intervals of the BI expected due to the minimum and maximum models at \qbb\ of $^{76}$Ge span the range between 17.6 and 23.8$\cdot$\pIIbi. This range includes the systematic uncertainty due to different source location assumptions.

 Predictions for the number of events in the blinded region around \qbb\ have
 been made.  It could be shown that the expected background is flat in a
 region of $\approx$~200~keV around \qbb\ and that no significant peak like structures are
 expected in the blinded energy region. The background model and an
 interpolation of a fit to data from a 200~keV energy window into the
 blinded energy window give compatible results.  

 The BI interpolated into the region of interest are\\
 (1.75$^{+0.26}_{-0.24}$)$\cdot$\tctsper\ for the coaxial detectors and
 (3.6$^{+1.3}_{-1.0}$)$\cdot$\tctsper\ for the BEGe detectors.  The
 statistical uncertainty on the BI prediction from interpolation is of the
 same size as the systematic uncertainty from the choice of the background
 model.

 The BI obtained from interpolation of the spectrum in a 200~keV
  window around $Q_{\beta\beta}$ will be used in the
 \onbb\ analysis of the Phase~I data.
Given the expected background rate  without pulse shape discrimination and assuming no signal, the sensitivity for the {\it GOLD-coax} data set is 
 $T_{1/2}^{0\nu} >\,1.9\cdot 10^{25}$~yr (90\,\% C.L.) using a
  profile likelihood fit and
  $T_{1/2}^{0\nu} >\,1.7\cdot 10^{25}$~yr (90\,\% C.I.) using
  a Bayesian analysis.


\section*{Acknowledgments}
 The \gerda\ experiment is supported financially by
   the German Federal Ministry for Education and Research (BMBF),
   the German Research Foundation (DFG) via the Excellence Cluster Universe,
   the Italian Istituto Nazionale di Fisica Nucleare (INFN),
   the Max Planck Society (MPG),
   the Polish National Science Centre (NCN),
   the Foundation for Polish Science (MPD programme),
   the Russian Foundation for Basic Research, and
   the Swiss National Science Foundation (SNF).
 The institutions acknowledge also internal financial support.

The \gerda\ collaboration thanks the directors and the staff of the LNGS
for their continuous strong support of the \gerda\ experiment.

\end{document}